\RequirePackage{ifpdf}
\ifpdf
\else
\PassOptionsToPackage{dvipdfmx}{graphicx}
\PassOptionsToPackage{dvipdfmx}{hyperref}
\fi

\documentclass[12pt,a4paper]{article}
\pdfoutput=1 
\usepackage{jheppub}
\usepackage{graphics, color}
\usepackage{bm}
\usepackage{subfigure}
\usepackage[utf8]{inputenc}

\allowdisplaybreaks[1]

\def \dd {\mathrm{d}}
\def\ie{{\it i.e.} }
\def\eg{{\it e.g.} }

\title{
Superradiance and black resonator strings encounter helical black strings
}

\author[1]{\'Oscar J.C.~Dias,}
\author[2]{Takaaki~Ishii,}
\author[3]{Keiju~Murata,}
\author[4]{Jorge~E.~Santos,}
\author[5]{Benson~Way}

\affiliation[1]{STAG research centre and Mathematical Sciences, University of Southampton, University Road, Southampton SO17 1BJ, UK}
\affiliation[2]{Department of Physics, Rikkyo University, Nishi-Ikebukuro, Tokyo 171-8501, Japan}
\affiliation[3]{Department of Physics, College of Humanities and Sciences, Nihon University, Sakurajosui, Tokyo 156-8550, Japan}
\affiliation[4]{DAMTP, Centre for Mathematical Sciences, University of Cambridge, Wilberforce Road, Cambridge CB3 0WA, UK}
\affiliation[5]{Departament de F\'{i}sica Qu\`{a}ntica i Astrof\'{i}sica, Institut de Ci\`{e}ncies del Cosmos\\
Universitat de Barcelona, Mart\'{i} i Franqu\`{e}s, 1, E-08028 Barcelona, Spain}
\emailAdd{ojcd1r13@soton.ac.uk}
\emailAdd{ishiitk@rikkyo.ac.jp}
\emailAdd{murata.keiju@nihon-u.ac.jp}
\emailAdd{jss55@cam.ac.uk}
\emailAdd{benson@icc.ub.edu}

\abstract{%
We construct a cohomogeneity-1 helical black string in six-dimensional Einstein gravity. The helical solution branches from the onset of the gravitational superradiant instability of the equal-spinning Myers-Perry black string.  The isometry group of the helical black string is $\mathbb{R}_T \times U(1)_Z \times SU(2)$, where the first two are helical isometries generated by linear combinations of time translation, shifts along the string, and rotation, each of which is individually broken by the superradiant instability. The helical black string is stationary, non-axisymmetric, 
and has nonzero horizon velocity despite the absence of momentum in the string direction. The entropy of the helical black string is higher than that of the Myers-Perry black string, but lower than cohomogeneity-2 ``black resonator strings'' (recently found) when the solutions overlap in the microcanonical ensemble. The entropy of the helical black string approaches zero when the horizon velocity along the string reaches its maximum given by the speed of light. Nevertheless, we find no evidence for the existence of regular horizonless solutions in this limit.
}

\preprint{RUP-23-1}

\begin{document}
\maketitle

\section{Introduction}
\label{sec:introduction}

Black strings \cite{Horowitz:1991cd} have been studied over the past several decades as models for understanding the behaviour of black holes in higher dimensions. One of the most striking properties of black strings is the Gregory-Laflamme instability \cite{Gregory:1993vy,Gregory:1994bj}, where black strings are unstable under perturbations that break translation symmetry along the string direction. The Gregory-Laflamme instability also affects other higher dimensional black holes such as black rings \cite{Emparan:2001wn,Santos:2015iua,Figueras:2015hkb} and ultraspinning Myers-Perry black holes \cite{Dias:2009iu,Dias:2010maa,Dias:2010eu,Dias:2010gk,Dias:2011jg,Emparan:2003sy,Dias:2014cia,Emparan:2014pra,Dias:2015nua}. The nonlinear dynamics of the Gregory-Laflamme instability was studied in \cite{Lehner:2010pn,Figueras:2022zkg}, where evidence suggests that the instability leads to a violation of the weak cosmic censorship.

Rotating black strings also exhibit the superradiant instability. The Kaluza-Klein circle creates an effective mass that confines perturbations of a rotating black string around its horizon and can (but does not always) induce a superradient instability, even in asymptotically flat spacetimes \cite{Marolf:2004fya,Cardoso:2004zz,Cardoso:2005vk,Dias:2006zv}. While the instability was first identified in the five-dimensional Kerr black strings, the situation is quite different in higher dimensions. For instance, in single-spinning Myers-Perry black strings \cite{Cardoso:2005vk} no instability was found for massless scalar fields. However, it was later shown in \cite{Dias:2022mde} that gravitational perturbations (unlike massive scalar fields) do lead to an instability. There, an interplay between the Gregory-Laflamme and superradiant instabilities was also discussed.

The nature of the superrariant instability has been well studied in asymptotically anti-de Sitter space (AdS) \cite{Hawking:1999dp,Reall:2002bh,Cardoso:2004hs,Kunduri:2006qa,Murata:2008xr,Kodama:2009rq,Dias:2011at,Dias:2013sdc,Cardoso:2013pza,Choptuik:2017cyd}. In this context, it was shown that helically-symmetric black holes called ``black resonators'' are nonlinear back-reactions of superradiant gravitational modes \cite{Dias:2015rxy}.  Because these solutions typically have few symmetries, it is not easy to study their properties.  Fortunately, in the five-dimensional equal-spinning case, cohomogeneity-1 black resonators were found \cite{Ishii:2018oms}, and their linear perturbations were also studied \cite{Ishii:2020muv}. The superradiant instabilities for matter fields also lead to cohomogeneity-1 resonating black holes \cite{Dias:2011at,Ishii:2019wfs,Ishii:2021xmn}.

In this manuscript, in a similar manner as the construction of cohomogeneity-1 black resonators in AdS, we will construct cohomogeneity-1 deformed black string solutions branching from the superradiant instability of six-dimensional equal-spinning Myers-Perry black strings.\footnote{Similar to the Kaluza-Klein spacetime considered here, a five-dimensional cohomogeneity-1 geometry as nonlinear extension of Kaluza-Klein modes in Poincar\'{e} AdS space with a $S^1$ direction has also been obtained in \cite{Garbiso:2020dys}.} Since such cohomogeneity-1 solutions have helical symmetries formed by the linear combination of the time translation, shift along the string, and rotation, we will call them \textit{helical black strings}. The existence of black strings with helical symmetry has been first observed in~\cite{Emparan:2009vd} using the blackfold effective worldvolume theory. In our recent paper \cite{Dias:2022str}, we obtained black resonator string solutions which also branch from the same onset of the superradiant instability. However, black resonator strings are cohomogeneity-2 non-stationary solutions, while helical black strings are cohomogeneity-1 stationary solutions. All three solutions (Myers-Perry black strings, helical black strings and black resonator strings) compete in the microcanonical ensemble for asymptotically Kaluza-Klein solutions.

This manuscript is organized as follows. In sections~\ref{sec:MP-sring} and \ref{sec:fluctuation}, we review the equal-spinning Myers-Perry black string and its superradiant instability as studied in detail in \cite{Dias:2022mde}. Here, we put emphasis on introducing the rotating frame at infinity and also discuss the isometries preserved by the superradiant perturbation of relevance to us. In section~\ref{sec:coh1bs}, setups for constructing cohomogeneity-1 helical black strings are introduced. To check our results and further explore the phase space, we will find helical black strings in the spherical gauge (section~\ref{sec:ansatzSphericalGauge}) and, alternatively, in the Einstein-DeTurck gauge (section~\ref{sec:ansatzEdT}). To complement the numerical analysis in the setup of section~\ref{sec:coh1bs}, in section~\ref{sec:perturbative}, we describe the perturbative construction of helical black strings, an analysis that is however valid only in the vicinity of the superradiant onset. In section~\ref{sec:result}, we present our nonlinear numerical results and discuss the physical properties of helical black strings. We also discuss their competition with the black resonator strings of  \cite{Dias:2022str}. We give our conclusions in section~\ref{sec:conclusion}. Technical details for some of our calculations are provided in several appendices.

\section{The Myers-Perry black string}\label{sec:MP-sring}

The Myers-Perry (MP) black hole \cite{Myers:1986un} extends the Kerr solution to higher dimensions.  In five dimensions, it can be parameterized by a mass radius parameter $r_0$ and two angular momenta parameters $a_1$ and $a_2$ \cite{Myers:1986un,Hawking:1999dp}. In general, this is a cohomogeneity-2 solution with the isometry group $\mathbb{R}_t \times U(1)^2$ but, in the equal-spinning case $a_1=a_2 \equiv a$, the solution has the enhanced isometry group $\mathbb{R}_t \times U(2)$ ($\mathbb{R}_t$ denotes time translation) and thus it has cohomogeneity-1 (\ie it depends nontrivially only on a single coordinate) \cite{Gibbons:2004js,Gibbons:2004uw}.

We are interested in the associated 6-dimensional rotating black string (with equal angular momenta) that asymptotes to ${\cal M}^{1,4}\times \mathbb{R}_z$ or ${\cal M}^{1,4}\times S^1$ if the string's direction is compactified (${\cal M}^{1,4}$ is 5-dimensional Minkowski space). This black string is obtained by adding an extended flat direction $z$ to the 5-dimensional Myers-Perry black hole.  In this section, we introduce this solution and briefly discuss its properties.

The metric of the 6-dimensional equal-spinning Myers-Perry black string (MPBS), which solves $R_{AB}=0$ (with $R_{AB}$ being the Ricci tensor), can be given by\footnote{The radial coordinate used here can be related to the standard Boyer-Lindquist radial coordinate of \cite{Myers:1986un} through $r^2 \to r^2+a^2$. We shall use notation where capital Latin indices $(A,B,\dots)$ run over the 6-dimensional coordinates, small Latin indices $(a,b,\dots)$ run over the 5-dimensional coordinates except the radial one, and ($i,j,\dots$) are used for $SU(2)$ indices.}
\begin{equation} \label{MPstring}
 {\mathrm d}s^2 _{\rm MP \,string}= -\frac{F}{H}\,{\mathrm d}t^2 +\frac{{\mathrm d}r^2}{F} + r^2 \left[ H \left( \frac{\sigma_3}{2} -\frac{\Omega}{H}\, {\mathrm d}t \right)^2+ {\mathrm d}s^2 _{\mathbb{C}\mathrm{P}^1} \right] + \mathrm{d}z^2 \, ,
 \end{equation}
where
\begin{equation}\label{MPfns}
F(r)= 1-\frac{r_0^2}{r^2}+\frac{a^2r_0^2}{r^4}\,, \qquad H(r)=1+ \frac{a^2 r_0^2}{r^4}\, , \qquad
\Omega=\frac{a \,r_0^2}{r^4}\, ,
\end{equation}
and ${\mathrm d}s^2 _{\mathbb{C}\mathrm{P}^1}=\frac{1}{4}\left( \sigma_1^2 +\sigma_2^2\right)$ is the metric of the complex projective space $\mathbb{C}\mathrm{P}^1$ (isomorphic to the 2-sphere $S^2$).\footnote{In $D=2N+3$ dimensions, the equal-spinning Myers-Perry black hole has a homogeneously squashed $S^{2N+1}$ written as an $S^1$ fibred over $\mathbb{C}\mathrm{P}^{N}$. The Fubini-Study metric for $\mathbb{C}\mathrm{P}^{1}$ happens to be the more familiar metric for $S^2$. For further details see  \cite{Gibbons:2004js,Gibbons:2004uw,Kunduri:2006qa,Dias:2010eu}.}
We have defined the $SU(2)$-invariant 1-forms $\sigma_{i}$ ($i=1,2,3$) on $S^3$ as
\begin{equation}\label{Eulerforms}
\begin{split}
\sigma_1 &= -\sin(2\psi) \, \mathrm{d}\theta + \cos(2\psi)\sin\theta \,\mathrm{d}\phi\, ,\\
\sigma_2 &= \cos(2\psi) \,\mathrm{d}\theta + \sin(2\psi)\sin\theta \,\mathrm{d}\phi\, ,\\
\sigma_3 &= 2\,\mathrm{d}\psi + \cos\theta \,\mathrm{d}\phi  \, ,
\end{split}
\end{equation}
where $(\theta,\phi,\psi)$ denote the Euler angles of the $S^3$ with the ranges chosen as $0\leq \theta < \pi $, $0\leq \phi <2\pi$, and $0\leq \psi <2\pi$. These satisfy the Maurer-Cartan equation $\mathrm{d}\sigma_i = \frac{1}{2} \epsilon_{i j k} \sigma_j \wedge \sigma_k$.

This solution has an event horizon at $r=r_+$ (given by the largest real root of $f$) with the Killing horizon generator $K=\partial_t+\Omega_H \partial_\psi$,
where $\Omega_H\equiv \Omega(r_+)/H(r_+)=a/r_+^2$ is the horizon angular velocity. The mass radius parameter can then be expressed as
\begin{equation}
r_0=\frac{r_+^2}{\sqrt{r_+^2-a^2}} = \frac{r_+}{\sqrt{1 - \Omega_H^2 r_+^2}} \, .
\end{equation}

The angular velocity parameter of the MPBS is bounded from above by regularity as $r_0 \geq a$. This translates into the condition for the horizon angular velocity as $\Omega_H r_+ \le 1/\sqrt{2}$. The MPBS is extremal (i.e. with zero temperature) at the upper limit of the angular velocity: $\Omega_H^\mathrm{ext} r_+ = 1/\sqrt{2}$.

The isometry group of the MPBS is $\mathbb{R}_t \times \mathbb{R}_z \times U(1)_\psi \times SU(2)$, whose six Killing vectors are listed as follows. The metric \eqref{MPstring} is clearly invariant under $\partial_t$ and $\partial_z$. The 1-forms \eqref{Eulerforms} are invariant under the operation of the following $SU(2)$ generators,
\begin{equation}
\begin{split}
\xi_1 &= \cos\phi\,\partial_\theta +
\frac{1}{2}\frac{\sin\phi}{\sin\theta}\,\partial_\psi -
\cot\theta\sin\phi\,\partial_\phi\, ,\\
\xi_2 &= -\sin\phi\,\partial_\theta +
\frac{1}{2}\frac{\cos\phi}{\sin\theta}\partial_\psi -
\cot\theta\cos\phi\,\partial_\phi\, ,\\
\xi_3 &= \partial_\phi\, .
\end{split}
\label{lkv}
\end{equation}
These generators satisfy the $SU(2)$ commutation relation $[\xi_i,\xi_j] = \epsilon_{ijk} \xi_k$ and leave $\sigma_i$ invariant as $\pounds_{\zeta i} \sigma_j = 0$, where $\pounds$ denotes the Lie derivative. Finally, the metric is also invariant under $\partial_\psi$ which mixes $(\sigma_1,\sigma_2)$:
$\pounds_{\frac{1}{2}\partial_\psi} \sigma_1 = -\sigma_2$ and $\pounds_{\frac{1}{2}\partial_\psi} \sigma_2 = \sigma_1$.
The round $S^2 \cong \mathbb{C}\mathrm{P}^1$, ${\mathrm d}s^2 _{\mathbb{C}\mathrm{P}^1}=\frac{1}{4}\left( \sigma_1^2 +\sigma_2^2\right)$, is invariant under $\partial_\psi$.

\section{Superradiant instability of the Myers-Perry black string for non-axisymmetric perturbations}
\label{sec:fluctuation}

\subsection{Decoupled tensor perturbations}
\label{sec:tensor}

As we have studied in detail in the companion paper \cite{Dias:2022mde}, the MPBS exhibits a superradiant instability for decoupled tensor gravitational perturbations. Here, we focus on the perturbation with the lowest nontrivial azimuthal quantum number, which preserves the $SU(2)$ isometry (whereas it is broken for higher azimuthal quantum numbers \cite{Dias:2022mde}). In this section, we first recall the relevant superradiant  perturbation of the background metric \eqref{MPstring}, which is written in the {\it non-rotating frame} at infinity. We then introduce the {\it rotating frame} at infinity, which is convenient for constructing helical solutions in following sections.

The $U(1)_\Psi$-charges of the MPBS are fundamental for the decoupling of the tensor gravitational perturbations from other sectors of perturbations.  To classify perturbations using these charges, it is convenient to use
\begin{equation}
\sigma_\pm = \frac{1}{2} \left( \sigma_1 \mp \mathsf{i} \sigma_2 \right) = \frac{1}{2} e^{\mp 2 \mathsf{i} \psi} \left( \mp\,\mathsf{i}\,\mathrm{d}\theta + \sin \theta \, \mathrm{d}\phi \right)\, .
\label{nonrot_sigma_pm}
\end{equation}
These satisfy $\pounds_{(\mathsf{i}/2) \partial_\psi} \sigma_{\pm} = \pm \sigma_\pm$, which means that $\sigma_\pm$ have charges $\pm 1$ with respect to the $U(1)_\Psi$. Because of this property, the following charge-2 perturbation of the MPBS is decoupled:\footnote{\label{foot:sign-m}\cite{Dias:2022mde} considered a perturbation of the form $h_{MN} dx^M dx^N = e^{-\mathsf{i} \omega  t + \mathsf{i} k z} r^2 \delta \eta(r) \sigma_-^2$, corresponding to the azimuthal quantum number $m=2$, whereas the perturbation \eqref{MPBS_fluc} corresponds to $m=-2$. These are physically equivalent and differ in the direction of the Kaluza-Klein momentum. 
}
\begin{equation}\label{MPBS_fluc}
h_{MN} \mathrm{d}x^M \mathrm{d}x^N = e^{-\mathsf{i} \omega  t + \mathsf{i} k z} r^2 \delta \eta(r) \sigma_+^2 \, .
\end{equation}
This perturbation has the lowest nontrivial azimuthal quantum number in the $\psi$-direction and preserves the $SU(2)$ symmetry. The linearized equation for the above perturbation is given by
\begin{equation}
\delta \eta'' + \left( \frac{F'}{F} + \frac{3}{r} \right) \delta \eta' + \frac{1}{F} \left( \frac{8}{r^2} - \frac{16}{r^2 H} + \frac{(\omega+4\Omega/H)^2}{f} - k^2 \right) \delta \eta=0\, .
\label{fluc_eq}
\end{equation}
Note the plus sign in $\omega+4\Omega/H$ in \eqref{fluc_eq} because we consider the perturbation with respect to $\sigma_+^2$
\eqref{MPBS_fluc}; see also footnote \ref{foot:sign-m}.

\subsection{Rotating frame at infinity}
\label{sec:rotframe}

We find it convenient to work in the rotating frame at infinity which we introduce in this section. The metric \eqref{MPstring} was written in the standard non-rotating frame at infinity, where $\Omega(r)|_{r \to \infty} \to 0$ and the angular velocity of the black hole is read off from the rotation at the horizon, $\Omega_H=\Omega(r_+)/H(r_+)$. We now redefine the angular coordinate $\psi$ in a $t$-dependent manner so that the rotation is carried by spatial infinity and the horizon has zero angular velocity. In addition, the $\psi$-dependence of $\sigma_\pm$ \eqref{nonrot_sigma_pm} suggests that we can also absorb the $z$-dependence in \eqref{MPBS_fluc} by a redefinition of $\psi$.

We use the upper case $(T,Z,\Psi)$, together with $(r,\theta,\phi)$, for the coordinates in the rotating frame. The transformation from the non-rotating to the rotating frames can be given by
\begin{equation}
T = t\, , \quad Z = z\, , \quad \Psi =\psi - \Omega_H t - \frac{k}{4} z\, .
\label{nonrot2rot4MP}
\end{equation}
The dual vectors to these coordinates are
\begin{equation}
\partial_T = \partial_t + \Omega_H \partial_\psi\, , \quad
\partial_Z = \partial_z + \frac{k}{4} \partial_\psi\, , \quad
\partial_\Psi = \partial_\psi\, .
\label{nonrotdualvectorsMP}
\end{equation}
Let $\Sigma_i$ denote the $SU(2)$ invariant 1-forms in the rotating frame, which are obtained by replacing $\psi$ by $\Psi$ in \eqref{Eulerforms}:
\begin{equation}
\begin{split}
\Sigma_1 &= - \sin(2\Psi) \,\mathrm{d}\theta + \cos(2\Psi)\sin\theta\,\mathrm{d}\phi\, , \\
\Sigma_2 &= \cos(2\Psi)\, \mathrm{d}\theta + \sin(2\Psi)\sin\theta \,\mathrm{d}\phi\, , \\
\Sigma_3 &= 2\mathrm{d}\Psi + \cos\theta \,\mathrm{d}\phi\, .
\end{split}
\label{su2_1forms}
\end{equation}
In the rotating frame at infinity, the MPBS metric \eqref{MPstring} becomes
\begin{equation}
 \mathrm{d}s^2 = -f(r) \mathrm{d}T^2 + \frac{\mathrm{d}r^2}{g(r)} + \frac{r^2}{4} \left[ \Sigma_1^2 + \Sigma_2^2 + \beta(r) \left( \Sigma_3 + 2 h(r) \mathrm{d}T +\frac{k}{2}\mathrm{d}Z\right)^2 \right] + \mathrm{d}Z^2\, ,
\label{MPBS_ds2}
\end{equation}
where
\begin{equation}
\begin{split}
f = \frac{g}{\beta}\, , \quad
g=F(r)\, \quad
h=\Omega_H-\frac{\Omega(r)}{H(r)}\, , \quad
\beta = H(r)
\label{MPBS_sol}
\end{split}
\end{equation}
with $F(r), \, H(r), \, \Omega(r)$ given in \eqref{MPfns}.
In \eqref{MPBS_sol}, one can see that the rotating frame at infinity satisfies $h(r_+)=0$. This means that $\partial_T$ coincides with the  Killing horizon generator: $K=\partial_T = \partial_t + \Omega_H \partial_\psi$. The rotation is carried by the asymptotic infinity as $h(r)|_{r \to \infty}=\Omega_H$.

\subsection{Onset of superradiant instability}
\label{sec:onset}

In the rotating frame at infinity, the perturbation \eqref{MPBS_fluc} is rewritten as
\begin{equation}
e^{-\mathsf{i} \omega  t + \mathsf{i} k z} r^2 \delta \eta(r) \sigma_+^2  = e^{-\mathsf{i} \widehat{\omega}  t} r^2 \delta \eta(r) \Sigma_+^2\, ,
\label{MPBS_fluc_rot}
\end{equation}
where the frequency parameter is shifted as $\omega = \widehat{\omega}-4 \Omega_H$. The perturbation \eqref{MPBS_fluc_rot} is manifestly $Z$-independent on the right hand side. The perturbation equation \eqref{fluc_eq} has the same form under the frame change aside from the shift. In the new variables \eqref{MPBS_sol}, it reads
\begin{equation}
\delta \eta'' + \left( \frac{g'}{g} + \frac{3}{r} \right) \delta \eta' + \frac{1}{g} \left( \frac{8}{r^2} - \frac{16}{r^2 \beta} + \frac{( \widehat{\omega}-4h)^2}{f} - k^2 \right) \delta \eta=0\,.
\label{fluc_eq_rot}
\end{equation}
As studied in \cite{Dias:2022mde}, the solutions to this equation, when $\omega$ is varied as a parameter, give an instability for a bounded range of $k$ -- the superradiant instability. In this paper, we are particularly interested in the onset of the instability, as it locates where new black string solutions branch from the MPBS.

A nice feature of using the rotating frame at infinity is that $ \widehat{\omega}=0$ at the onset of instability. Typically, $\textrm{Im}\, \widehat{\omega}=0$ at the onset of instability but, additionally, we can show that $\textrm{Re}\, \widehat{\omega}=0$ as follows. At the horizon, the perturbation satisfies the ingoing wave boundary condition
\begin{equation}
\delta \eta(r) \sim (r - r_+)^{-\mathsf{i} \widehat{\omega}/(2 \kappa)}\, ,
\label{fluc_hrzn}
\end{equation}
where $\kappa=\sqrt{f'g'}/2|_{r=r_+}$ is the surface gravity ( $'$ denotes an $r$-derivative). Meanwhile, at asymptotic infinity $r \to \infty$, the perturbation behaves as
\begin{equation}
\delta \eta(r) \sim \frac{1}{r^{3/2}} \, e^{-\sqrt{k^2 - ( \widehat{\omega}-4\Omega_H)^2} \, r}\, ,
\label{fluc_asymp}
\end{equation}
where we assume $k^2 - ( \widehat{\omega}-4\Omega_H)^2 > 0$. The Wronskian for the linearized system \eqref{fluc_eq} can be given by
\begin{equation}
 W=r^3\sqrt{fg\beta}(\delta \eta^\ast \delta \eta'-\delta \eta \delta \eta^\ast{}')\, .
\end{equation}
When $\textrm{Im}\, \widehat{\omega}=0$, the Wronskian is conserved along the $r$-direction, $\mathrm{d}W/\mathrm{d}r=0$. This can be checked by using the perturbation equation \eqref{fluc_eq}. With the behaviours (\ref{fluc_hrzn}) and (\ref{fluc_asymp}), the Wronskian can be evaluated at the horizon and infinity as $W(r=r_+)\propto \widehat{\omega}$ and $W(r=\infty)=0$, respectively. Then, it follows that not only $\textrm{Im}\, \widehat{\omega}$ but also $\textrm{Re}\, \widehat{\omega}$ vanish at the onset of instability.

At $ \widehat{\omega}=0$, the perturbation equation \eqref{fluc_eq} has nontrivial  linear mode solutions only for specific values of parameters $(k r_+, \Omega_H r_+)$, corresponding to the onset of an instability. With $ \widehat{\omega}=0$, the ingoing solution (\ref{fluc_hrzn}) is replaced with a regular solution (normalised as $\delta \eta(r_+)=1$),
\begin{equation}
\delta \eta(r) = 1 - \frac{(1-\Omega_H^2 r_+^2 )\{(k^2-16 \Omega_H^2) r_+^2  + 8\}}{2 r_+ (1-2\Omega_H^2 r_+^2)}  (r - r_+) + \cdots\, .
\label{fluc_hrzn_reg}
\end{equation}
With this boundary condition and the regular asymptotic behaviour (\ref{fluc_asymp}), we numerically solve \eqref{fluc_eq} and find linear mode solutions for specific values of $(k r_+, \Omega_H r_+)$. These correspond to the onset of superradiant instability of the MPBS. The analysis has been done in detail in \cite{Dias:2022mde}. The location of the onset will be plotted together with our results later.

\subsection{Isometries of the perturbed Myers-Perry black string}
\label{sec:isopert}

With a non-zero Kaluza-Klein wavenumber $k$, we assume that the black string is compactified with periodicity
\begin{equation}
L \equiv 2\pi/k \, .
\label{L_for_MPBS}
\end{equation}
In the non-rotating frame coordinates, the isometry group of the MPBS is  $\mathbb{R}_t \times U(1)_z \times U(1)_\psi \times SU(2)$, and the spacetime is asymptotically Kaluza-Klein ${\cal M}^{1,4} \times S^1$.

The perturbation \eqref{MPBS_fluc} breaks some isometries of the MPBS. At first sight, it manifestly breaks the three symmetries $\mathbb{R}_t, \, U(1)_z$ and $U(1)_\psi$ in the non-rotating frame at infinity. However, some of their linear combinations (which can be identified with the isometries in the rotating frame at infinity) can be preserved. Hence, it is more appropriate to argue isometry breaking in the rotating frame at infinity, where we have the isometries $\mathbb{R}_T, \, U(1)_Z, \, U(1)_\Psi$ for the MPBS. In our preceding paper \cite{Dias:2022str}, after back-reaction, the perturbation is extended to nonlinear black resonator strings, where both $U(1)_\Psi$ and $U(1)_Z$ are broken. In this paper, we focus on different nonlinear solutions that break only the $U(1)_\Psi$ and preserve the $U(1)_Z$. Thanks to the $U(1)_Z$, such solutions are described by a cohomogeneity-1 metric.

Summing \eqref{MPBS_fluc_rot} and its complex conjugate (as well as multiplying them by a normalization factor), we obtain a real gravitational perturbation at the onset of instability,
\begin{equation}
h_{MN} \mathrm{d}x^M \mathrm{d}x^N = \frac{r^2}{2} \delta \eta(r) (\Sigma_+^2+\Sigma_-^2)=\frac{r^2}{4} \delta \eta(r) (\Sigma_1^2-\Sigma_2^2)\, .
\label{MPBS_fluc2}
\end{equation}
This perturbation is obviously invariant under $\partial_T$, $\partial_Z$ as well as the generators of $SU(2)$, while $U(1)_\Psi$ is broken. Thus, the MPBS perturbed by \eqref{MPBS_fluc2} admits the isometry group $\mathbb{R}_T\times U(1)_Z \times SU(2)$. In the original non-rotating frame at infinity, the onset perturbation \eqref{MPBS_fluc2} takes the form
\begin{equation}
h_{MN} \mathrm{d}x^M \mathrm{d}x^N = \frac{r^2}{2} \delta \eta(r) (e^{-4\mathsf{i}\Omega_Ht}e^{\mathsf{i}kz}\sigma_+^2+e^{4\mathsf{i}\Omega_Ht}e^{-\mathsf{i}kz}\sigma_-^2)\, .
\label{MPBS_fluc2_nonrot}
\end{equation}
It is obvious that $\partial_t$ and $\partial_z$ are no longer Killing vectors independently. Their helical combinations with $\partial_\psi$ as in \eqref{nonrotdualvectorsMP} generate the $\mathbb{R}_T$ and $U(1)_Z$ isometries.

The perturbed MPBS spacetime is also invariant under two discrete transformations $P_1$ and $P_2$ defined by
\begin{align}
 P_1(T,\Psi,\theta,\phi,Z)&=(-T,-\Psi,\theta,-\phi,-Z)\, ,\\
 P_2(T,\Psi,\theta,\phi,Z)&=(T,\Psi+\pi/2,\theta,\phi,Z)\, .
\end{align}
The 1-forms $(\mathrm{d}T,\mathrm{d}Z,\Sigma_1,\Sigma_2,\Sigma_3)$ are transformed by $P_1$ and $P_2$ as
\begin{equation}
\begin{split}
&P_1(\mathrm{d}T,\mathrm{d}Z,\Sigma_1,\Sigma_2,\Sigma_3)=(-\mathrm{d}T,-\mathrm{d}Z,-\Sigma_1,\Sigma_2,-\Sigma_3)\, , \\
&P_2(\mathrm{d}T,\mathrm{d}Z,\Sigma_1,\Sigma_2,\Sigma_3)=(\mathrm{d}T,\mathrm{d}Z,-\Sigma_1,-\Sigma_2,\Sigma_3)\, .
\end{split}
\label{P1P2}
\end{equation}
Both the perturbation~(\ref{MPBS_fluc2}) and background~(\ref{MPBS_ds2}) are invariant under $P_1$ and $P_2$.

Notably, the spacetime perturbed by \eqref{MPBS_fluc2} is still stationary.\footnote{A spacetime is stationary when it admits a Killing vector field that becomes timelike near asymptotic infinity.} Indeed, let us consider a linear combination of the Killing vectors of the perturbed MPBS: $\zeta\equiv~c_T \partial_T + c_Z \partial_Z$. Near the spatial infinity, the norm of this Killing vector becomes
\begin{equation}\label{normKVFinf}
 \zeta^2= (g_{MN}+h_{MN})\zeta^M \zeta^N=\frac{(4c_T\Omega_H + c_Z k)^2}{16} r^2 +(-c_T^2+c_Z^2) + \mathcal{O}(r^{-2})\, .
\end{equation}
While we include $h_{MN}$ of \eqref{MPBS_fluc2} in \eqref{normKVFinf} to describe the metric of the perturbed spacetime, $h_{MN}$  is actually exponentially decaying near the asymptotic infinity and does not affect the asymptotic power series of $\zeta^2$. Choosing $c_T=k$ and $c_Z=-4\Omega_H$, we obtain $\zeta^2 = -(k^2-16\Omega_H^2) + \mathcal{O}(r^{-2})$, where $k^2-16\Omega_H^2>0$ at the onset of instability \cite{Dias:2022mde}. Hence, the Killing vector $\zeta=k \partial_T -4\Omega_H \partial_Z$ is timelike at asymptotic infinity, and the perturbed MPBS is stationary.

\section{Cohomogeneity-1 helical black strings}
\label{sec:coh1bs}

To check our results as much as we can and further explore the phase space, we will find helical black strings in the spherical gauge (section~\ref{sec:ansatzSphericalGauge}) and, alternatively, in the Einstein-DeTurck gauge (section~\ref{sec:ansatzEdT}).

\subsection{Helical black string ansatz in the spherical gauge}
\label{sec:ansatzSphericalGauge}

\subsubsection{Metric ansatz}
\label{sec:ansatz}

We wish to construct a new family of black hole solutions with the isometry group $\mathbb{R}_T\times U(1)_Z \times SU(2)$, branching from the onset of the superradiant instability. To write a metric ansatz that nonlinearly extends the perturbation \eqref{MPBS_fluc2}, we also assume the discrete isometries $P_1$ and $P_2$. With these isometries, we introduce the following cohomogeneity-1 metric ansatz:\footnote{We can also consider the helicity-flipped metric by $k(r)\mathrm{d}Z \to -k(r)\mathrm{d}Z$. That has the same physical properties along with the opposite direction of Kaluza-Klein momentum.}
\begin{multline}
\mathrm{d}s^2 = -f(r) \mathrm{d}T^2 + \frac{\mathrm{d}r^2}{g(r)} + \frac{r^2}{4} \left[ \eta(r) \Sigma_1^2 + \frac{1}{\eta(r)} \Sigma_2^2 \right. \\
\left. + \beta(r) \left( \Sigma_3 + 2 h(r) \mathrm{d}T + \frac{k(r)}{2} \mathrm{d}Z \right)^2 \right] + \gamma(r) \left( \mathrm{d}Z + q(r) \mathrm{d}T \right)^2\, .
\label{screw_metric}
\end{multline}
We have used the freedom of the redefinition of the radial coordinate to fix the gauge such that the product of the metric coefficients of $\sigma_1^2$ and $\sigma_2^2$ is $r^4/16$. In the absence of a better nomenclature, we call this as the {\it spherical gauge} because the area of the $S^2$ base space is then simply proportional to $r^2$.
The metric ansatz is not invariant under a $U(1)_\Psi$ shift if $\eta(r)\neq1$. Instead, the ansatz has a discrete symmetry
\begin{equation}
\eta(r)\to \frac{1}{\eta(r)} \, ,
\label{etatrans}
\end{equation}
because the coordinate transformation $\Psi\to\Psi+\pi/2$ flips $\sigma_1$ and $\sigma_2$ as $\sigma_1\to -\sigma_2$ and $\sigma_2\to \sigma_1$.

The event horizon $r=r_+$ is located at the largest root of $f(r)=g(r)=0$. The generator of the horizon is given by
\begin{equation}
 K=\partial_T - (4h_0-k_0 q_0) \partial_\Psi - q_0 \partial_Z\, ,
\label{horgenerator}
\end{equation}
where $h_0\equiv h(r=r_+)$, $k_0\equiv k(r=r_+)$, and $q_0\equiv q(r=r_+)$ (see also \eqref{Xexp}).

The perturbation equation at the onset of instability (i.e.~\eqref{fluc_eq_rot} at $ \widehat{\omega}=0$) must be reproduced by a small fluctuation of \eqref{screw_metric} around the MPBS as we check next. For the MPBS \eqref{MPBS_ds2}, $f(r),\,g(r),\,h(r)$, $\beta(r)$ are given in \eqref{MPBS_sol}, and the other field variables are $\eta(r)=\gamma(r)=1$, $k(r)=k$, $q(r)=0$. Introducing a linear perturbation as $\eta(r)=1+\delta\eta(r)$, we obtain the perturbation equation \eqref{fluc_eq_rot} for $ \widehat{\omega}=0$, as it should be.

We will construct solutions with $\eta(r) \neq 1$ in the metric ansatz \eqref{screw_metric}. From the symmetries of the spacetime, we call such solutions \emph{helical black strings}. This is a nomenclature that was first proposed in \cite{Emparan:2009vd}. Because of the cohomogeneity-1 ansatz, the Einstein field equations are reduced to coupled ODEs. These are given by
\begin{align}
f'&=
\frac{1}{r \left(r \gamma  \beta'+r \beta  \gamma'+6 \beta  \gamma \right)}
\Bigg[
\frac{1}{16} r^4 \beta ^2 \left(f k'^2-\gamma \left(4 h'-k' q\right)^2\right)-r^2 \beta  \gamma ^2 q'^2
\nonumber \\ & \phantom{=\ }
+ f \beta  \gamma  \left\{\frac{8}{g} \left(\eta+\frac{1}{\eta }-\frac{\beta}{2}\right)+ \frac{r^2 \eta'^2}{\eta ^2}+12\right\}
- f \left(r \beta'+6\beta\right) \left(r \gamma'+4\gamma\right)
\nonumber \\ & \phantom{=\ }
+ \frac{1}{4g}\left(\eta-\frac{1}{\eta }\right)^2 \left\{r^2 \beta  \gamma (4 h-k q)^2-f \left(r^2 k^2 \beta+16 \gamma \right)\right\}
\Bigg]\, ,
\label{EOM_f}
\\
g' &= \frac{4}{r}\left( \eta+\frac{1}{\eta} - \beta -g \right) - \frac{(f\beta\gamma)'g}{f\beta\gamma}\, ,
\label{EOM_g}
\\
h''&=
\frac{1}{2} h'\left(-\frac{f'}{f}-\frac{g'}{g}-\frac{3 \beta'}{\beta}+\frac{\gamma'}{\gamma}-\frac{10}{r}\right)
+\frac{1}{4} k'q' \nonumber \\
& \phantom{=\ }
+\frac{1}{4} \left(\frac{f'}{f}-\frac{q \gamma  q'}{f}-\frac{\gamma'}{\gamma}\right) \left(4 h'-k' q\right)
+\frac{4 h \left(\eta-\frac{1}{\eta }\right)^2}{r^2 g \beta}\, ,
\label{EOM_h}
\\
k''&=
\frac{1}{2} k'\left(-\frac{f'}{f}-\frac{g'}{g}-\frac{3 \beta'}{\beta}
+\frac{\gamma'}{\gamma}-\frac{10}{r}\right)-\frac{\gamma  q' \left(4 h'-k' q\right)}{f}
+\frac{4 k \left(\eta-\frac{1}{\eta}\right)^2}{r^2 g \beta}\, ,
\label{EOM_k}
\\
q''&=
\frac{1}{2} q'\left(\frac{f'}{f}-\frac{g'}{g}-\frac{\beta'}{\beta}-\frac{3 \gamma'}{\gamma}-\frac{6}{r}\right)
-\frac{r^2 \beta  k'\left(4 h'-k' q\right)}{16 \gamma}
-\frac{k \left(\eta-\frac{1}{\eta }\right)^2 (4 h-k q)}{4 g \gamma}\, ,
\label{EOM_q}
\\
\eta''&=
\frac{1}{2} \eta' \left(-\frac{f'}{f}-\frac{g'}{g}+\frac{2 \eta'}{\eta}-\frac{\beta'}{\beta}-\frac{\gamma'}{\gamma}-\frac{6}{r}\right) \nonumber \\
& \phantom{=\ }
+\frac{\eta}{g} \left(\eta-\frac{1}{\eta}\right) \left[\frac{1}{4} \left(\eta+\frac{1}{\eta}\right) \left(\frac{16}{r^2 \beta}+\frac{k^2}{\gamma}-\frac{(4 h-k q)^2}{f}\right)-\frac{4}{r^2}\right]\, ,
\label{EOM_alpha}
\\
\beta''&=
\frac{1}{2} \beta'\left(-\frac{f'}{f}-\frac{g'}{g}+\frac{\beta'}{\beta}-\frac{\gamma'}{\gamma}-\frac{6}{r}\right)
-\frac{4 \left(\eta+\frac{1}{\eta }\right) \beta}{r^2 g} \nonumber \\
& \phantom{=\ }
+\beta ^2 \left(\frac{r^2}{16} \left(\frac{k'^2}{\gamma}-\frac{\left(4 h'-k' q\right)^2}{f}\right)+\frac{8}{r^2 g}\right)
-\frac{4 \left(\eta-\frac{1}{\eta }\right)^2}{r^2 g}\, ,
\label{EOM_beta}
\\
\gamma''&=
\frac{1}{2} \gamma'\left(-\frac{f'}{f}-\frac{g'}{g}-\frac{\beta'}{\beta}+\frac{\gamma'}{\gamma}-\frac{4}{r}\right)
-\frac{1}{16} r^2 \beta  k'^2
-\frac{\gamma ^2 q'^2}{f}
-\frac{k^2 \left(\eta-\frac{1}{\eta }\right)^2}{4 g} \nonumber \\
& \phantom{=\ }
+\frac{\gamma}{r^2}  \left(\frac{r f'}{f}+\frac{r g'-4 \left(\eta+\frac{1}{\eta }\right)+4 \beta}{g}+\frac{r \beta'}{\beta}+4\right)\, .\label{EOM_gamma}
\end{align}
These will be solved with suitable boundary conditions at the horizon $r=r_+$ and spatial infinity $r=\infty$.

\subsubsection{Boundary conditions at the horizon and infinity}
\label{sec:rhser}

First, let us discuss the boundary conditions for the black hole horizon at $r=r_+$. Let $X \equiv (f,g,h,k,q,\eta,\beta,\gamma)$ denote all field variables collectively. The fields can be expanded near the horizon as a Taylor series
\begin{equation}
X(r) = \sum_{n=0}^\infty  X_n (r - r_+)^n\, ,
\label{Xexp}
\end{equation}
where $f_0 = g_0  = 0$ for the black hole horizon. Coefficients $X_n$ are determined order by order when this series expansion is substituted into the equations of motion (\ref{EOM_f}-\ref{EOM_gamma}). At the leading order, the nontrivial equation is given by
\begin{equation}
(4 h_0 - k_0 q_0)(\eta_0-1)=0\, .
\end{equation}
When $\eta_0=1$,  we recover the MPBS. Other solutions with $\eta_0 \neq 1$ are possible if
\begin{equation}
4 h_0 - k_0 q_0 = 0\, .
\label{horizon_4hkq_relation}
\end{equation}
The horizon generator \eqref{horgenerator} of such solutions is
\begin{equation}
K=\partial_T - q_0 \partial_Z\, .
\label{horgenerator_4khq_imposed}
\end{equation}
We assume \eqref{horizon_4hkq_relation} is solved as $h_0=k_0 q_0/4$, leaving $k_0$ and $q_0$ as free parameters that will be fixed by the equations of motion subject to the two boundary conditions. Continuing to higher orders in the asymptotic expansion, we find that 8 parameters $(f_1,h_1,k_0,q_0,q_1,\eta_0,\beta_0,\gamma_0)$ remain undetermined in the asymptotic analysis near the horizon. Other (subleading) coefficients are completely fixed by these leading order coefficients (see appendix~\ref{app:horexp} for details). Because of the coordinate freedom~(\ref{etatrans}), we can assume $\eta_0\leq 1$ without loss of generality.

The asymptotic analysis near the horizon also tells us that the event horizon remains a Killing horizon. Although $\partial_\Psi$ is no longer a Killing vector if $\eta \neq 1$, the coefficient in front of $\partial_\Psi$ in (\ref{horgenerator}) actually vanishes by (\ref{horizon_4hkq_relation}), and the horizon generator is given by \eqref{horgenerator_4khq_imposed}.

At spatial infinity, we impose the asymptotically locally flat condition with {\it no} Lorentz boost ({\it i.e.}~with {\it zero} momentum, as defined in~\eqref{thermo_P_def}, along the $z$-direction, $P=0$):
\begin{equation} \label{asympcond}
 f,\eta,\beta,\gamma \to 1\, ,\quad q\to 0\, , \quad r^2 q\to 0 \qquad (r\to\infty)\, .
\end{equation}
Then, $g\to 1$ automatically follows from the equations of motion. This imposes 6 conditions on the coefficients in the asymptotic expansion in $r \to \infty$, whose detailed analysis is given in appendix~\ref{app:uvser}. Note that we require that both the leading term of $q$ as well as its next-to-leading term vanish. As we will see in section~\ref{sec:charges}, $r^2 q|_{r=\infty}$ is related to the amount of the Lorentz boost along the string. The condition \eqref{asympcond} corresponds to imposing no Lorentz boost ($P=0$).

To summarize our asymptotic analyses, in units where $r_+=1$, there are 8 free parameters at the horizon and 6 conditions at spatial infinity that these coefficients have to obey. Hence, the helical black string is a 2-parameter family. One might naively think that 6 parameters at the horizon need to be tuned. In practice, however, the number of the tuning parameters can be reduced to 2 because of 4 residual coordinate degrees of freedom, whose details are explained in appendices~\ref{app:scaling0} and \ref{app:tech}.

\subsubsection{Non-rotating frame at infinity}
\label{sec:nonrot}

The cohomogeneity-1 metric ansatz \eqref{screw_metric} is given in the rotating frame at infinity.
With the boundary condition (\ref{asympcond}), the asymptotic form of the metric \eqref{screw_metric} is
\begin{align}
\mathrm{d}s^2 &= -\mathrm{d}T^2 + \mathrm{d}r^2 + \frac{r^2}{4} \left[ \Sigma_1^2 + \Sigma_2^2  + \left( \Sigma_3 + 2 h_\infty \mathrm{d}T + \frac{1}{2} k_\infty \mathrm{d}Z \right)^2 \right] + \mathrm{d}Z^2\, ,
\label{asymp_metric_rot}
\end{align}
where $h_\infty \equiv h(r=\infty)$ and $k_\infty \equiv k(r=\infty)$ are non-zero in general. We assume that the black string is compactified with the length scale set by
\begin{equation}
L \equiv 2\pi/k_\infty \, ,
\label{kklengthscale}
\end{equation}
which is reduced to \eqref{L_for_MPBS} for the MPBS.

We will use the non-rotating frame at infinity when we discuss physical quantities such as conserved charges. The transformation to the non-rotating frame at infinity, whose coordinates are denoted by $(t,z,\psi)$, is given by
\begin{equation}
t = T\, , \quad z = Z\, , \quad \psi =\Psi + h_\infty T + \frac{k_\infty}{4} Z\, .
\label{rot2nonrot}
\end{equation}
For the MPBS, this is nothing but the inverse of \eqref{nonrot2rot4MP}. The dual vectors are related as
\begin{equation}
\partial_T = \partial_t + h_\infty \partial_\psi\, , \quad
\partial_Z = \partial_z + \frac{k_\infty}{4} \partial_\psi\, , \quad
\partial_\Psi = \partial_\psi\, .
\label{nonrotdualvectors}
\end{equation}
For MPBS, this coincides with \eqref{nonrotdualvectorsMP}. The asymptotic metric in the non-rotating frame at infinity reads
\begin{equation}
\mathrm{d}s^2 = -\mathrm{d}t^2 + \mathrm{d}r^2 + \frac{r^2}{4} \left( \sigma_1^2 + \sigma_2^2 + \sigma_3^2 \right) + \mathrm{d}z^2\, .
\label{asymp_metric_nonrot}
\end{equation}

When $\eta(r) \neq 1$, {\it i.e.}~when the $U(1)_\Psi$ is broken, the frame change \eqref{rot2nonrot} introduces explicit $(t,z)$ dependence in the bulk metric. This is because of the relation of the $SU(2)$-invariant 1-forms between the two frames,
\begin{equation}
\Sigma_\pm = e^{\pm i \Theta(t,z)} \sigma_\pm\, ,
\end{equation}
where
\begin{equation}
\Theta(t,z) \equiv 2h_\infty t + \frac{1}{2} k_\infty z\, .
\label{Theta_ty}
\end{equation}
The asymptotic metric does not have $(t,z)$-dependence because $\Sigma_1^2 + \Sigma_2^2 = \sigma_1^2 + \sigma_2^2$, as seen in \eqref{asymp_metric_nonrot}. The bulk metric has a part that is not proportional to $\Sigma_1^2 + \Sigma_2^2$ if $\eta(r) \neq 1$. It is transformed as
\begin{equation}
\begin{split}
&\eta \, \Sigma_1^2 + \frac{1}{\eta} \Sigma_2^2 \\
&= 2 \left( \eta + \frac{1}{\eta} \right) \Sigma_+ \Sigma_- + \left(\eta- \frac{1}{\eta} \right) \left( \Sigma_+^2 + \Sigma_-^2 \right) \\
&= \frac{1}{2} \left( \eta + \frac{1}{\eta} \right) \left( \sigma_1^2 + \sigma_2^2 \right) + \left(\eta- \frac{1}{\eta} \right) \left( \frac{1}{2} \cos (2\Theta) \left( \sigma_1^2 - \sigma_2^2 \right) + \sin (2\Theta) \sigma_1 \sigma_2 \right)\,.
\end{split}
\label{dSigma12}
\end{equation}
The last bracket, proportional to $(\eta - 1/\eta)$ vanishes for the MPBS ($\eta=1$), while it gives explicit $(t,z)$-dependence when $\eta \neq 1$. Then, continuous shifts of $t$ and $z$ are no longer independent isometries. The metric is periodic in $t$ and $z$. The periodicities of $t$ and $z$ are $\pi/(2 h_\infty)$ and $2\pi/k_\infty$, respectively. These are mixed with $\psi$-shifts to form helical isometries $\partial_T$ and $\partial_Z$ as in \eqref{nonrotdualvectors}.

We could consider alternative coordinates in which the bulk metric does not depend on ``time'' (that would be different from $t$) while the coordinates are non-rotating at infinity. We will comment on such coordinates in section~\ref{sec:timelike}.

\subsubsection{Thermodynamic quantities}
\label{sec:charges}

Requiring the canonically normalised asymptotic conditions~(\ref{asympcond}) except the last one (this condition  $r^2q|_{r\to\infty} \to 0$ corresponds to setting $c_q=0$), we obtain the asymptotic form of the metric components at spatial infinity as
\begin{equation}
\begin{alignedat}{4}
f &= 1 + \frac{c_f}{r^2} + \cdots\, , \quad &
h &= h_\infty + \frac{c_h}{r^4} + \cdots\, , \quad &
k &= k_\infty + \frac{c_k}{r^4} + \cdots\, , \quad &
q &= \frac{c_q}{r^2} + \cdots\, ,\\
\eta &= 1 + \cdots\, , \quad &
\beta &= 1 + \frac{c_\beta}{r^4} + \cdots\, , \quad &
\gamma &= 1 + \frac{c_\gamma}{r^2} + \cdots\, , &&
\end{alignedat}
\label{asymp_0th_power_sol}
\end{equation}
where the subleading terms of $\eta$ are suppressed exponentially. See appendix~\ref{app:uvser} for more details, where the asymptotic expansion is naively given in the form \eqref{asymp_0th_power_sol_app}, and then we set $(f_\infty,\gamma_\infty,q_\infty)=(1,1,0)$ in \eqref{asymp_0th_power_sol_app} by the scaling symmetries (\ref{scaling1}-\ref{scaling4}) (see also \eqref{condbyscaling}, which are nothing but the first two conditions in \eqref{asympcond}).
We do not impose $c_q=0$ at this point; it is instructive to keep track of $c_q$ for a while and see how it is related to a physical quantity of the string. Eventually, we will set $c_q=0$ when we compare results by selecting the Lorentz boost to be such that there is no momentum ($P=0$).

Using the asymptotic solutions \eqref{asymp_0th_power_sol}, we derive the Brown-York quasi-local stress tensor \cite{Brown:1992br,Kraus:1999di} (see appendix~\ref{app:counterterm} for details). The result is given in the non-rotating frame at infinity as
\begin{equation}
\begin{split}
&16\pi G_6 r^3 T_{ab}\mathrm{d}x^a \mathrm{d}x^b\\
&=
-(3 c_f+c_\gamma)\,\mathrm{d}t^2
+4 c_q \, \mathrm{d}t \mathrm{d}z
+ 4 c_h \,\mathrm{d}t \, \sigma_3
+ (3 c_\gamma + c_f)\,\mathrm{d}z^2
+ c_k \, \mathrm{d}y \,\sigma_3\\
&\phantom{= \ }+2 c_\beta \sigma_3^2 +\frac{1}{6}(3 c_f^2 + 3 c_\gamma^2 + 2 c_f c_\gamma + 4 c_q^2-16 c_\beta)\mathrm{d}\Omega_3^2\, ,
\end{split}
\label{Tij_result_can}
\end{equation}
where $a,b=t,z,\theta,\phi,\psi$ and $\mathrm{d} \Omega_3^2 = (\sigma_1^2+\sigma_2^2+\sigma_3^2)/4$ is the metric of a unit $S^3$.

Conserved charges can be defined by integrating the quasi-local stress tensor. Let $\xi_a$ be an asymptotic Killing vector at $r \to \infty$. On a constant time slice for the asymptotic time $t$, the conserved charge associated to $\xi_a$ can be defined as
\begin{equation}\label{charge}
Q[\xi] = \lim_{r \to \infty} \int \mathrm{d}\Omega_3 \mathrm{d}Z \, r^3 \xi^a T_{ta}\, .
\end{equation}
Note that $\partial_t, \, \partial_\psi, \, \partial_z$ are not Killing vectors in the bulk if $\eta \neq 1$, but they are so at asymptotic infinity \eqref{asymp_metric_nonrot}, i.e. they are asymptotically Killing vector fields. Associated to these asymptotically Killing vectors are the energy $E$, angular momenta $J$, and momentum $P$ along the string  given by
\begin{align}
E &= \lim_{r \to \infty} \int \mathrm{d}\Omega_3 \mathrm{d}Z \, r^3 T_{tt} = -\frac{\pi L}{8 G_6} (3 c_f+c_\gamma)\, , \label{thermo_M_def}\\
J &= - \lim_{r \to \infty} \int \mathrm{d}\Omega_3 \mathrm{d}Z \,  r^3 T_{t(2\psi)} = - \frac{\pi L}{4 G_6}c_h\, , \label{thermo_J_def}\\
P &= -\lim_{r \to \infty} \int \mathrm{d}\Omega_3 \mathrm{d}Z \, r^3 T_{tz} = -\frac{\pi L}{4 G_6}c_q\, , \label{thermo_P_def}
\end{align}
where $\int \mathrm{d} \Omega_3 = 2 \pi^2$ and $L \equiv \int \mathrm{d}Z$.\footnote{In accordance with the work of \cite{Myers:1986un}, $J$ is defined for $T_{t(2\psi)} \mathrm{d} t \, \mathrm{d}(2\psi)$ because our $\psi$ is related to the Boyer-Lindquist angular coordinates $\phi_{1,2}$ as $\mathrm{d}(2\psi)=\mathrm{d}\phi_1+\mathrm{d}\phi_2$.}

The tension of the string is given by \cite{Townsend:2001rg}
\begin{equation}\label{thermo_Ten_def}
T_z = - \lim_{r \to \infty} \int \mathrm{d}\Omega_3 \, r^3 T_{zz} = -\frac{\pi}{4 G_6} (3 c_\gamma + c_f)\, .
\end{equation}
Note that the definition for $T_z$ does not include the $z$-integral. The tension of the MPBS satisfies
\begin{equation}
\frac{T_z L}{E} = \frac{1}{3}\, .
\label{TenLE4MPBS}
\end{equation}
We will see that this relation does not hold if $\eta \neq 1$.

The temperature $T_H$ and entropy $S_H$ of the black string can be obtained from the field variables at the horizon as
\begin{equation}
T_H = \left. \frac{\sqrt{f' g'}}{4 \pi} \right|_{r=r_+}\,, \quad
S_H = \frac{1}{4G_6}\int \mathrm{d} \Omega_3 dz \left. r^3 \sqrt{\beta \gamma}\right|_{r=r_+}\, . \label{thermo_THSH}
\end{equation}

The angular velocity of the horizon $\Omega_H$ and the horizon velocity along the string $v_H$ can identified from the Killing horizon generator in the non-rotating frame at infinity. By using \eqref{nonrotdualvectors}, the horizon generator (\ref{horgenerator_4khq_imposed}) is rewritten as
\begin{equation}
K = \partial_t + \left( h_\infty - \frac{k_\infty q_0}{4} \right) \partial_\psi - q_0 \partial_z\, ,
\end{equation}
where the coefficients $(h_\infty,k_\infty,q_0)$ are read off from numerical data in the rotating frame at infinity. Comparing this with $K=\partial_t + \Omega_H \partial_\psi + v_H \partial_z$, we find
\begin{equation}
\Omega_H = h_\infty - \frac{k_\infty q_0}{4} \,, \quad v_H = -q_0\, .
\label{formula_for_OmHvH}
\end{equation}

The quantities obtained above satisfy thermodynamic relations. The first law of thermodynamics is given by\footnote{The first law for the Schwarzschild black string \cite{Townsend:2001rg,Harmark:2003eg} was generalized to include boost \cite{Kastor:2007wr}. This can be straightforwardly generalized to the MPBS.}
\begin{equation}\label{first_law}
\mathrm{d}E = T \mathrm{d}S + 2\Omega_H \mathrm{d}J + v_H \mathrm{d} P + T_z^\mathrm{eff} \mathrm{d}L\, ,
\end{equation}
where the effective tension is introduced as \cite{Kastor:2007wr}\footnote{\label{foot:Tension}The helical black strings that we will find have $P=0$ $-$ see  discussions of~\eqref{asympcond} or~\eqref{NoPcond} $-$ and thus, for our solutions, one effectively has $T_z^\mathrm{eff} = T_z$.}
\begin{equation}\label{EffT:def}
T_z^\mathrm{eff} \equiv T_z + \frac{v_H P}{L}\, .
\end{equation}
That is to say, $T_z$ is the conserved charge \eqref{thermo_Ten_def} associated to $\partial_z$ but the thermodynamic potential conjugate to the string length $L$ is $T_z^\mathrm{eff}$ (not $T_z$) which differs from $T_z$ when there is momentum along the string.
The Smarr formula can be derived by using the first law and dimensional analysis as
\begin{align}\label{Smarr}
E & = \frac{4}{3}(T S +2 \Omega_H J) + v_H P + \frac{1}{3} T_z^\mathrm{eff} L \nonumber\\
& = \frac{4}{3}(T S + 2\Omega_H J + v_H P ) + \frac{1}{3} T_z L\, ,
\end{align}
These relations can be explicitly checked for the boosted MPBS \eqref{boost_MPBS_components} analytically. We can also check them for numerical solutions.  From the Smarr relation, the relation
\begin{equation}
\left|1- \frac{4(TS+\Omega_H J+v_H P)+L T_z}{3E}\right|
\label{smarr_ratio}
\end{equation}
must vanish.
We use this for monitoring our numerical results, that fail to satisfy the Smarr relation when the numerical errors increase. This happens especially when the deformation $(1-\eta_0)$ is increased. We stop data generation if \eqref{smarr_ratio} is not satisfied with an accuracy $0.1\%$.

We measure the thermodynamic quantities in units of $L$ as
\begin{align}\label{dimensionlessThermo}
& \mathcal{E}\equiv \frac{E}{L^3}\,, \qquad \mathcal{J}\equiv \frac{J}{L^4}\,, \qquad \mathcal{T}_z\equiv \frac{T_z}{L^2}\,,\qquad  \mathcal{P}\equiv \frac{P}{L^3}\,,   \nonumber \\
& \tau_H \equiv T_H L\,, \qquad \sigma_H\equiv \frac{S_H}{L^4}\,,\qquad  \omega_H\equiv \Omega_H L \,, \qquad v_H\,.
\end{align}
That is to say, the thermodynamic quantities relevant for physical discussions are the {\it dimensionless} quantities \eqref{dimensionlessThermo}.
In an abuse of language but for simplicity, onwards  we will omit the prefix ``dimensionless-''  when referring to the dimensionless thermodynamic quantities \eqref{dimensionlessThermo}. For example, the dimensionless-energy $\mathcal{E}$ will be simply called the energy.
For the MPBS, the angular momenta are bounded from above as
\begin{equation}
G_6\mathcal{J} \le \frac{2^{3/2}}{3^{3/2} \pi^{1/2}} (G_6\mathcal{E})^{3/2} \, ,
\end{equation}
which follows from $a \leq r_0$. The equality is satisfied in the extremal MPBS (zero temperature). We can write the extremal MPBS's $\mathcal{J}$ and $\sigma_H$ as functions of $\mathcal{E}$,
\begin{equation}
G_6\mathcal{J}^\mathrm{ext} = \frac{2^{3/2}}{3^{3/2} \pi^{1/2}} (G_6\mathcal{E}^\mathrm{ext})^{3/2}\, , \quad
G_6 \sigma_H^\mathrm{ext} = \frac{2^{5/2} \pi^{1/2}}{3^{3/2}} (G_6\mathcal{E}^\mathrm{ext})^{3/2}\, ,
\label{themo_ext_J}
\end{equation}
and these quantities will be shown later in  Fig.~\ref{fig:thermoJ}.

We can show that the parameter region of the angular velocity $\omega_H$ is bounded for helical black strings. From \eqref{asymp_m}, the presence of the exponentially decaying asymptotic solution ($\mu>0$) implies
\begin{equation}
k_\infty - 4 h_\infty>0 \, ,
\end{equation}
where we have imposed $f_\infty=\gamma_\infty=1$ and $q_\infty=0$. The exponential tail ceases to exist if $k_\infty=4h_\infty$. Rewriting this equality by using \eqref{kklengthscale} and \eqref{formula_for_OmHvH}, we obtain
\begin{equation}
\omega_H = \frac{\pi(1+v_H)}{2} \, .
\label{omegaH_allowed}
\end{equation}
Because the parameter region of $v_H$ can be $-1 < v_H < 1$, where $|v_H|=1$ corresponds to the speed of light, it follows that the range of $\omega$ for the helical black string is $0 < \omega_H < \pi$. The angular velocity will be maximal $\omega_H \to \pi$ when $v_H \to 1$ (and minimal $\omega_H \to 0$ if $v_H \to -1$). We will come back to this when discussing the results of Fig.~\ref{fig:thermoJ}.

\subsection{Helical black string ansatz in the Einstein-DeTurck gauge}
\label{sec:ansatzEdT}

In the previous subsection~\ref{sec:ansatzSphericalGauge} we described how we can search for helical black strings using the spherical gauge ansatz~\eqref{screw_metric}. To further test our numerical results, and to explore some corners of the parameter space more easily, in this subsection we describe an alternative/independent approach to construct the helical black strings. This time we use the Einstein-DeTurck gauge with associated ansatz \eqref{ansatz} that we describe below. Of course, the physical properties of the two numerical/gauge constructions agree. Moreover, in section~\ref{sec:perturbative} we will use the Einstein-DeTurck gauge formulation to find a perturbative description of the helical black strings (alternatively, we could have done this perturbative analysis in the spherical gauge).

Without fixing the gauge, a most general ansatz that  describes an asymptotically ${\cal M}^{1,4}\times S^1$ rotating string with the isometries of a helical black string $-$ isometry group $\mathbb{R}_\tau\times \mathbb{R}_x \times SU(2)$ and discrete isometries $P_1$ and $P_2$ $-$ can be written as (in the rotating frame at infinity)
\begin{align}\label{ansatz}
{\mathrm ds}^2&=r_+^2\bigg\{ -\frac{y^2 \mathcal{F} \,q_1}{\mathcal{H}}\,{\mathrm d}\tau^2
+\frac{4 q_2}{(1-y^2)^4 \mathcal{F}}\,{\mathrm d}y^2 +q_5\left(\frac{\widetilde{L}}{2}{\mathrm d}x+q_8 y^2(1-y^2) {\mathrm d}\tau \right)^2
\nonumber\\
&\qquad\quad\: +\frac{1}{(1-y^2)^2}
\bigg[ q_3 \mathcal{H} \Bigg(\frac{ \Sigma_3}{2}+ \frac{y^2 \mathcal{W}}{\mathcal{H}}
\left[ 1+ (1-y^2)^3 q_6\right]{\mathrm d}\tau \nonumber\\
&\qquad\quad
+\frac{\pi}{4}\big[ 1+(1-y^2)^3 q_7 \big] dx \Bigg)^2
+q_4 \frac{1}{4}\left( q_9 \Sigma_1^2 +\frac{ \Sigma_2^2}{q_9}\right) \bigg] \bigg\},
\end{align}
where $q_j=q_j(y)$, for $j=1,2,3,\cdots, 9$, are nine functions of a radial coordinate $y$ (to be  discussed below), and all coordinates are adimensional: $\tau$ is a time coordinate, $x$ is the direction along which the string $-$ with physical length $L$  $-$ extends, and $ \Sigma_{1,2,3}$ are the $SU(2)$ left-invariant 1-forms of $S^3$, in the rotating frame, as  defined in \eqref{su2_1forms}. Finally, in \eqref{ansatz} one has:
\begin{align}\label{ansatz2}
\mathcal{F}(y)&=\left(2-y^2\right) \left(1-\frac{\widetilde{a} ^2}{1-\widetilde{a} ^2}(1-y^2)^2\right)\,,\nonumber\\
\mathcal{H}(y)&=1+\frac{\widetilde{a} ^2}{1-\widetilde{a} ^2}(1-y^2)^4\,,\nonumber\\
\mathcal{W}(y)&=\widetilde{a}  \left(2-y^2\right) \left[1+\left(1-y^2\right)^2\right],
\end{align}
where $\widetilde{a}=a/r_+$.

In \eqref{ansatz},  $x\in[-1,1]$ is the periodic coordinate along the string direction. The radial coordinate $y$ is  compact: one has $y\in[0,1]$ with $y=0$ being the  horizon location and $y=1$ the asymptotic boundary. The ansatz  \eqref{ansatz}  describes black strings with horizon generator
\begin{equation}\label{SKV}
K=\partial_\tau \,.
\end{equation}

To further understand the motivation for the ansatz  \eqref{ansatz}, note that doing the coordinate and field redefinitions
\begin{align}\label{CoordTransfNonRotatingFrame}
& \tau =\frac{1}{r_+}\frac{t-v \,z}{\sqrt{1-v^2}}\,, \quad y=\sqrt{1-\frac{r_+}{r}}\,,\quad\Psi =\psi -\frac{2a}{r_+}\,  \tau -\frac{\pi}{2}\,x \,, \quad  x=\frac{2}{\widetilde{L}}\,\frac{1}{r_+}\frac{z- v \,t }{\sqrt{1-v^2}}\,;  \nonumber \\
& \mathcal{F}=\left( 1-\frac{r_+}{r}\right)^{-1} F\,,\qquad \mathcal{H}=H\,,  \qquad
\mathcal{W}=\left( 1-\frac{r_+}{r}\right)^{-1}\left(\frac{a}{r_+}\,H-r_+ \Omega \right)\,,
\end{align}
while also setting $q_{1,2,3,4,5,9}=1$ and $q_{6,7,8}=0$, takes \eqref{ansatz} into
the MPBS \eqref{MPstring}-\eqref{MPfns} with an additional boost $v$ along the string direction.
The relation between the (dimensionful) string length $L$ and the dimensionless parameter $\widetilde{L}$ introduced in \eqref{ansatz} is
\begin{equation}\label{dimensionlessL}
 \widetilde{L}= \frac{L}{r_+} \,  \left(\sqrt{1-v^2} + v\,\frac{2}{\pi} \frac{a}{r_+} \frac{L}{r_+}\right)^{-1}\,.
\end{equation}
So, in the absence of boost, $v=0$, one has  $x=\frac{2}{\widetilde{L}}\,z$, where $z\in[-L/2,L/2]$ is the coordinate used in \eqref{MPstring} and $\widetilde{L}=L/r_+$.

 The ansatz  \eqref{ansatz} is in the rotating frame, {\it i.e.}~it describes solutions that have rotation and velocity along the string direction at infinity,  but have vanishing angular velocity and  velocity at the horizon.  However, the coordinate transformation \eqref{CoordTransfNonRotatingFrame} brings the system to a frame $-$ that we call the non-rotating frame $-$ that has no rotation nor velocity at infinity.  Consequently, in this frame, solutions of  \eqref{ansatz}   have angular velocity $\Omega_H=\frac{2 \widetilde{a}}{r_+}  \sqrt{1-v^2}$ and velocity $v_H=v$  at the horizon, and the horizon generator \eqref{SKV} now reads\footnote{Strictly speaking, \eqref{SKV} and \eqref{SKVnonRotatingFrame} differ by an (irrelevant) global factor of $r_+/\sqrt{1-v^2}$.}
\begin{align}\label{SKVnonRotatingFrame}
&K=\partial_t+ \Omega_H\, \partial_\psi + v_H\,\partial_z\,,\quad \\
& \hbox{with} \quad \Omega_H=\frac{2 a}{r_+^2}  \sqrt{1-v^2} \quad \hbox{and} \quad v_H=v\,. \nonumber
\end{align}
Later, we will choose $v$ such that there is also no momentum at infinity along the string direction. Essentially, the latter is proportional to a subleading decay of $g_{tz}$. Thermodynamics quantities will be computed in this non-rotating frame.

Unlike the $S^2$ or spherical gauge ansatz \eqref{screw_metric} that we used in section~\ref{sec:ansatzSphericalGauge}, the DeTurck ansatz \eqref{ansatz} leaves the gauge unconstrained (the gauge is fixed {\it \`a posteriori} after solving the equations of motion \eqref{EdeT}).
In particular, the system has a nonlinear symmetry that leaves the metric \eqref{ansatz} invariant when we shift the Euler angle $\Psi$ by a constant $\Psi_0$, $\Psi \to \Psi+\Psi_0$, and there is a choice of $\Psi_0$ that allows to set the cross term $\Sigma_1 \Sigma_2$ to zero. Perturbations along $\mathbb{C}\mathrm{P}^{1}$ have thus only two degrees of freedom that at nonlinear level are described by the functions $q_4$ and $q_9$.

To discuss the relation between the spherical gauge \eqref{screw_metric}  and DeTurck \eqref{ansatz} ansatz\"e, note that \eqref{screw_metric} corresponds to taking \eqref{ansatz} and setting
\begin{align}\label{relationAnsatze}
& q_1=\frac{r\,f}{r-r_+} \frac{\mathcal{H}}{\mathcal{F}} , \qquad q_2=\left(1-\frac{r_+}{r}\right)\frac{\mathcal{F}}{g},\qquad q_3=\frac{\beta}{\mathcal{H}}  \,, \quad q_4=1, \quad q_5=\gamma, \nonumber \\
& q_6=\frac{r^3\left[r_+ r \,h \,\mathcal{H}-(r-r_+)\mathcal{W}\right]}{r_+^3(r-r_+)\mathcal{W}},\qquad    q_7=\frac{r^3}{r_+^3} \left(\frac{\widetilde{L} r_+}{2 \pi }\,k -1\right),  \nonumber \\
& q_8=\frac{r}{r_+}\left(1-\frac{r_+}{r}\right)^{-1}q,  \qquad q_9=\eta,
\end{align}
while doing the coordinate transformations $\tau=\frac{T}{r_+}$, $y=\sqrt{1-\frac{r_+}{r}}$ and $x=\frac{2}{\widetilde{L}}\frac{Z}{r_+}$.

Further notice that linearizing   \eqref{ansatz} about the MPBS one gets a perturbation that is described by \eqref{MPBS_fluc2} with $\delta\eta\equiv \delta q_9$ (after imposing the traceless-transverse gauge that sets $q_4=1$). Thus,  \eqref{ansatz} is a good ansatz to study the nonlinear back-reaction of the superradiant onset mode discussed previously. This is possible because this onset mode is regular both at the future and past event horizons.

 With the exception of the MPBS limit, solutions of \eqref{ansatz} are not time independent nor axisymmetric since the associated $\mathbb{R}_t$ and $U(1)_\psi$ symmetries that exist in the equal angular momenta MPBS are broken in \eqref{ansatz} when $q_9\neq 1$.\footnote{This becomes evident when we return to the frame that does not rotate at infinite and rewrite
 \begin{equation*}
q_9 \Sigma_1^2+\frac{1}{q_9}\Sigma_2^2
 =\frac{1}{2} \left(q_9-\frac{1}{q_9}\right)
 \big[ \left(\sigma_1^2-\sigma_2^2\right) \cos (4 \Omega_H t )+2 \sigma_1 \sigma_2 \sin (4 \Omega_H t )\big]+\frac{1}{2} \left(q_9+\frac{1}{q_9}\right) \left(\sigma_1^2+\sigma_2^2\right)
\end{equation*}
which explicitly depends on $t$ when $q_9\neq 1$. Moreover, this term also depends explicitly on $\psi$, when  $q_9\neq 1$, as can be seen when we expand $\left(\sigma_1^2-\sigma_2^2\right)= \cos (2 \psi) \left(-\mathrm{d}\theta^2+\sin ^2\theta\mathrm{d}\phi ^2\right)-2 \sin (2 \psi ) \sin\theta \mathrm{d}\theta \mathrm{d}\phi $ in terms of the Euler angles. Further recall that $\left(\sigma_1^2+\sigma_2^2\right)$ is the line element of $\mathbb{C}\mathrm{P}^{1}\simeq S^2$.} However, the helical black string solutions of  \eqref{ansatz}  do preserve the translation invariance along the string direction $x$.
Thus, \eqref{ansatz}  has  $\mathbb{R}_\tau \times \mathbb{R}_x \times SU(2)$ symmetries with the $\mathbb{R}_\tau$ (time-periodic) isometry generated by the helical horizon generator \eqref{SKV} and $\mathbb{R}_x$ by the Killing vector field $\partial_x$. The helical black strings we will find are stationary spacetimes since we will conclude that the helical Killing vector field \eqref{SKV} is everywhere timelike at the asymptotic boundary, $|\partial_\tau|_{y=1}<0$.

It is important to emphasize that the MPBS and the helical black strings that we find here are not the only asymptotically $\mathcal{M}^{1,4}\times S^1$ equal angular momenta solutions of Einstein gravity. Indeed, there are also black resonator string solutions recently found in \cite{Dias:2022str}. Like the helical black strings, black resonator strings are time-periodic solutions but not time symmetric nor axisymmetric.\footnote{They are the string counterparts of the black resonators in AdS with spherical horizon topology \cite{Dias:2011at,Dias:2015rxy,Ishii:2018oms,Ishii:2020muv,Ishii:2021xmn}.} But, unlike the helical strings, resonator strings further break translational invariance along the string direction. Ultimately, this justifies why black resonator strings are cohomogeneity-2 solutions  \cite{Dias:2022str} while helical black strings are cohomogeneity-1 solutions.
Interestingly, the helical strings share another important feature with resonators strings that is ultimately responsible for the interesting fact that they encounter each other in phase diagrams. Indeed, helical and resonator strings merge with MP strings along the same 1-parameter family of solutions. This merger is described by the onset superradiant mode described by  \eqref{MPBS_fluc}  or \eqref{MPBS_fluc2}. This superradiant onset mode is the same for the resonator and helical systems, as discussed in detail in \cite{Dias:2022mde}.

In this section, because our ansatz \eqref{ansatz} itself has not fixed a gauge, we find the helical string solutions using the Einstein-DeTurck formalism \cite{Headrick:2009pv,Figueras:2011va,Wiseman:2011by,Dias:2015nua}.
This formalism formulates the Einstein equation into a manifestly elliptic form without {\it \`a priori} gauge fixing. The DeTurck gauge-fixing  is not an algebraic gauge-fixing condition. Instead,  the DeTurck gauge-fixing is itself a differential equation for the (unknown) metric $g$ given a choice of reference background $\overline g$ that is  solved simultaneously with the gravitational differential equations. Consequently, the differential DeTurck gauge-fixing condition
is imposed by the end of the computation, \ie after solving the full set of equations.
The Einstein-DeTurck formulation of the gravitational equations requires a choice of reference metric $\overline g$, which must have the same causal structure and contain the same symmetries of the desired solution (it can have other symmetries).  For the reference metric we choose the MPBS which, as discuss above, is given by \eqref{ansatz} with  $q_{1,2,3,4,5,9}=1$ and $q_{6,7,8}=0$. The DeTurck method modifies the Einstein equation $R_{AB}=0$ into
\begin{equation}\label{EdeT}
R_{AB}-\nabla_{(A}\xi_{B)}=0\;,\qquad \xi^A \equiv g^{C D}[\Gamma^A_{\:C D}-\overline{\Gamma}^A_{\:C D}]\;,
\end{equation}
where $\Gamma$ and $\overline{\Gamma}$ define the Levi-Civita connections for $g$ and $\bar g$, respectively. Unlike $R_{AB}=0$ with our ansatz, this equation yields a well-posed elliptic boundary value problem. Indeed, it was proved in \cite{Figueras:2011va} and \cite{Figueras:2016nmo} that static and stationary (with $t-\psi$ symmetry) solutions to \eqref{EdeT} necessarily satisfy  the DeTurck gauge-fixing condition $\xi^A=0$, and hence are also solutions to $R_{AB}=0$. Note that the results of  \cite{Figueras:2011va,Figueras:2016nmo}  apply to asymptotically flat and asymptotically AdS spacetimes and, of relevance here, to asymptotically Kaluza-Klein backgrounds.

We now have to discuss the physical boundary conditions of the problem.
At the asymptotic boundary, $y=1$, we impose as a Dirichlet condition that our solutions must approach the reference metric. At $y=0$, we demand a regular bifurcate Killing horizon generated by $\partial_\tau$. This amounts to imposing Neumann boundary conditions, $q'_j(0)=0$, $j=1,\cdots 9$.

We are now ready to solve the Einstein-DeTurck differential equations subject to the above boundary conditions. We will do this within perturbation theory (to higher order) in section~\ref{sec:perturbative}. But we will also solve the nonlinear problem numerically. For that,
we will use a standard Newton-Raphson algorithm and discretise the Einstein-DeTurck equations using pseudospectral collocation (with Chebyshev-Gauss-Lobatto nodes). The resulting algebraic linear systems are solved by LU decomposition. These methods are detailed in the review \cite{Dias:2015nua}.

The helical black strings depend on three dimensionful parameters that we could take to be the horizon radius $r_+$, the rotation parameter $a$ and the length $L$.  For convenience, we will opt here to work in units of $r_+$ by using the scaling symmetry
\begin{equation}\label{scalingSym}
\{\tau,y,\theta,\phi,\Psi\}\to \{\tau,y,\theta,\phi,\Psi\},\qquad \{r_+,a,L\}\to \left\{\lambda \, r_+, \lambda\, a, \lambda L \right\}, \qquad \{ q_{j} \}\to \{ q_{j} \}\,,
\end{equation}
which rescales the line element as $\mathrm{d}s^2\to \lambda^2 \,\mathrm{d}s^2$, namely $g_{AB}\to \lambda^2 \, g_{AB}$ but leaves the equations of motion invariant (since the affine connection $\Gamma^C_{\phantom{C}AB}$, and the Riemann ($R^A_{\phantom{A}BCD}$) and thus Ricci ($R_{AB}$) tensors are left invariant). We can use this scaling symmetry to fix the horizon radius to $r_+ \equiv 1$ so that helical strings are parametrized by only two dimensionless ratios, $\widetilde{a}=a/r_+$ and $\widetilde{L}$ defined in terms of $L/r_+$ in \eqref{dimensionlessL}.

However, it is more appropriate to measure thermodynamic/physical quantities in units of $L$ as in \eqref{dimensionlessThermo} with $ v_H=v$.  The (dimensionful) energy $E$, angular momenta $J$, tension $T_z$ and associated effective tension \eqref{EffT:def} (see also footnote~\ref{foot:Tension}), momentum $P$, temperature $T_H$, entropy $S_H$,  angular velocity $\Omega_H$ and velocity $v_H$ are computed from \eqref{ansatz}  after moving to the non-rotating frame via  \eqref{CoordTransfNonRotatingFrame} and using the counterterm formalism of Appendix~\ref{app:counterterm}  \cite{Kraus:1999di,Mann:2005yr,Kleihaus:2009ff}  and \eqref{charge}-\eqref{thermo_Ten_def}. In terms of the functions introduced in  \eqref{ansatz}, the associated dimensionless thermodynamic quantities \eqref{dimensionlessThermo} then read:
\begin{eqnarray}\label{ThermoHelDless}
&&\mathcal{E}=\frac{1}{G_6 \widetilde{L}^2} \, \frac{(\pi -2 \widetilde{a} \widetilde{L} v)^2}{32 \pi  \left(1-v^2\right)^2}
\Big( \frac{4 \left(3-v^2\right)}{1-\widetilde{a} ^2}+3 q_2''(1)+q_5''(1)+8 v q_8'(1)
-v^2 \left[q_1''(1)+3 q_2''(1)\right]\!\!
 \Big), \nonumber \\
&& \mathcal{J}=\frac{1}{G_6 \widetilde{L}^4}\,\frac{(\pi -2 \widetilde{a} \widetilde{L} v)^3}{32 \pi ^2 \left(1-v^2\right)^2}
 \left( \frac{4 \widetilde{a} \widetilde{L}}{1-\widetilde{a} ^2} + 2 \widetilde{a} \widetilde{L} q_6'(1)-\pi  v q_7'(1) \right), \nonumber\\
&& \mathcal{T}_z^\mathrm{eff}=\frac{1}{G_6 \widetilde{L}^2}\, \frac{(\pi -2 \widetilde{a} \widetilde{L} v)^2}{32 \pi  \left(1-v^2\right)^2} \Big( \frac{4 \left(1-3 v^2\right)}{1-\widetilde{a} ^2}+q_1''(1)-v^2 q_5''(1)-8 v q_8'(1) +3 \left(1-v^2\right) q_2''(1)\! \Big),
\nonumber \\
&& \mathcal{P}=\frac{1}{G_6  \widetilde{L}^2} \,\frac{(\pi -2 \widetilde{a}  \widetilde{L}v)^2}{32 \pi  \left(1-v^2\right)^2}
\Big( \frac{4 v}{1-\widetilde{a} ^2} +2 \left(1+v^2\right) q_8'(1) -\frac{1}{2} v \left[q_1''(1)-q_5''(1)\right] \Big),
\nonumber \\
&&\tau_H=\frac{\widetilde{L}}{2 \pi} \, \frac{\pi  \left(1-v^2\right)}{\pi -2 \widetilde{a} \widetilde{L} v} \,\frac{1-2 \widetilde{a} ^2}{\sqrt{1-\widetilde{a} ^2}}\,,\nonumber \\
&& \sigma_H=\frac{1}{G_6 \widetilde{L}^3}\, \frac{(\pi -2 \widetilde{a} \widetilde{L} v)^3}{2 \pi  \sqrt{1-\widetilde{a} ^2} \left(1-v^2\right)^2}
 \sqrt{q_3(0) q_4(0)^2 q_5(0)}\,,\nonumber\\
 && \omega_H=2 \widetilde{a} \,\widetilde{L} \,\frac{\pi  \left(1-v^2\right)}{\pi -2 \widetilde{a} \widetilde{L} v} \,,\nonumber \\
 && v_H=v\,.
\end{eqnarray}

Starting from the first law \eqref{first_law} and Smarr relation \eqref{Smarr} for the dimensionful  thermodynamic quantities,
one can show (following the steps of, \eg, \cite{Dias:2022str}) that the first law and Smarr relations for the dimensionless quantities \eqref{ThermoHelDless} (a.k.a. Gibbs-Duhem and Euler relations) are:
\begin{align}
& \dd  \mathcal{E}=  \tau_H\,\dd \sigma_H+ 2\,\omega_H \, \dd \mathcal{J} +v_H\dd \mathcal{P}\,, \label{1stLawDless}
\\
& \mathcal{E}=\frac{4}{3}\left( \tau_H \,\sigma_H +  2\,\omega_H \, \mathcal{J} +\frac{3}{4}v_H \mathcal{P}+ \frac{1}{4}\mathcal{T}_z^\mathrm{eff}  \right).\label{GibbsDuhemEuler}
\end{align}
We will  use these  two relations to check our results.

We are interested on helical black strings that have {\it no} momentum $ \mathcal{P}$ along the string. This fixes the boost velocity parameter $v$ to be:\footnote{In the spherical gauge ansatz case, this condition was imposed in~\eqref{asympcond}.}
\begin{eqnarray}\label{NoPcond}
 \mathcal{P}=0  \quad \Rightarrow  \quad &&  v=-\frac{8 \left(1-\widetilde{a} ^2\right) q_8'(1)}
 {\chi+\sqrt{\chi^2-64 \left(1-\widetilde{a} ^2\right)^2 q_8'(1)^2}}  \\
&&\hbox{with $\chi=8-\left(1-\widetilde{a} ^2\right) \left[q_1''(1)-q_5''(1)\right]$}.   \nonumber
\end{eqnarray}
Note that although there is no linear momentum along the string,  the horizon linear velocity is non-vanishing, $v_H=v$.

When we set $q_{1,2,3,4,5,9}=1$ and $q_{6,7,8}=0$  (and thus $v=0$) in \eqref{ThermoHelDless} we recover the dimensionless thermodynamic quantities of the MP string. The latter is a 2-parameter family of solutions parametrized by $(\widetilde{L},\widetilde{a})$, and the extremal MPBS is a 1-parameter family of solutions with $\widetilde{a}=1/\sqrt{2}$ parametrized by $\widetilde{L}$.

We can now discuss the strategies we will use to numerically generate the 2-parameter space of helical black strings.
The above discussion naturally invites us to follow one of two strategies (we will use both):
\begin{enumerate}
\item We can choose to generate lines of helical strings that have the same dimensionless rotation $\widetilde{a}$  as the MPBS they bifurcate from. The dimensionless length $\widetilde{L}$ is varying along these lines of solutions.
\item Alternatively, we can choose to generate lines of helical strings that have the same dimensionless length $\widetilde{L}$ as the MP string they bifurcate from (\ie  we fix the length to be $L=2\pi /k_{(0)}$ where $k_{(0)}$ is the zero mode wavenumber for the superradiant instability of the MP string that was found in \cite{Dias:2022str} and discussed previously).
\end{enumerate}
We use these two strategies to span the 2-dimensional phase space parameter of helical black strings. We will present our results in section~\ref{sec:result}.

\section{Perturbative construction of helical black strings}
\label{sec:perturbative}

In the previous subsection~\ref{sec:ansatzEdT} we have set up the boundary value problem (BVP) that will allow us to find the helical black strings in the Einstein-DeTurck gauge~\eqref{ansatz} that obey the boundary conditions discussed below \eqref{EdeT}. This nonlinear BVP can be solved in full generality  using numerical methods outlined below \eqref{EdeT} \cite{Dias:2015nua}. Alternatively, we can find the helical black strings using the spherical gauge ansatz~\eqref{screw_metric} of section~\ref{sec:ansatzSphericalGauge}.
We will find these nonlinear solutions using both formulations and discuss the full phase diagram of solutions in section~\ref{sec:result}.

Meanwhile, in the present section, we complement these nonlinear numerical analyses with a nonlinear perturbative analysis that finds helical strings in the region of the phase diagram near their merger with the MPBS. For that we choose to work in the Einstein-DeTurck gauge ansatz \eqref{ansatz} of subsection~\ref{sec:ansatzEdT} (although, with no disbenefit, we could have used the spherical gauge ansatz~\eqref{screw_metric} of subsection~\ref{sec:ansatzSphericalGauge}).
This perturbative analysis will already provide valuable physical properties of the system. Additionally, these perturbative results will also be important to test the numerical results of section \ref{sec:result}. We  solve the  BVP in perturbation theory up to fifth order in the expansion parameter where we can distinguish the thermodynamics of the helical and MP strings.

We follow a perturbative approach  developed in \cite{Dias:2017coo} (to find vacuum lattice branes), in \cite{Bea:2020ees} (to find AdS lumpy branes) and, more recently, to find black resonator strings in \cite{Dias:2022str}. This perturbative scheme has its roots in \cite{Gubser:2001ac,Wiseman:2002zc,Sorkin:2004qq} (to explore the existence of vacuum non-uniform black strings associated to the Gregory-Laflamme instability of the Schwarzschild black string). More concretely, our perturbative strategy has two main steps:
\begin{enumerate}

\item At linear ($n=1$) order in perturbation theory, we find the locus in the phase space of MP strings where a superradiant onset mode, namely a mode that is marginally stable, exists.
In practice, we  identify this locus by finding the critical length $\widetilde{L}=\widetilde{L}_{(0)}$ (wavenumber $\widetilde{k}_{(0)}=2\pi/\widetilde{L}_{(0)}$) above (below) which MP strings become locally unstable (stable). This linear analysis was already performed in \cite{Dias:2022str} and is briefly reviewed below.

\item The second step is to extend perturbation theory to higher orders, $n\geq 2$, and construct the helical black strings that bifurcate  (in a phase diagram of solutions) from the superradiant onset curve of MP strings.
\end{enumerate}

We adopt a perturbation scheme that is consistent with our nonlinear ansatz \eqref{ansatz}  and we will linearize the nonlinear Einstein-DeTurck equations of motion \eqref{EdeT} to get the perturbative equations of motion (EoM). At linear order in perturbation theory about the MPBS, we consider perturbations of the form $q_j(y)=\mathcal{Q}_j+\epsilon \, q_j^{(1)}(y)$ where $\mathcal{Q}_{1,2,3,4,5,9}=1$, $\mathcal{Q}_{6,7,8}=0$ describe the background solution and $q_j^{(1)}(y)$ its relevant linear perturbations.\footnote{The superscript $\! ^{(n)}\!$ here and henceforth always denotes the order $n$ of the perturbation theory, not order of derivatives.}  Here, $\epsilon\ll 1$ is the amplitude of the linear perturbation and, ultimately, it will be the expansion parameter of our perturbation theory to higher order.
Consistent with the discussion of section~\ref{sec:isopert}, we want to consider perturbations of the form \eqref{MPBS_fluc2} that break the $U(1)_\Psi$ symmetry of the MP string. This means that, at linear order, the only metric component that is perturbed is $q_9(y)$ with a deformation of the form
\begin{equation}\label{PT:n1}
q_9^{(1)}(y)=\mathfrak{q}_9^{(1)}(y)\,,\qquad q^{(1)}_{1,2,3,4,5,6,7,8}(y)=0.
\end{equation}
The physical length $L$ of the periodic coordinate is given in terms of the wavenumber $k$ of the perturbation by $L=2\pi/k$, and it will change with high-order corrections in perturbation theory. Since the equations of motion depend on $L$, this relation $L=2\pi/k$ introduces the onset mode wavenumber $k$ in the problem. We denote the leading-order contribution to $\widetilde{k}$ as $\widetilde{k}_{(0)}$.

In these conditions, the linearized equations of motion for $q_9^{(1)}(y)$  becomes a quadratic eigenvalue problem for $\widetilde{\kappa}\equiv  \sqrt{\widetilde{k}_{(0)}^2-4 \widetilde{\Omega}_H^2}$ (where  $\widetilde{\Omega}_H=\Omega_H/r_+$  and  $\widetilde{k}=k r_+$) that we solve to ultimately get the {\it leading-order}  wavenumber $\widetilde{k}_{(0)}$. Actually, this eigenvalue problem was already solved in \cite{Dias:2022str} since it determines the onset of the superrradiant instability on the MPBS.

Next, we want to climb the perturbation ladder to higher order, $n\geq 2$.
Before we do so, we must fix the expansion parameter of our perturbation scheme unambiguously.  We define $\mathfrak{q}_9^{(1)}(y)=\left(1-y^2\right)^{3/2} e^{-\sqrt{\widetilde{k}_{(0)}^2-4 \widetilde{\Omega}_H^2}/(1-y^2)} \widehat{\mathfrak{q}}_9(y)$ and choose $ \widehat{\mathfrak{q}}_9|_{y=1}\equiv 1$. The solution of the linear order eigenvalue problem (including, of course, its boundary conditions) then determines the horizon value $\widehat{\mathfrak{q}}_9|_{y=0}\equiv \widehat{\mathfrak{q}}_9^{H}$. We then require that higher order perturbations  do {\it not} change the value at the horizon of $q_9$. This procedure uniquely fixes the expansion parameter $\epsilon$.

Although the sector of perturbations we look at only excites $q_9(y)$ at linear order as in \eqref{PT:n1}, at higher orders the back reaction of the linear mode perturbs all metric components.
So, to find the solution at order $\mathcal{O}(\epsilon^n)$, we expand the metric functions and wavenumber in powers of $\epsilon$:
\begin{subequations} \label{PTexpansion}
\begin{eqnarray}
&& q_j(y)=\mathcal{Q}_j+\sum_{n=1}^{\infty} \epsilon^n \,\mathfrak{q}_j^{(n)}(y); \label{PTexpansionA} \\
&& \widetilde{k}=\sum_{n=1}^{\infty} \epsilon^{n-1} \widetilde{k}^{(n-1)}\equiv \widetilde{k}_{(0)}+\sum_{n=2}^{\infty} \epsilon^{n-1} \widetilde{k}^{(n-1)},\quad \hbox{with} \:\: \widetilde{L}=\frac{2\pi}{\widetilde{k}}, \label{PTexpansionB}
\end{eqnarray}
\end{subequations}
where $\mathcal{Q}_j$ ($j=1,\cdots,9$), already given above \eqref{PT:n1}, describe the background MPBS solution.

We have already found the $n=1$ contribution, \eqref{PT:n1} and $\widetilde{k}^{(0)}\equiv \widetilde{k}_{(0)}$, of this expansion by solving a homogeneous quadratic eigenvalue problem.  The expansion \eqref{PTexpansion} is now such that at order $\mathcal{O}(\epsilon^n)$, we must solve the BVP to find the coefficients $\{ \widetilde{k}^{(n-1)}, \mathfrak{q}_j^{(n)}\}$. Further note that, as explained above, our choice of perturbation scheme is such that the length $\widetilde{L}$ is corrected at each  order $n$. That is, one has
\begin{equation}
 \widetilde{L}= \widetilde{L}_{(0)}+\sum_{n=2}^{\infty} \epsilon^{n-1}  \widetilde{L}^{(n-1)}\,,
\end{equation}
  where the coefficients $\widetilde{L}^{(n-1)}$ can be read straightforwardly from \eqref{PTexpansionB}.

At order $\mathcal{O}(\epsilon^n)$, $n\ge 2$, the perturbative equations are no longer homogeneous. Instead, they describe an inhomogeneous BVP with a source  ${\cal S}_j^{(n)}$. Not surprisingly, this source  is a function of the lower order solutions $\{\widetilde{k}^{(i-1)}, \mathfrak{q}_j^{(i)} \}$, $i=1,\ldots,n-1$ (and their derivatives): ${\cal S}_j^{(n)}(\widetilde{k}^{(i-1)}, \mathfrak{q}_j^{(i)}$).
Actually, because $q_9$ is the only field that is  excited at linear order, it turns out that for $n\geq 2$ the perturbative equation for $q_9$ decouples from the set  of 8 coupled ODEs for $q_{1,\cdots,8}$.
More concretely, the structure of the perturbative equation of motion is
 \begin{eqnarray}
&& {\cal L}_H \, \mathfrak{q}_j^{(n)}={\cal S}_j^{(n)}, \quad \hbox{if $n\geq 2$ and $j=1,\cdots, 8$.}\label{highEoMgeneral}\\
&& {\cal L}_{H,9} \, \mathfrak{q}_{9}^{(n)}=\widetilde{k}^{(n-1)}\frac{8 \widetilde{k}_{(0)}  \mathfrak{q}_9^{(1)} }{(1-y^2)^4 f} + {\cal S}_9^{(n)}, \quad \hbox{if $n\geq 2$ and $j=9$.}\label{highEoMspecial}
\end{eqnarray}
The  differential operators ${\cal L}_H$ and ${\cal L}_{H,9}$ that describe the associated homogeneous system of equations are the same at each order $\mathcal{O}(\epsilon^n)$.  That is, they only depend on the MPBS $\mathcal{Q}_j$ we expand about and $\widetilde{k}_{(0)}$.
It follows that the complementary functions of the homogeneous system are the same at each order $\mathcal{O}(\epsilon^n)$, $n\geq 2$. But, we also need to find the particular integral of the inhomogeneous system and this is different for each $n$ since the sources ${\cal S}_j^{(n)}$ differ.
We now have to solve \eqref{highEoMgeneral} for $\mathfrak{q}_{j\leq 8}^{(n)}(y)$ and
\eqref{highEoMspecial} to find the eigenvalue $\widetilde{k}^{(n-1)}$ and $\mathfrak{q}_{9}^{(n)}(y)$.
We impose the boundary conditions discussed below \eqref{EdeT}, namely we impose vanishing asymptotic Dirichlet boundary conditions $\mathfrak{q}_j^{(n)}|_{y=1}=0$ $-$ since the full solution \eqref{PTexpansion} must approach the DeTurck reference MPBS solution $-$ and Neumann conditions at the horizon, $\mathfrak{q}_j^{(n)\,\prime}|_{y=0}=0$.

We complete this perturbation scheme up to order $\mathcal{O}(\epsilon^5)$: this is the order required to find a deviation between the relevant thermodynamics of the helical and MPBS, as it will be found when obtaining \eqref{EntropyDiff}.

\begin{figure}[th]
\centering
\includegraphics[width=.46\textwidth]{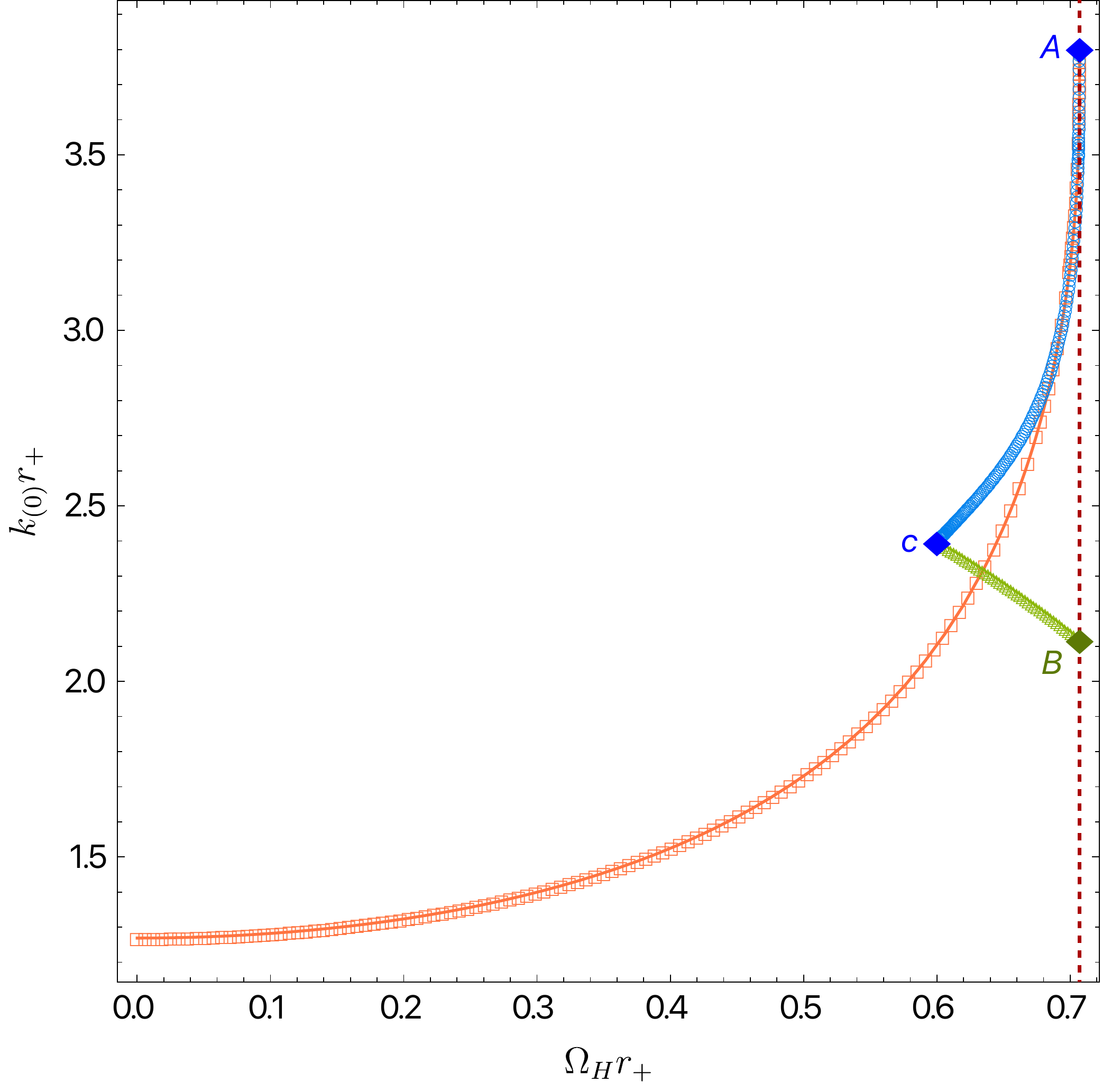}\hspace{1cm}
\includegraphics[width=.46\textwidth]{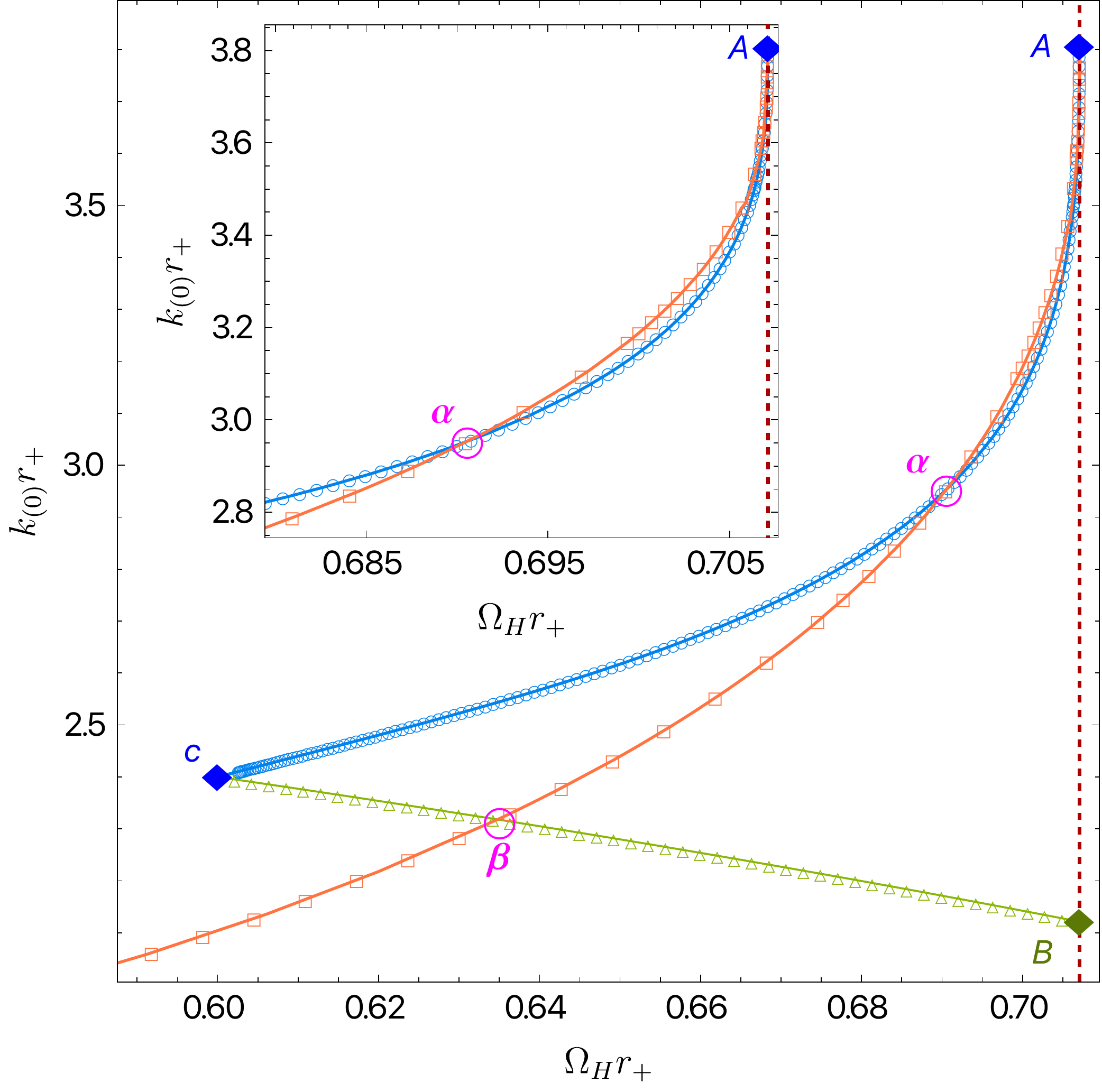}
\caption{Superradiant instability (for azimuthal number $m=-2$; see footnote \ref{foot:sign-m}) and Gregory-Laflamme instability of Myers-Perry black strings with parameters $\widetilde{k}_{(0)}$, $\widetilde{\Omega}_H$.  Superradiant instability occurs inside the triangular region $ABc$, and Gregory-Laflamme instability occurs below the curve marked by orange squares. The Gregory-Laflamme onset curve intersects with the edge of the unstable superradiant region at points $\widetilde{a}$ and $\beta$. For reference, $\widetilde{\Omega}_H|_c=3/5$ and the vertical dashed line at $\widetilde{\Omega}_H = 1/\sqrt{2}$ is extremality.
} \label{Fig:zeroModeGL}
\end{figure}

Having described the perturbation scheme, we are now ready to discuss the properties that can be extracted from the perturbative analysis.
The leading wavenumber $\widetilde{k}_{(0)}( \widetilde{\Omega}_H)$ (already computed in \cite{Dias:2022str}) is described by the blue curve $Ac$ of Fig.~\ref{Fig:zeroModeGL}. Borrowing the details found in \cite{Dias:2022str},  the MPBS is unstable for  $ \widetilde{\Omega}_H|_c \leq \widetilde{\Omega}_H \leq 1/\sqrt{2}$ with $\widetilde{\Omega}_H|_c=3/5$ and $ \widetilde{k}_{\star}( \widetilde{\Omega}_H) \leq \widetilde{k}\leq \widetilde{k}_{(0)}( \widetilde{\Omega}_H)$ where  \cite{Dias:2022str}
 \begin{equation}\label{Cutoff}
\widetilde{k}_\star^{(m)} (\widetilde{\Omega}_H)=3\sqrt{1-\widetilde{\Omega}_H^2}
 \end{equation}
is the green curve $Bc$ in Fig.~\ref{Fig:zeroModeGL}. The instability shuts down below this cut-off curve $Bc$ because the superradiant modes are no longer bounded states with exponentially decaying behaviour at the asymptotic infinity (see \cite{Dias:2022str} for details).
So, the MPBS is unstable inside the region bounded by the closed curve $ABc$ in Fig.~\ref{Fig:zeroModeGL}. For completeness, in Fig.~\ref{Fig:zeroModeGL} we also show as the orange curve the zero mode wavenumber $\widetilde{k}_{(0)}\big|_{\hbox{\tiny GL}}(\widetilde{\Omega}_H)$ of the Gregory-Laflamme instability of the MPBS.  The MPBS is unstable below this onset curve which exists for any rotation, $0 \leq \widetilde{\Omega}_H \leq 1/\sqrt{2}$ (see \cite{Dias:2022str} for details).

\begin{figure}[th]
\centering
\includegraphics[width=.45\textwidth]{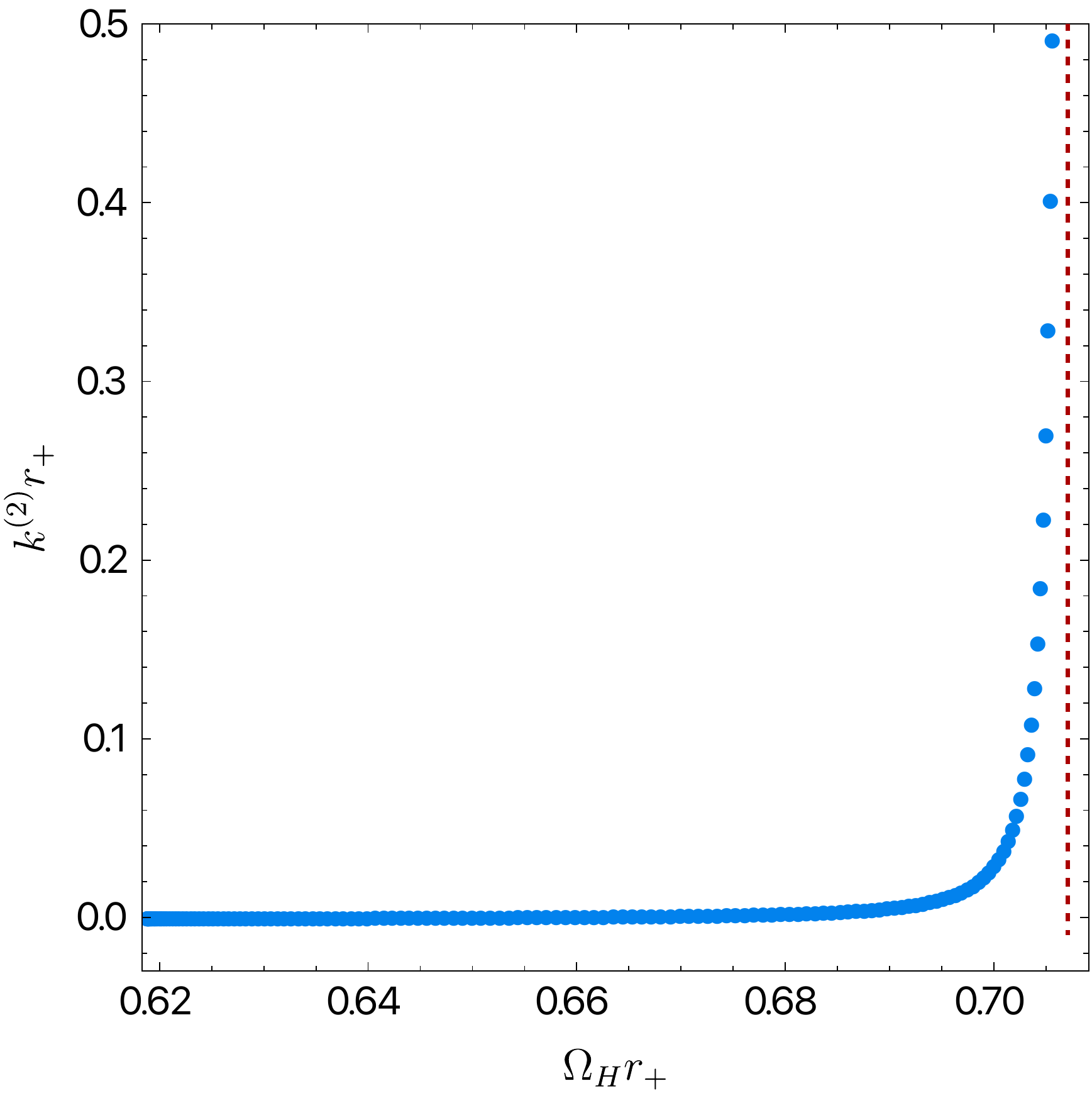}\hspace{1cm}
\includegraphics[width=.465\textwidth]{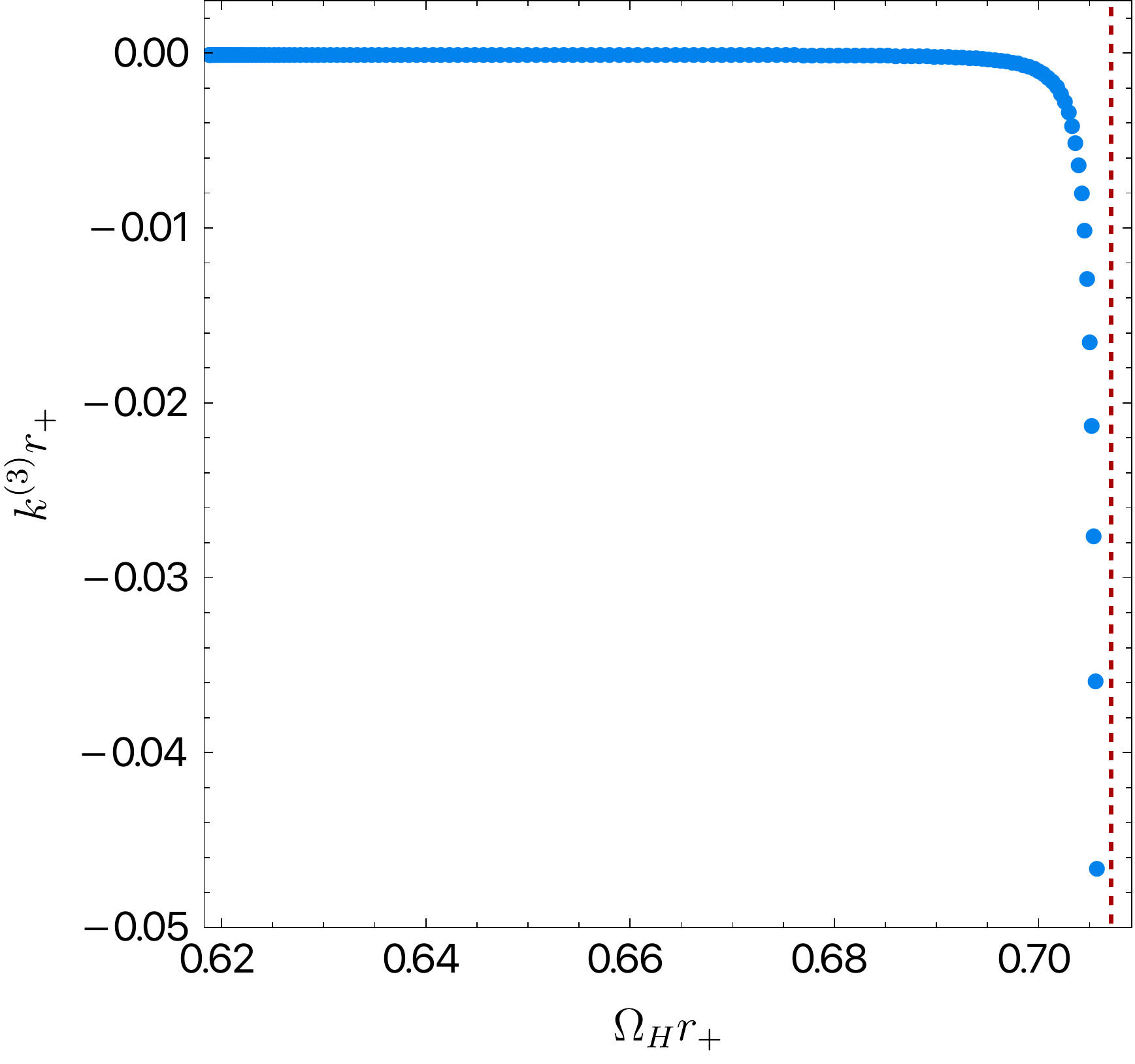}\hspace{1cm}
\includegraphics[width=.46\textwidth]{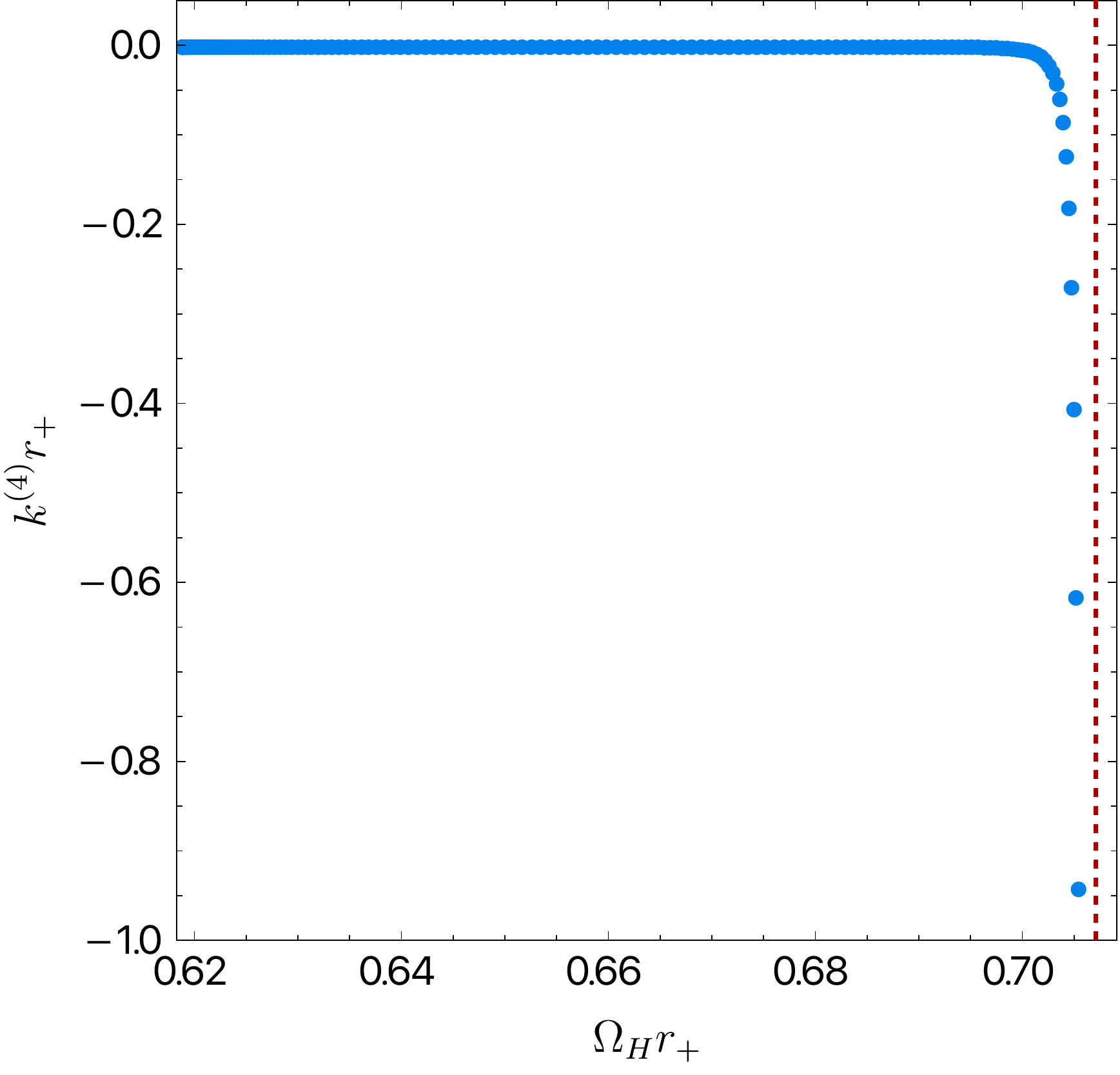}
\caption{Wavenumber corrections $\widetilde{k}^{(2)}$ (top-left panel), $\widetilde{k}^{(3)}$ (top-right panel) and $\widetilde{k}^{(4)}$ (bottom panel), as defined in \eqref{PTexpansionB}, as a function of the dimensionless angular velocity of the MP string $\widetilde{\Omega}_H=\Omega_H r_+=\widetilde{a}$ ($\widetilde{\Omega}_H|_c \leq \widetilde{\Omega}_H \leq 1/\sqrt{2}$). The vertical red dashed line represents the extremal configuration with $\widetilde{\Omega}_H=1/\sqrt{2}$. Recall that the leading wavenumber $\widetilde{k}^{(0)}$ is given in Fig.~\ref{Fig:zeroModeGL} and we find that $\widetilde{k}^{(1)}=0$.} \label{Fig:k2k3k4}
\end{figure}

Moving to order $n\geq 2$, we find that the next-to-leading order wavenumber correction vanishes, $\widetilde{k}^{(1)}=0$. This has important consequences that we discuss later. On the other hand, the wavenumber corrections $\widetilde{k}^{(2)}$, $\widetilde{k}^{(3)}$ and $\widetilde{k}^{(4)}$, as defined in \eqref{PTexpansionB}, are plotted in Fig.~\ref{Fig:k2k3k4}. The fact that these higher order quantities grow large as one approaches $\widetilde{\Omega}_H=\widetilde{\Omega}_H|_c$ and $\widetilde{\Omega}_H=1/\sqrt{2}$ tells us that our perturbation theory breaks down in these regions. We will also come back to this below.

Once we have found all corrections $\mathfrak{q}_j^{(n)}(y)$ and $\widetilde{k}^{(n-1)}$ up to $n=5$, we can reconstruct the nine fields $q_j(y)$ using \eqref{PTexpansion}. We can then substitute  these fields in the thermodynamic formulas \eqref{ThermoHelDless} of  section~\ref{sec:ansatzEdT} to obtain all the thermodynamic quantities of the system up to $\mathcal{O}(\epsilon^5)$.

\begin{figure}[t!]
\centerline{
\includegraphics[width=.70\textwidth]{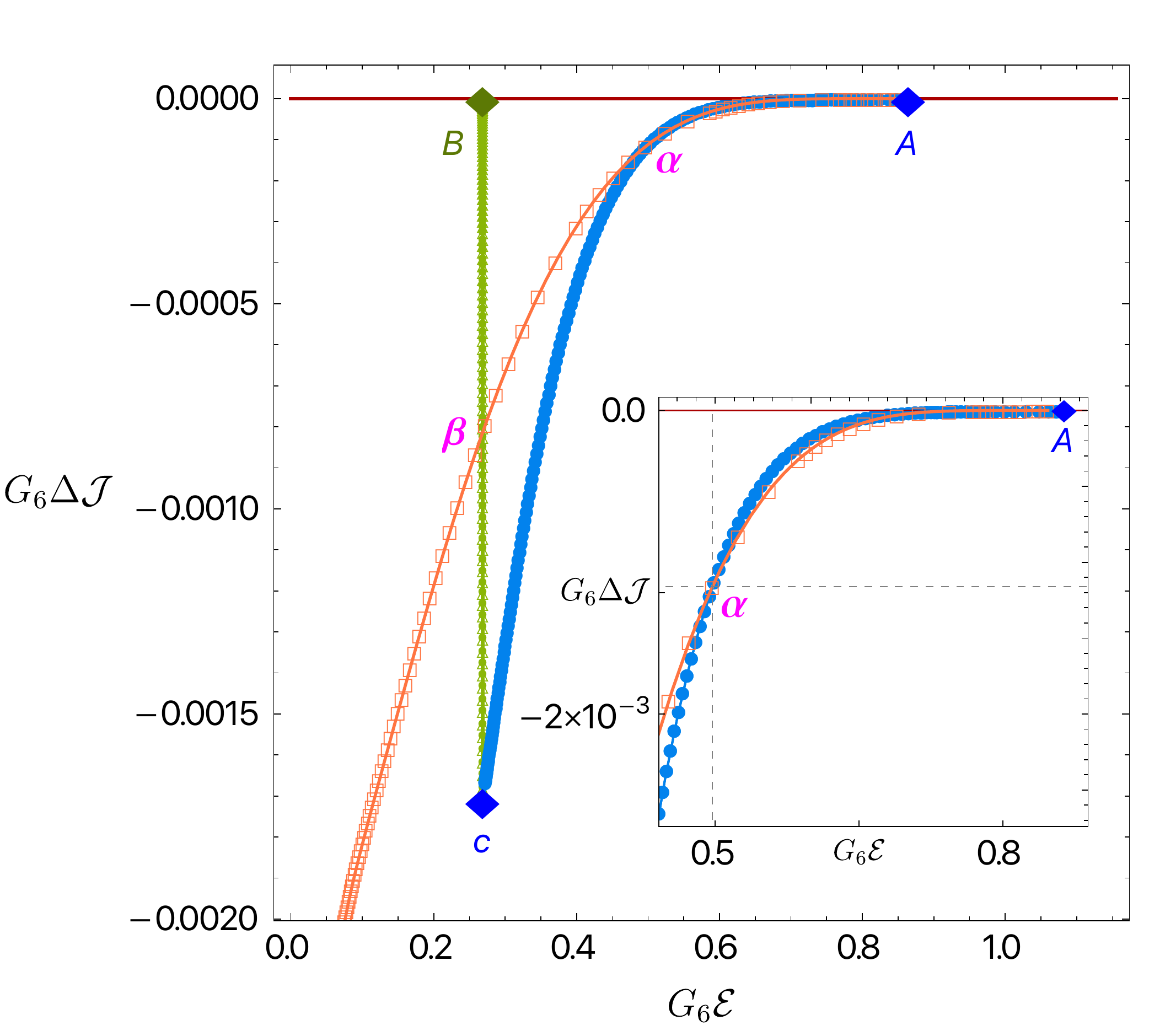}
}
\caption{Superradiant instability (for $m=-2$; see footnote \ref{foot:sign-m}) and Gregory-Laflamme instability of Myers-Perry black strings with parameters $\mathcal E$, $\mathcal J$.  For presentation, we show the angular momentum difference with the extremal Myers-Perry black string $G_6 \Delta \mathcal{J}\equiv G_6(\mathcal{J}-\mathcal{J}_{\hbox{\tiny ext\, MP}})_{\hbox{\tiny same}\,\mathcal{E}}$. Superradiant instability occurs inside the triangular region $ABc$, and Gregory-Laflamme instability occurs above the curve marked by orange squares. The Gregory-Laflamme onset curve intersects with the edge of the unstable superradiant region at points $\widetilde{a}$ and $\beta$. The horizontal red line at $\Delta \mathcal{J}=0$ is extremality. For reference, $G_6 \mathcal{E}|_c=G_6 \mathcal{E}|_B=27/(32\pi)\simeq 0.268574$ and $G_6 \mathcal{E}|_A= 45/(16 \pi)\simeq 0.895247$.
}
\label{fig:stabilityDiag}
\end{figure}

Before discussing helical strings, it is useful to  recall that it follows from \eqref{ThermoHelDless} (with $q_{1,2,3,4,5,9}=1$ and $q_{6,7,8}=0$) that the thermodynamic quantities of the MPBS parametrized by $\widetilde{L},\widetilde{a}$ ($v_H=0$) are given by
\begin{eqnarray}\label{ThermoMPonset}
&&\mathcal{E}\big|_{\hbox{\tiny MP}}=\frac{1}{G_6}\frac{3 \pi}{8 \widetilde{L}^2} \frac{1}{1-\widetilde{a} ^2}\,, \qquad \mathcal{J}\big|_{\hbox{\tiny MP}}=\frac{1}{G_6}\frac{\pi}{4 \widetilde{L}^3}\, \frac{\widetilde{a}}{1-\widetilde{a} ^2}\,, \qquad \mathcal{T}_z\big|_{\hbox{\tiny MP}}=\frac{1}{G_6}\frac{\pi}{8 \widetilde{L}^2}\,\frac{1}{1-\widetilde{a} ^2},
\nonumber \\
&&\tau_H\big|_{\hbox{\tiny MP}}=\frac{\widetilde{L}}{2 \pi} \frac{1-2 \widetilde{a} ^2}{\sqrt{1-\widetilde{a} ^2}}\,, \qquad \omega_H\big|_{\hbox{\tiny MP}}= \widetilde{a} \,\widetilde{L}\,, \qquad \sigma_H\big|_{\hbox{\tiny MP}}=\frac{1}{G_6} \frac{\pi^2}{2 \widetilde{L}^3}\frac{1}{\sqrt{1-\widetilde{a} ^2}} \,.
\end{eqnarray}
We can represent these MPBS (and other solutions) in a  phase diagram $\mathcal{E}$-$\mathcal{J}$. Actually, to better differentiate the different families of solutions, it is convenient to plot instead $\mathcal{E}$ {vs} $\Delta \mathcal{J}$ where
\begin{equation}\label{DeltaJ}
\Delta \mathcal{J}\equiv (\mathcal{J}-\mathcal{J}_{\hbox{\tiny ext\, MP}})|_{\hbox{\tiny same}\,\mathcal{E}}
\end{equation}
describes the angular momentum difference between a given solution and the extremal MPBS, with $\mathcal{J}=\mathcal{J}_{\hbox{\tiny ext\, MP}}$ defined in \eqref{themo_ext_J} with the same $\mathcal{E}$. This phase diagram $\mathcal{E}$-$\Delta\mathcal{J}$ is displayed in Fig.~\ref{fig:stabilityDiag}. The horizontal red line with $\Delta \mathcal{J}=0$ represents the 1-parameter family of extremal MPBSs with $\widetilde{\Omega}_H=\widetilde{a}=1/\sqrt{2}$. It extends to arbitrarily large $\mathcal{E}$. Non-extremal MPBSs exist below this line.

Besides the extremal MPBS, there are other special 1-parameter families of black strings that play a relevant role in our discussion:
\begin{enumerate}
\item Replacing the superradiant onset mode $\widetilde{L}_{(0)}(\widetilde{\Omega}_H)=2\pi/\widetilde{k}_{(0)}(\widetilde{\Omega}_H)$ of Fig.~\ref{Fig:zeroModeGL} in  \eqref{ThermoMPonset}, 
 we get the thermodynamics of the 1-parameter family of the MPBS that are at the onset of the superradiant instability, as shown in the  blue disk curve $Ac$ of Fig.~\ref{fig:stabilityDiag}.

\item Recall that the MPBS is unstable if its length is longer that the superradiant onset mode length, $\widetilde{L}> \widetilde{L}_{(0)}(\widetilde{\Omega}_H)$, but smaller that the critical cut-off length $\widetilde{L}=\widetilde{L}_{\star}(\widetilde{\Omega}_H) =2\pi/\widetilde{k}_{\star}(\widetilde{\Omega}_H)$  with $\widetilde{k}_{\star}(\widetilde{\Omega}_H)$ given by \eqref{Cutoff}, \ie if they are above the green curve $Bc$ of  Fig.~\ref{Fig:zeroModeGL}. Inserting $\widetilde{L}=\widetilde{L}_{\star}(\widetilde{\Omega}_H)$ into  \eqref{ThermoMPonset} we get the  1-parameter family $Bc$ (vertical green curve) of MPBSs in the $\mathcal{E}$-$\Delta\mathcal{J}$ plane of Fig.~\ref{fig:stabilityDiag}.\footnote{There is a series of onset and cut-off superradiant curves alike the ones described in items 1 and 2 for $m=2$ modes, but this time for $m\geq 2$. The reader can find a discussion and phase diagrams for these cases in \cite{Dias:2022str}.}

\item For reference, since the MPBS also has the GL instability, in Fig.~\ref{fig:stabilityDiag} we also display the Gregory-Laflamme onset curve (orange squares). This curve is also obtained from \eqref{ThermoMPonset}  when we replace $\widetilde{L}= \widetilde{L}_{(0)}|_{\hbox{\tiny GL}}$, where the latter can be read from the orange square data of Fig.~\ref{Fig:zeroModeGL}.
The MPBS above this orange square curve, all the way up to the axis $\Delta \mathcal{J}=0$, are unstable to the Gregory-Laflamme instability.

\end{enumerate}

The superradiant onset curve $Ac$ in Fig.~\ref{fig:stabilityDiag} also represents where the MPBS and helical black strings merge. At order $\mathcal{O}(\epsilon)$, perturbation theory identifies this merger but it does not describe the properties of the helical strings as we move away from the merger line. For that, we need to proceed to higher order $\mathcal{O}(\epsilon^n)$ in the perturbation theory.
We do so up to $n=5$ and thus we get the thermodynamic description of  helical black  strings up to $\mathcal{O}(\epsilon^5)$. We can then compare it against the thermodynamics of the MPBS and find which of these two families is the preferred. We are particularly interested in the microcanonical ensemble (with all solutions having the same Kaluza-Klein circle size $L$), so the dominant  phase is the one with the highest dimensionless entropy $\sigma_H$ for a given  $(\mathcal{E},\mathcal{J})$ ({\it i.e.}~energy and angular momenta in units of the circle length $L$). Let $\mathcal{Q}_{\rm MP}$ and $\mathcal{Q}_{\rm hel}$ denote generic thermodynamic quantities $\mathcal{Q}$ for the MP string and  helical black string, respectively.
When comparing these two solutions in the microcanonical ensemble, we must use the same Kaluza-Klein circle size $L$ and require that the solutions have the same dimensionless energy and angular momenta:
\begin{equation}\label{microCondition}
 \mathcal{E}_{\rm hel}=\mathcal{E}_{\hbox{\tiny MP}}\,, \qquad \mathcal{J}_{\rm hel}=\mathcal{J}_{\hbox{\tiny MP}}.
\end{equation}
Given a helical black string with $(\mathcal{E}_{\rm hel},\mathcal{J}_{\rm hel})$, we must thus identify the parameters $(\widetilde{L}_{\hbox{\tiny MP}},\widetilde{a}_{\hbox{\tiny MP}})$ of the MP string whose  energy and angular momenta satisfy \eqref{microCondition}, with $\widetilde{L}_{\hbox{\tiny MP}}=L_{\hbox{\tiny MP}}/r_+$, namely:
\begin{align}
& \widetilde{L}_{\hbox{\tiny MP}} =\sqrt{3 \pi} \, \frac{\sqrt{\mathcal{E}_{\rm hel}}}{\Sigma_+} \,,
  \qquad
 \widetilde{a}_{\hbox{\tiny MP}}= \frac{1}{12 \sqrt{3 \pi}}\, \frac{\Sigma_+ \Sigma_-^2}{\mathcal{J}_{\rm hel} \mathcal{E}_{\rm hel}^{3/2}} \nonumber \\
& \hbox{with} \quad  \Sigma_\pm \equiv \sqrt{4\mathcal{E}_{\rm hel}^2 \pm  \sqrt{2}\sqrt{ 8\mathcal{E}_{\rm hel}^4-27 \pi  \mathcal{J}_{\rm hel}^2 \mathcal{E}_{\rm hel}}}
\end{align}
We can now place these quantities in \eqref{ThermoMPonset} with the identifications $ \widetilde{L}\to \widetilde{L}_{\hbox{\tiny MP}}$ and   $ \widetilde{a}\to \widetilde{a}_{\hbox{\tiny MP}}$
to find the thermodynamic quantities (in particular, the entropy $\sigma_H$) of the MPBS with the same energy and angular momenta as the helical string.
The thermodynamics of the MPBS is then expressed as a function of $(\mathcal{E}_{\rm hel},\mathcal{J}_{\rm hel})$. The latter, and thus the former as well, have an expansion in $\epsilon$.

We now compute the entropy  difference $\Delta \sigma_H=\left( \sigma_{H,\rm hel}-\sigma_{H,\hbox{\tiny MP}}\right)|_{{\rm same}\,(\mathcal{E},\mathcal{J})}$ between helical strings and the MPBS with the same energy and angular momenta. This yields $\Delta \sigma_H=c_{\Delta \sigma}^{(2)}\epsilon^2+ c_{\Delta \sigma}^{(3)}\epsilon^3+c_{\Delta \sigma}^{(4)}\epsilon^4$ where $c_{\Delta \sigma}^{(2)}$,  $c_{\Delta \sigma}^{(3)}$ and $c_{\Delta \sigma}^{(4)}$ are  functions of $\mathfrak{q}_j^{(n)}|_{y=0}$ ($n=2,3,4$). However, our solutions satisfy the first law of thermodynamics \eqref{1stLawDless}, as we explicitly verify. As a further check of our solutions we also verify that the Smarr relation \eqref{GibbsDuhemEuler} is obeyed.

This first law must be obeyed at each order in $\epsilon$ and thus it effectively gives three conditions (one at each order $\epsilon^{n-1}$, $n=2,3,4$) that we can use to express the second derivatives of $\mathfrak{q}_5^{(2)}|_{y=1}$, $\mathfrak{q}_5^{(3)}|_{y=1}$ and $\mathfrak{q}_5^{(4)}|_{y=1}$ as a function of other functions $\mathfrak{q}_j^{(n)}$ and their first derivatives evaluated at the horizon, $y=0$, or at $y=1$. When we do this, we simplify $c_{\Delta \sigma}^{(2)}$, $c_{\Delta \sigma}^{(3)}$ and $c_{\Delta \sigma}^{(4)}$. In particular, we find that $c_{\Delta \sigma}^{(2)}\equiv 0$ and $c_{\Delta \sigma}^{(3)}\equiv 0$ which justifies our need to extend the perturbation expansion up to $\mathcal{O}(\epsilon^5)$.

Altogether, after using the first law of \eqref{1stLawDless}, we find that
 \begin{eqnarray} \label{EntropyDiff}
\Delta \sigma_H&=&\left( \sigma_{H,\rm hel}-\sigma_{H,\hbox{\tiny MP}}\right)\big|_{{\rm same}\,(\mathcal{E},\mathcal{J})}\nonumber \\
&=&G_6\,c_{\Delta \sigma}^{(4)}\,\epsilon^4  +\mathcal{O}(\epsilon^6)
\end{eqnarray}
with
{\small
 \begin{eqnarray} \label{EntropyDiff2}
c_{\Delta \sigma}^{(4)}& =& \frac{\widetilde{k}_{(0)}^2}{384 \pi  \sqrt{1-\widetilde{a} ^2} \left(1-2 \widetilde{a} ^2\right)}\Bigg\{
\widetilde{k}_{(0)} \Bigg[
2 \left(1-\widetilde{a} ^2\right)^2 \left(3-2 \widetilde{a} ^2\right) \widetilde{a} ^2 q_6^{(2)\,\prime}(1){}^2\nonumber \\
& &
-8 \left(\widetilde{a} ^4-3 \widetilde{a} ^2+2\right) \widetilde{a} ^2 \Big(q_3^2(0)+2 q_4^2(0)+q_5^2(0)\Big) q_6^{(2)\,\prime}(1)\nonumber \\
& & -\left(4 \widetilde{a} ^4-10 \widetilde{a} ^2+1\right) \Big(q_3^2(0)+2 q_4^2(0)+q_5^2(0)\Big)^2
\Bigg]\nonumber \\
& &
-12 \widetilde{a}  \left(1-\widetilde{a} ^2\right) q_8^{(2)\,\prime}(1) \Big[2 \left(1-\widetilde{a} ^2\right) \widetilde{a} ^2 q_6^{(2)\,\prime}(1)+\left(1-2 \widetilde{a} ^2\right) \Big(q_3^2(0)+2 q_4^2(0)+q_5^2(0)\Big) \Big]
\nonumber \\
& &
-6 \widetilde{k}^{(2)} \Bigg[
2 \left(1-\widetilde{a} ^2\right) \widetilde{a} ^2 q_6^{(2)\,\prime}(1)+\left(1-2 \widetilde{a} ^2\right)\Big( q_3^2(0)+2 q_4^2(0)+q_5^2(0)\Big)
\Bigg]
\Bigg\},
\end{eqnarray}
}
where  $\mathfrak{q}_{6,8}^{(2)\,\prime}(1)$ stands for the first derivative of $\mathfrak{q}_{6,8}^{(2)}$ evaluated at $y=1$ and all other functions in \eqref{EntropyDiff2} are evaluated at $y=0$.\footnote{The entropy difference depends also on functions evaluated at the asymptotic boundary because we have subtracted the MPBS background that obeys  \eqref{microCondition} and because we used first law to get \eqref{EntropyDiff2}.}

\begin{figure}[t]
\centering
\includegraphics[width=.46\textwidth]{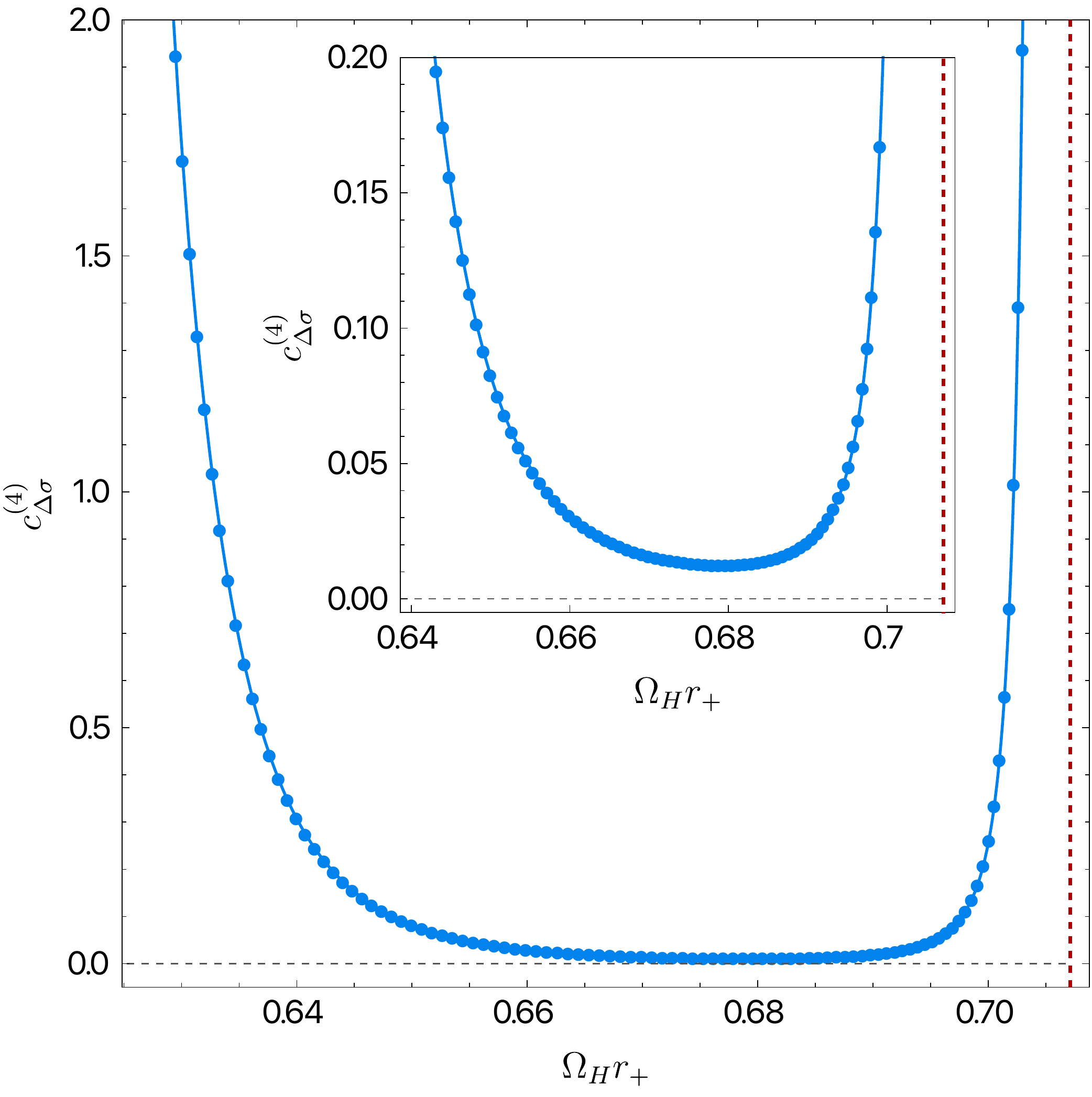}
\caption{Perturbative identification of the dominant microcanonical phase. The horizontal axis shows the dimensionless angular velocity $\widetilde{\Omega}_H$ of the MP string. The vertical axis displays the difference \eqref{EntropyDiff2} between the dimensionless entropy of the helical and the MPBS with the same $(\mathcal{E},\mathcal{J})$. Since $c_{\Delta \sigma}^{(4)}>0$, \ie $\Delta\sigma_H>0$, for any $\widetilde{\Omega}_H|_c \leq \widetilde{\Omega}_H \leq 1/\sqrt{2}$ (see inset plot that zooms in the region where $c_{\Delta \sigma}^{(4)}$ attains its lower values), black helical strings always dominate the microcanonical ensemble around the superradiant merger line.}
\label{Fig:DeltaSigma}
\end{figure}

We now need to evaluate the positivity of $c_{\Delta \sigma}^{(4)}$ to determine which phase is preferred. For that, recall that we are doing perturbation about the merger line $Ac$ of MP and helical black strings (see Fig.~\ref{Fig:zeroModeGL} or Fig.~\ref{fig:stabilityDiag}). This curve is parametrized by $\widetilde{\Omega}_H|_c \leq \widetilde{\Omega}_H \leq 1/\sqrt{2}$. Thus, to analyse the positivity of $\Delta\sigma_H$, we  just need to compute the coefficient  $c_{\Delta \sigma}^{(4)}$ in \eqref{EntropyDiff}-\eqref{EntropyDiff2} as a function of $ \widetilde{\Omega}_H $. This is done in Fig.~\ref{Fig:DeltaSigma}.
For any value  of $\widetilde{\Omega}_H|_c \leq \widetilde{\Omega}_H \leq 1/\sqrt{2}$, we find that $c_{\Delta \sigma}^{(4)}$ and thus $\Delta\sigma_H$ are positive quantities. It follows
that, in a neighbourhood of the merger line $Ac$, helical black strings that branch from the superradiant onset of MPBS have higher entropy  than the MPBS with the same length, energy and angular momenta.
That is to say, at least in a neighbourhood  of the merger line $Ac$ in the $\mathcal{E}$-$\mathcal{J}$ phase diagram, helical black  strings are the preferred phase in the microcanonical ensemble. To find if this is true non-perturbatively far from the merger line, we need to solve the full nonlinear gravitational equations: we do this in section \ref{sec:result}.

In Fig.~\ref{Fig:DeltaSigma} we see that the  entropy correction $c_{\Delta \sigma}^{(4)}\,\epsilon^4$ $-$ which should not be larger than $\mathcal{O}(\epsilon)$ $-$ is growing large as we approach both endpoints $c$ and $A$ of the merger line. So, the higher order perturbative theory breaks down in the vicinity of these points no matter how close to the merger line we are.

On last remark is in order. We find that all the thermodynamic quantities of helical black strings have a vanishing leading order correction ($n=1$), which is ultimately a consequence of the fact that  the leading wavenumber correction vanishes, $\widetilde{k}^{(1)}=0$. Consequently, the leading correction to the entropy difference \eqref{EntropyDiff} appears at fourth order, \ie $c_{\Delta \sigma}^{(3)}=0$. It follows that the perturbations $\pm |\epsilon|$ are physically equivalent and thus there is a single family (and not two) of helical black strings solutions that branch from the merger line $Ac$. There are other situations where two families branch from a merger, but this is not the case here\footnote{Black hole resonators that bifurcate from the onset of other superradiant systems (typically, in asymptotically AdS spaces \cite{Dias:2011at,Dias:2011tj,Dias:2015rxy,Dias:2016pma,Ishii:2018oms,Dias:2019fmz,Ishii:2020muv,Ishii:2021xmn} or in asymptotically flat backgrounds with bound states confined by a massive scalar \cite{Herdeiro:2014goa} or by a box \cite{Dias:2018yey,Dias:2021acy,Davey:2021oye}) also have a single branching family. The non-uniform strings that bifurcate from the Gregory-Laflamme onset also have this property \cite{Gubser:2001ac,Harmark:2002tr,Kol:2002xz,Wiseman:2002zc,Kol:2003ja,Harmark:2003dg,Harmark:2003yz,Kudoh:2003ki,Sorkin:2004qq,Gorbonos:2004uc,Kudoh:2004hs,Dias:2007hg,Harmark:2007md,Wiseman:2011by,Figueras:2012xj,Dias:2017coo}. However, there are also systems that have two families of solutions branching from the onset of the instability. This is, e.g., the case in ultraspinning unstable systems \cite{Emparan:2003sy,Dias:2009iu,Dias:2010maa,Dias:2010eu,Dias:2010gk,Dias:2011jg,Dias:2014cia,Emparan:2014pra,Dias:2014eua,Dias:2015nua}, in the GL instability of black branes/lattices \cite{Dias:2017coo}, and in the  GL-like instability of $AdS_5\times S^5$ \cite{Dias:2015pda,Dias:2016eto}.}.
This information is also useful when looking for nonlinear helical black strings: once we find one branch of solutions emerging from the merger line, we know there is not a second one.

\section{Results and discussion of physical properties}
\label{sec:result}

In this section, we describe the phase diagram of helical black strings and discuss their physical/thermodynamic properties. Recall that we find the nonlinear numerical solutions by solving the boundary value problem in the spherical gauge ansatz~\eqref{screw_metric} of section~\ref{sec:ansatzSphericalGauge} and/or the Einstein-DeTurck gauge ansatz \eqref{ansatz} of section~\ref{sec:ansatzEdT}. Moreover, we also compare these full nonlinear numerical results with the perturbative results of section~\ref{sec:perturbative}.

\subsection{Phase diagram and physical properties of helical black strings}
\label{sec:numsol}

\begin{figure}[b]
\centering
\includegraphics[width=.5\textwidth]{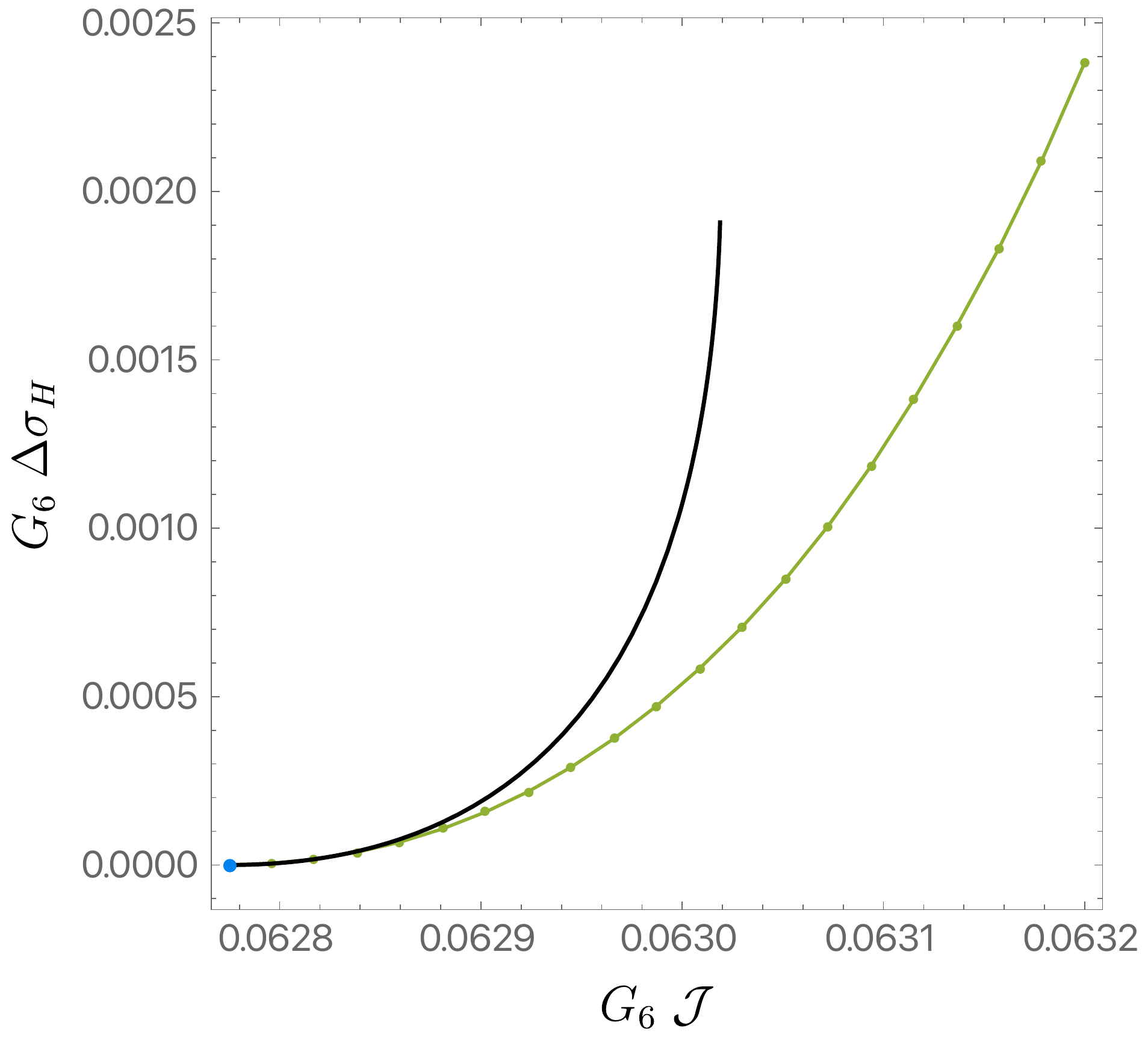}
\caption{Comparison between nonlinear helical string solutions and the perturbative result \eqref{EntropyDiff}-\eqref{EntropyDiff2} of section~\ref{sec:perturbative}.  The solid black line is the perturbative expansion~\eqref{EntropyDiff}, whereas the green disks are the full nonlinear results. Both curves were generated with $G_6 \mathcal{E}=0.35$ (recall that $\Delta\sigma_H$ is the difference with respect to the Myers-Perry black string at the same $\mathcal E$ and $\mathcal J$).  The Myers-Perry superradiant onset is the blue disk with $\Delta\sigma_H=0$ (it has $\tilde{\Omega}_H\simeq 0.64821697$).}
\label{fig:nearonsetPertNonLin}
\end{figure}

In Fig.~\ref{fig:nearonsetPertNonLin}, we show the entropy difference $\Delta \sigma_H$ between a black helical string and a Myers-Perry black string (at the same $\mathcal{E}$ and $\mathcal{J}$ as defined in \eqref{EntropyDiff}), for strings at constant $G_6 \mathcal{E}=0.35$. The black solid line shows the perturbative result \eqref{EntropyDiff}-\eqref{EntropyDiff2} constructed in section~\ref{sec:perturbative}, whereas the green disks give the fully nonlinear numerical data using either the spherical gauge ansatz~\eqref{screw_metric} of section~\ref{sec:ansatzSphericalGauge} or the Einstein-DeTurck gauge ansatz \eqref{ansatz} of section~\ref{sec:ansatzEdT}. As anticipated, the perturbative analysis provides a very good approximation near the merger with the MP black string (blue disk with $\Delta \sigma_H=0$). Further note that the fact that $\Delta \sigma_H\geq0$ indicates that black helical strings dominate the microcanonical ensemble over MPBS.

\begin{figure}[t]
\centering
\subfigure[$f$]{\includegraphics[scale=0.33]{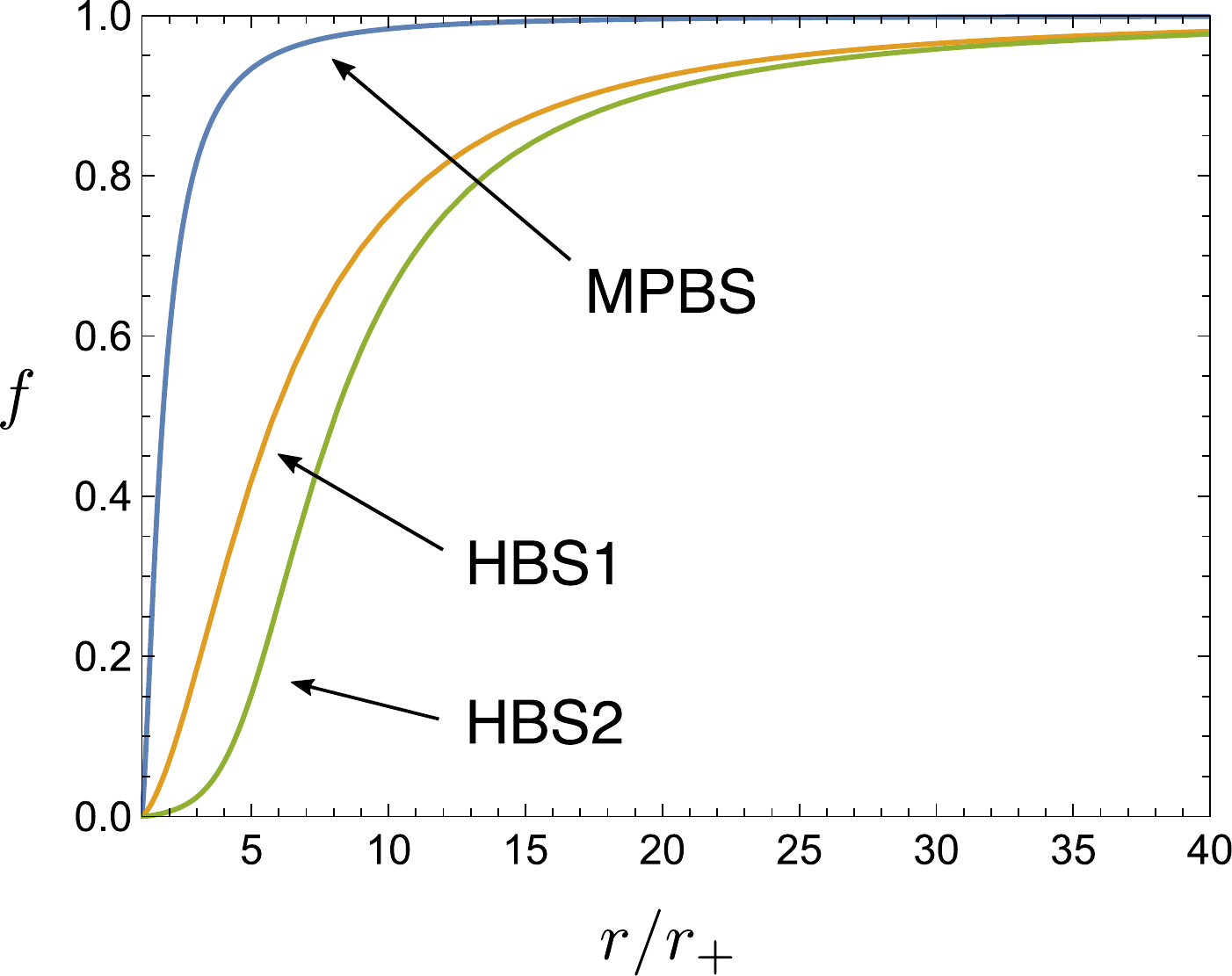}\label{fig:bulk_f}}
\subfigure[$g$]{\includegraphics[scale=0.33]{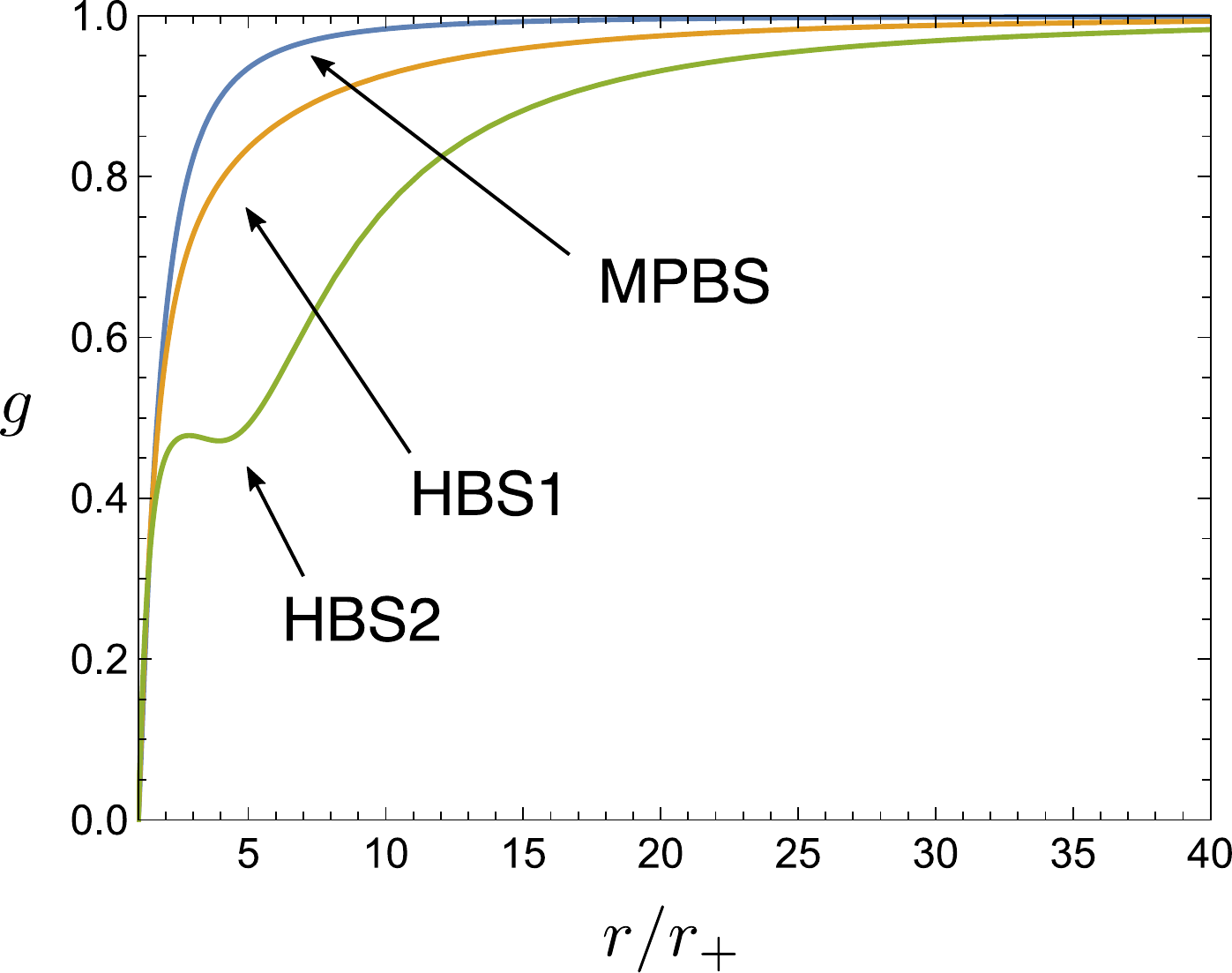}\label{fig:bulk_g}}
\subfigure[$h$]{\includegraphics[scale=0.33]{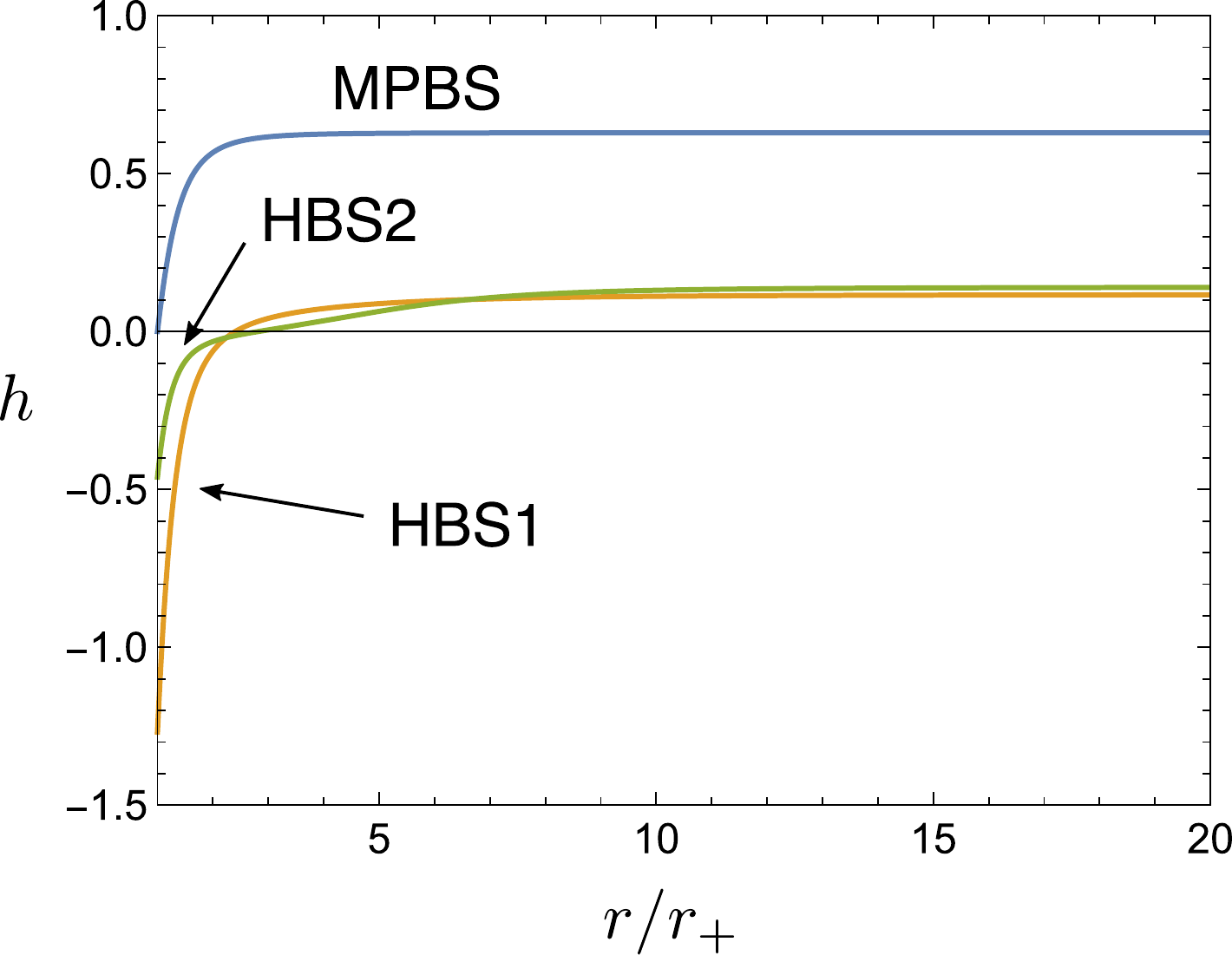}\label{fig:bulk_h}}
\subfigure[$k$]{\includegraphics[scale=0.33]{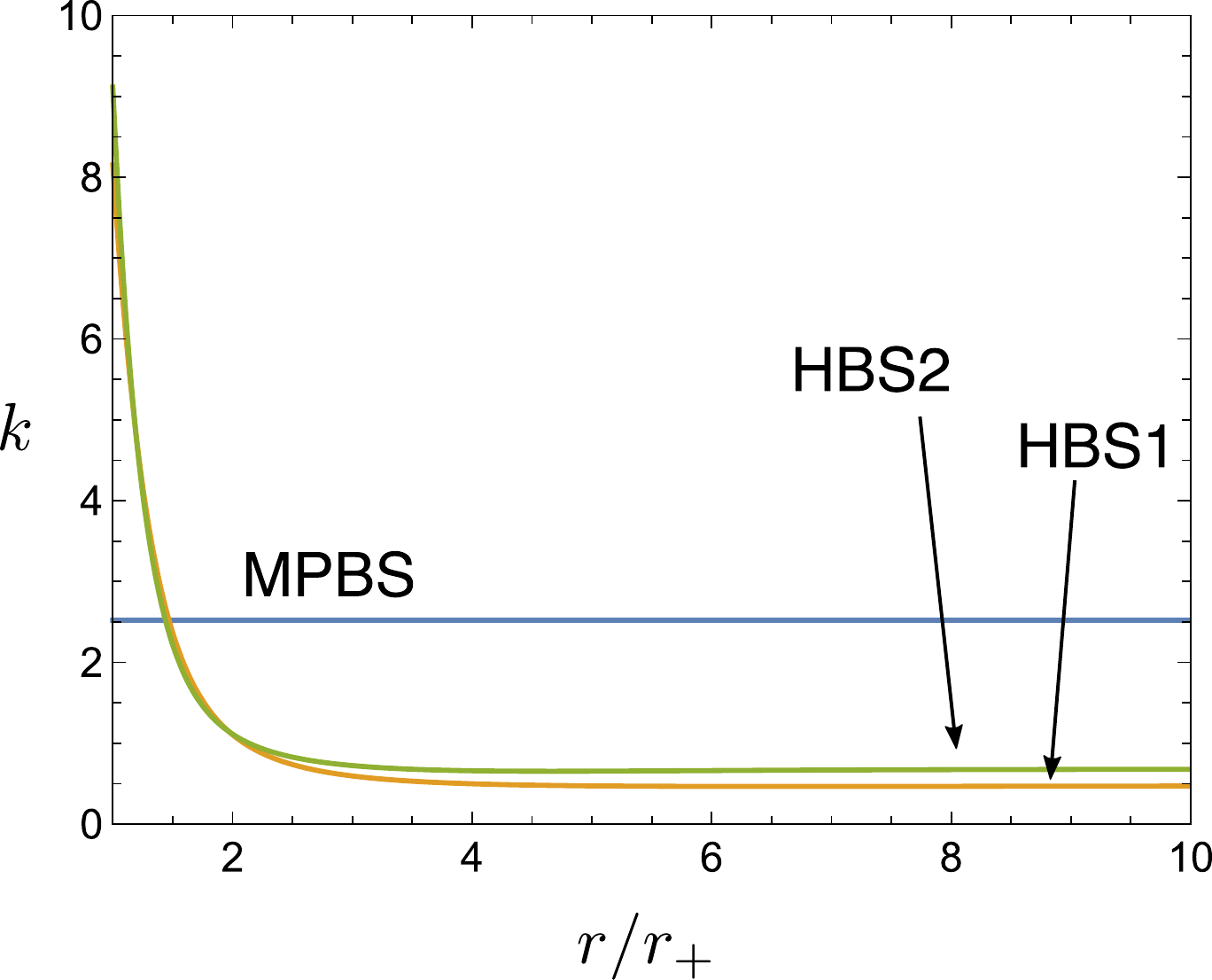}\label{fig:bulk_k}}
\subfigure[$q$]{\includegraphics[scale=0.33]{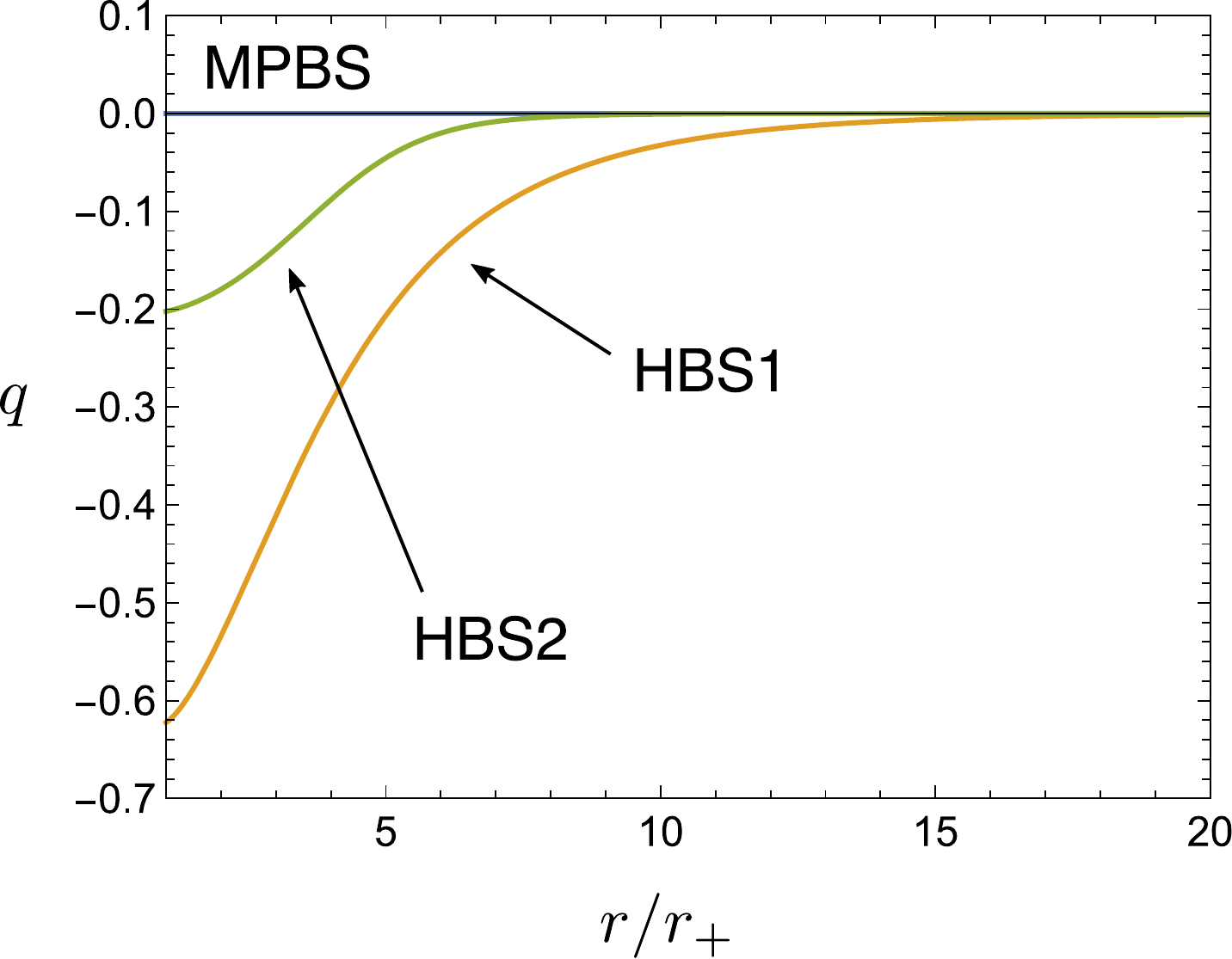}\label{fig:bulk_q}}
\subfigure[$\eta$]{\includegraphics[scale=0.33]{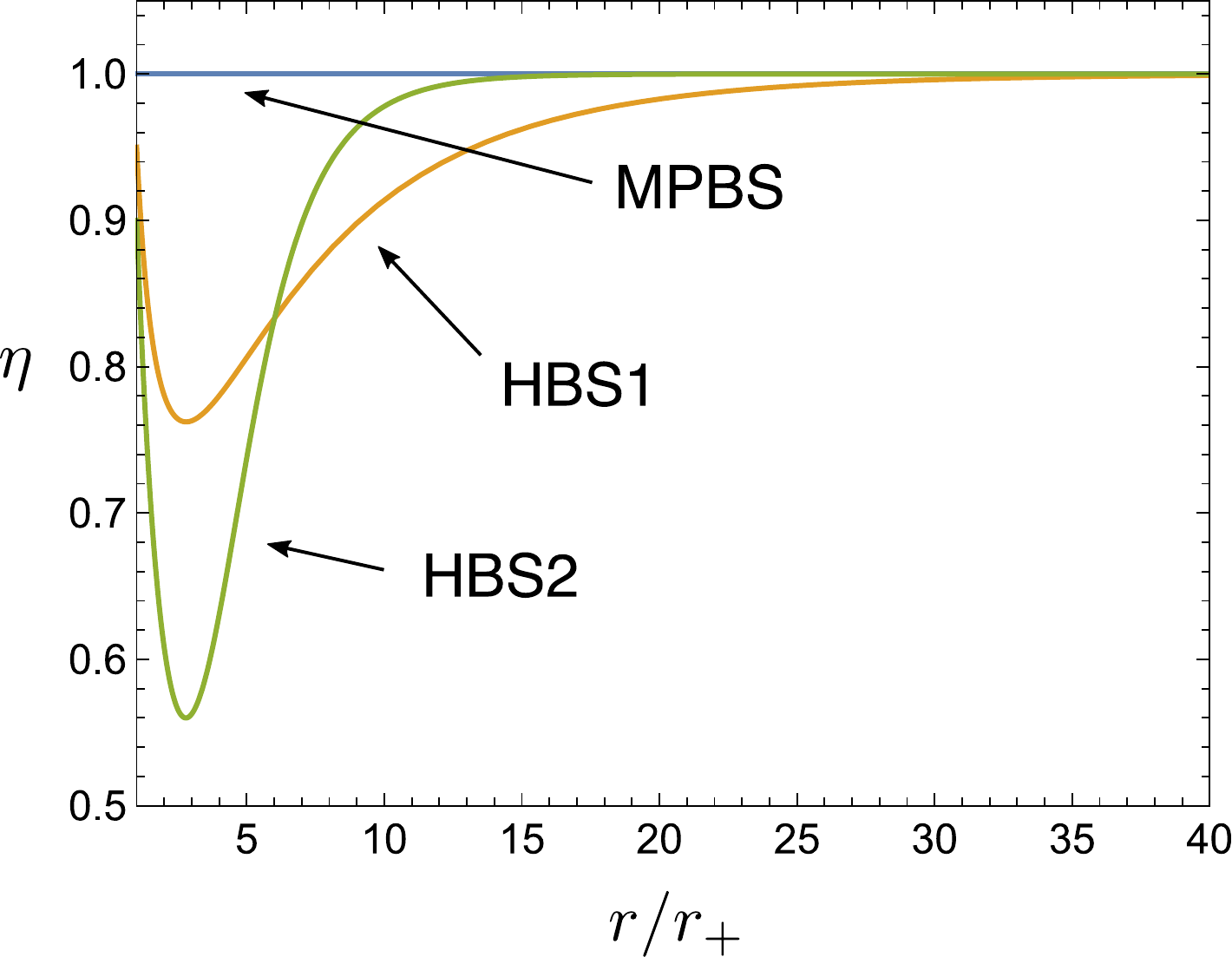}\label{fig:bulk_a}}
\subfigure[$\beta$]{\includegraphics[scale=0.33]{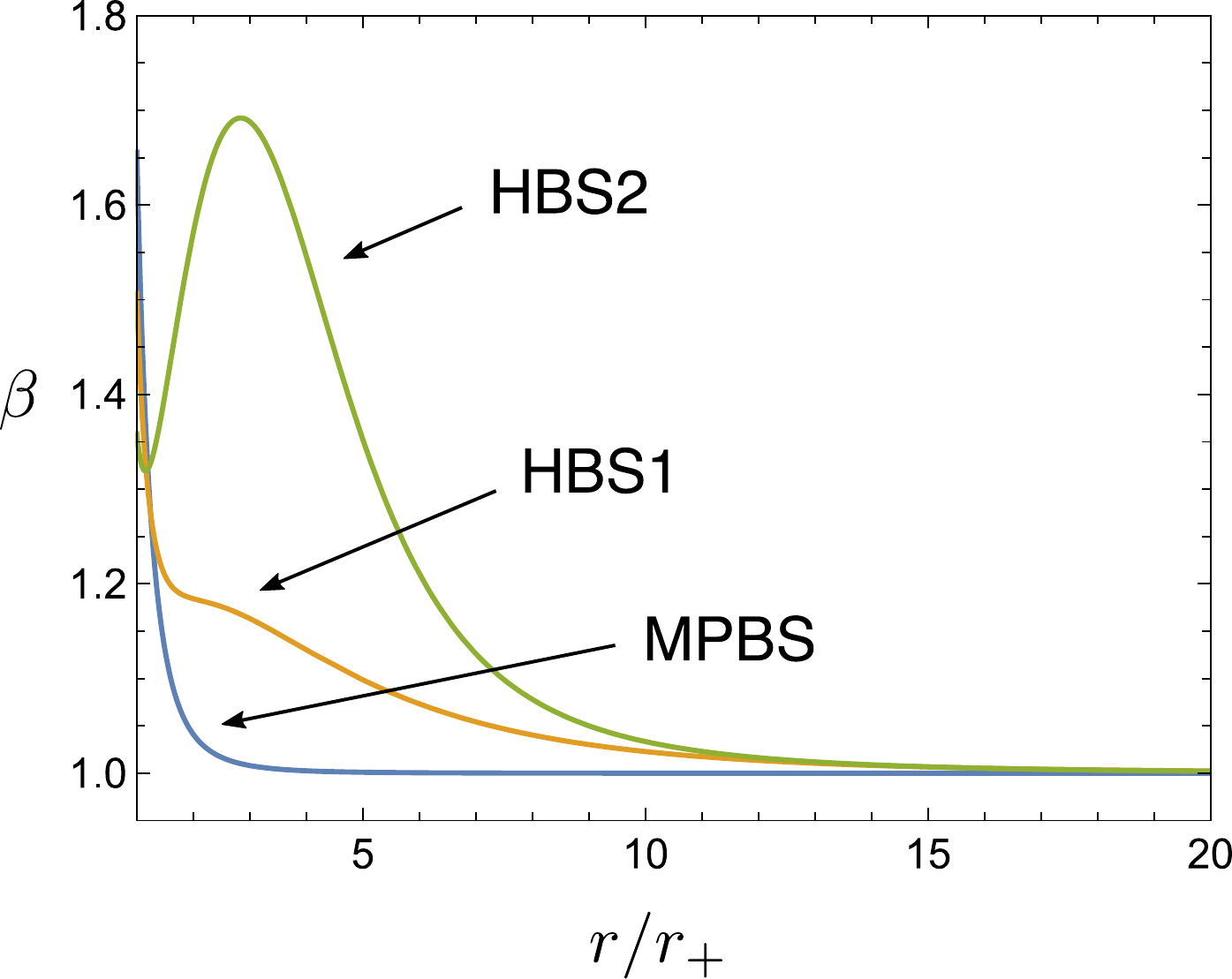}\label{fig:bulk_b}}
\subfigure[$\gamma$]{\includegraphics[scale=0.33]{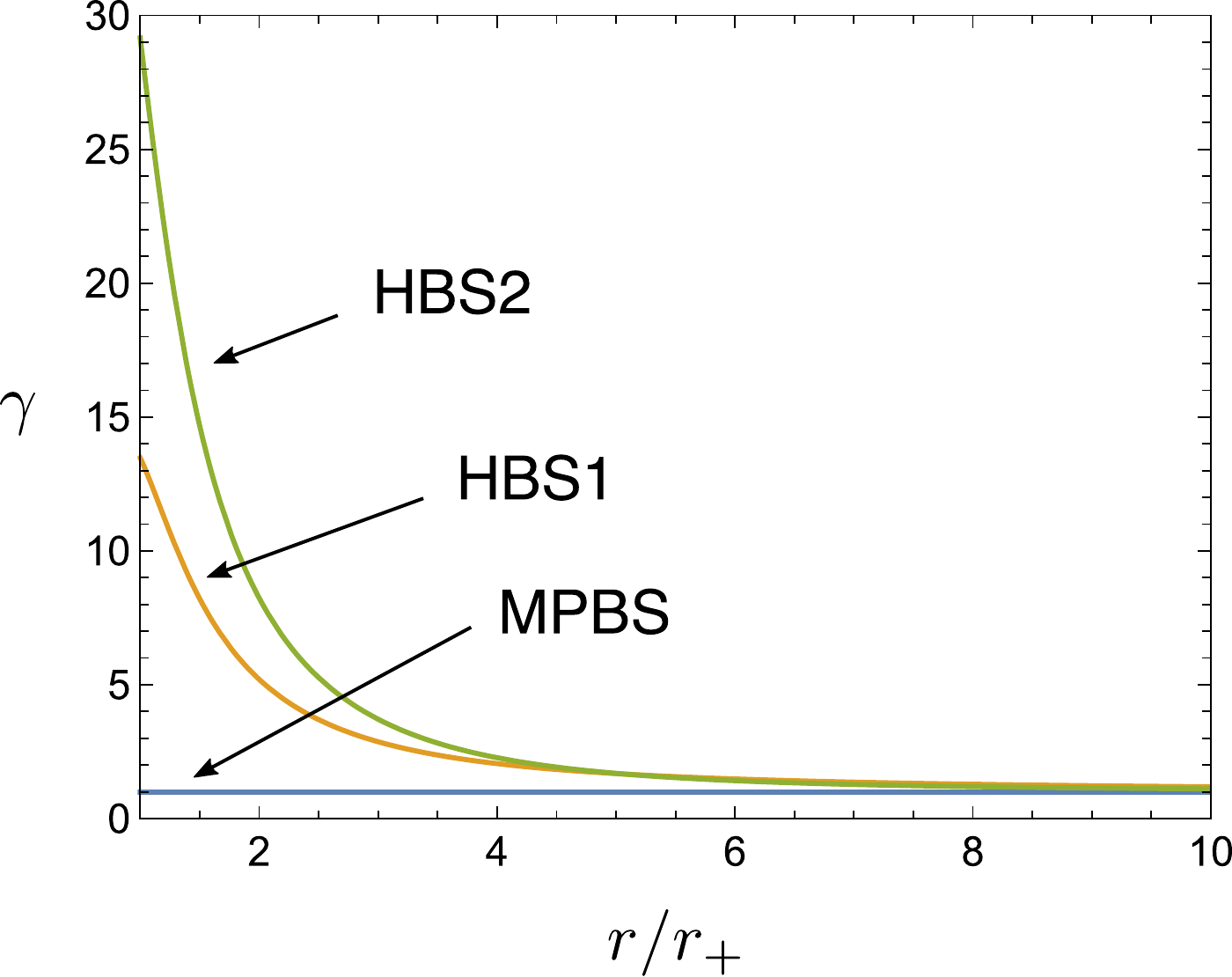}\label{fig:bulk_c}}
\subfigure[$\zeta^2$]{\includegraphics[scale=0.33]{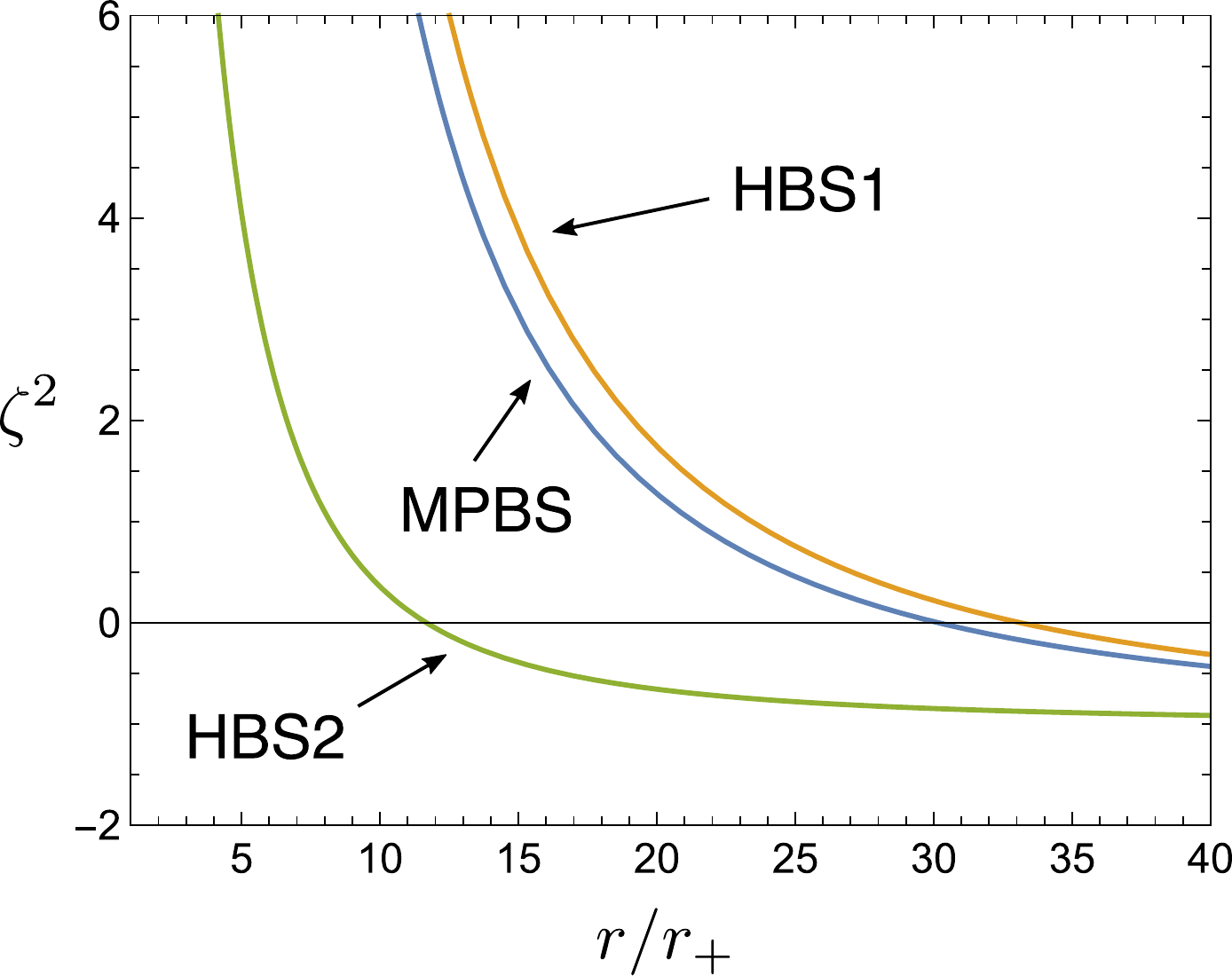}\label{fig:bulk_z}}
\caption{(a)-(h) bulk profiles of the metric fields, and (i) $\zeta^2$ (see section~\ref{sec:timelike}). Lines are for the MPBS at $\Omega_H/\Omega_H^\mathrm{ext}=0.89$ (blue) and helical black strings with $(G_6\mathcal{E},G_6\mathcal{J},G_6\sigma_H)=(0.170,0.0567,0.00977)$ (HBH1, orange) and $(0.481,0.215,0.0395)$ (HBH2, green).}
\label{fig:bulk}
\end{figure}

In the spherical gauge ansatz~\eqref{screw_metric} of section~\ref{sec:ansatzSphericalGauge},
helical black strings are characterized by having the metric function $\eta(r) \neq 1$ (with the MPBS having $\eta=1$). Some profiles of the metric component functions of \eqref{screw_metric} are shown in Figs.~\ref{fig:bulk_f}-\ref{fig:bulk_c}. We pick up three particular solutions: the MPBS/helical string at the onset of instability for $\Omega_H/\Omega_H^\mathrm{ext}=0.89$ (blue MPBS), and two helical black strings (orange HBS1, and green HBS2). Specifically, the HBH1 and HBH2 helical black strings have $(G_6\mathcal{E},G_6\mathcal{J},G_6\sigma_H) = (0.170,0.0567,0.00977)$ and $(0.481,0.215,0.0395)$, respectively. It can be seen that the deformation (w.r.t. MPBS) of the helical string metric typically takes place near the horizon (i.e, in a neighbourhood of $r/r_+\sim 1$).

Shown also in Fig.~\ref{fig:bulk_z} is the squared of the asymptotically timelike Killing vector $\zeta$ as discussed perturbatively in \eqref{normKVFinf} of section~\ref{sec:isopert} and given nonlinearly in \eqref{normKVFinf_nonlinear} of section~\ref{sec:timelike}. Note that $\zeta^2$ is finite at the horizon $r=r_+$, but the horizon value of $\zeta^2$ is outside the plotted region. The value is positive near the horizon, crosses zero at finite $r$, and becomes $\zeta^2  \to -1$ as $r \to \infty$. This shows that  helical black strings are stationary spacetimes.

\begin{figure}[t]
\centering
\subfigure[$(\mathcal{E},\Delta\mathcal{J};\omega_H)$.]{\includegraphics[scale=0.475]{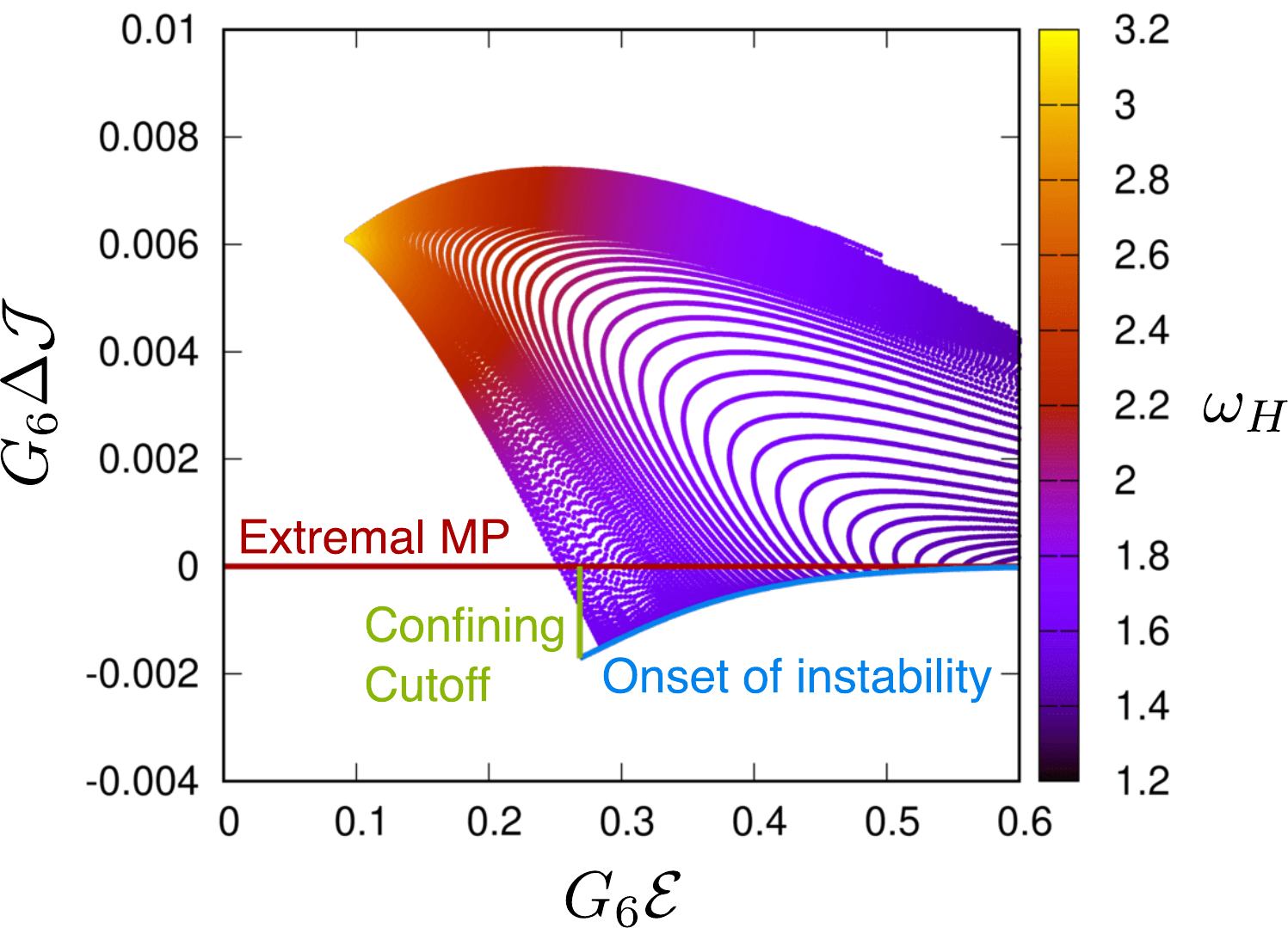}\label{fig:EJdiffOm}} \:\quad
\subfigure[$(\mathcal{E},\sigma_H;\tau_H)$.]{\includegraphics[scale=0.475]{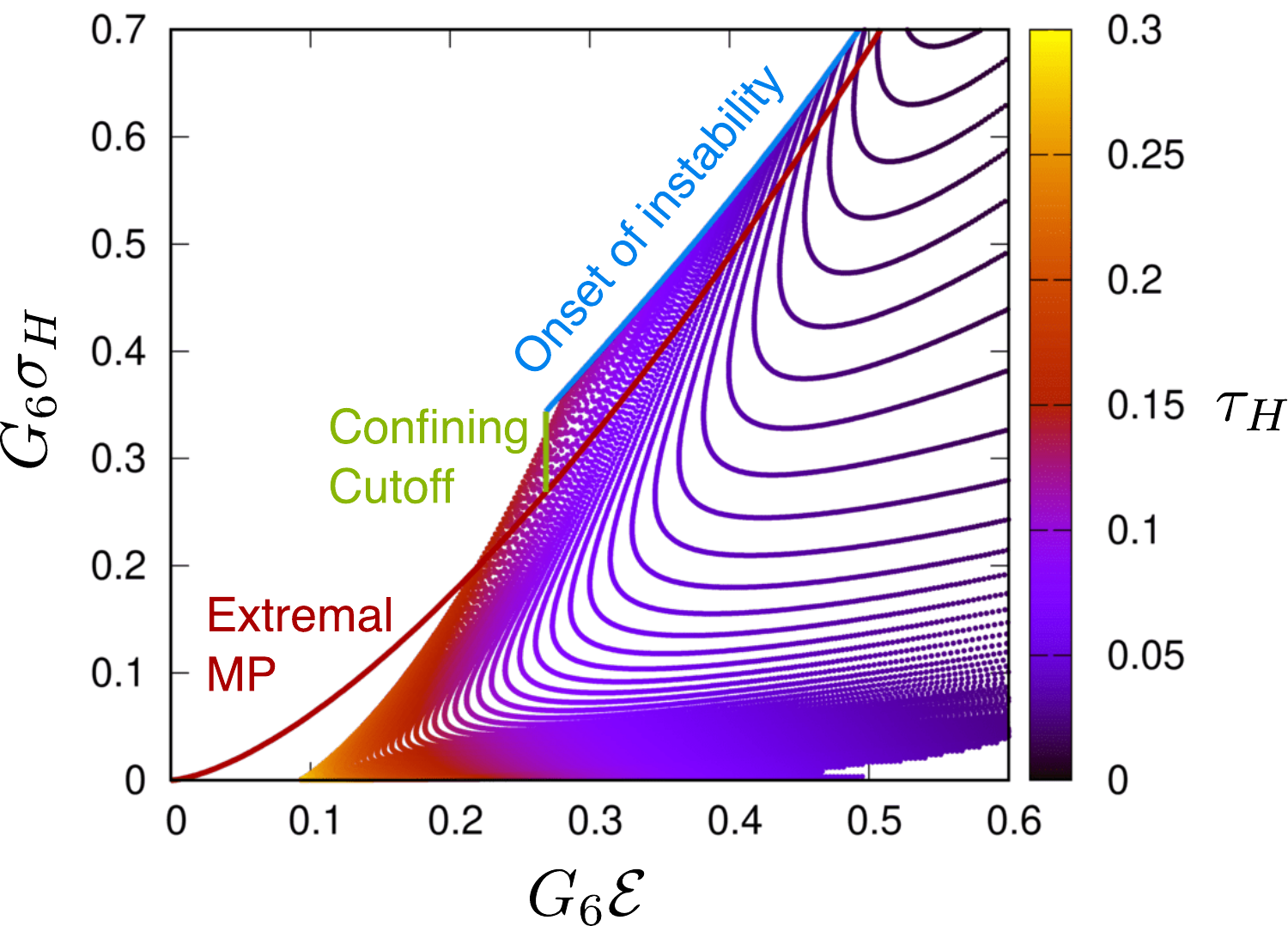}\label{fig:EST}}
\subfigure[$(\mathcal{E},\mathcal{T}_z/\mathcal{E};v_H)$.]{\includegraphics[scale=0.475]{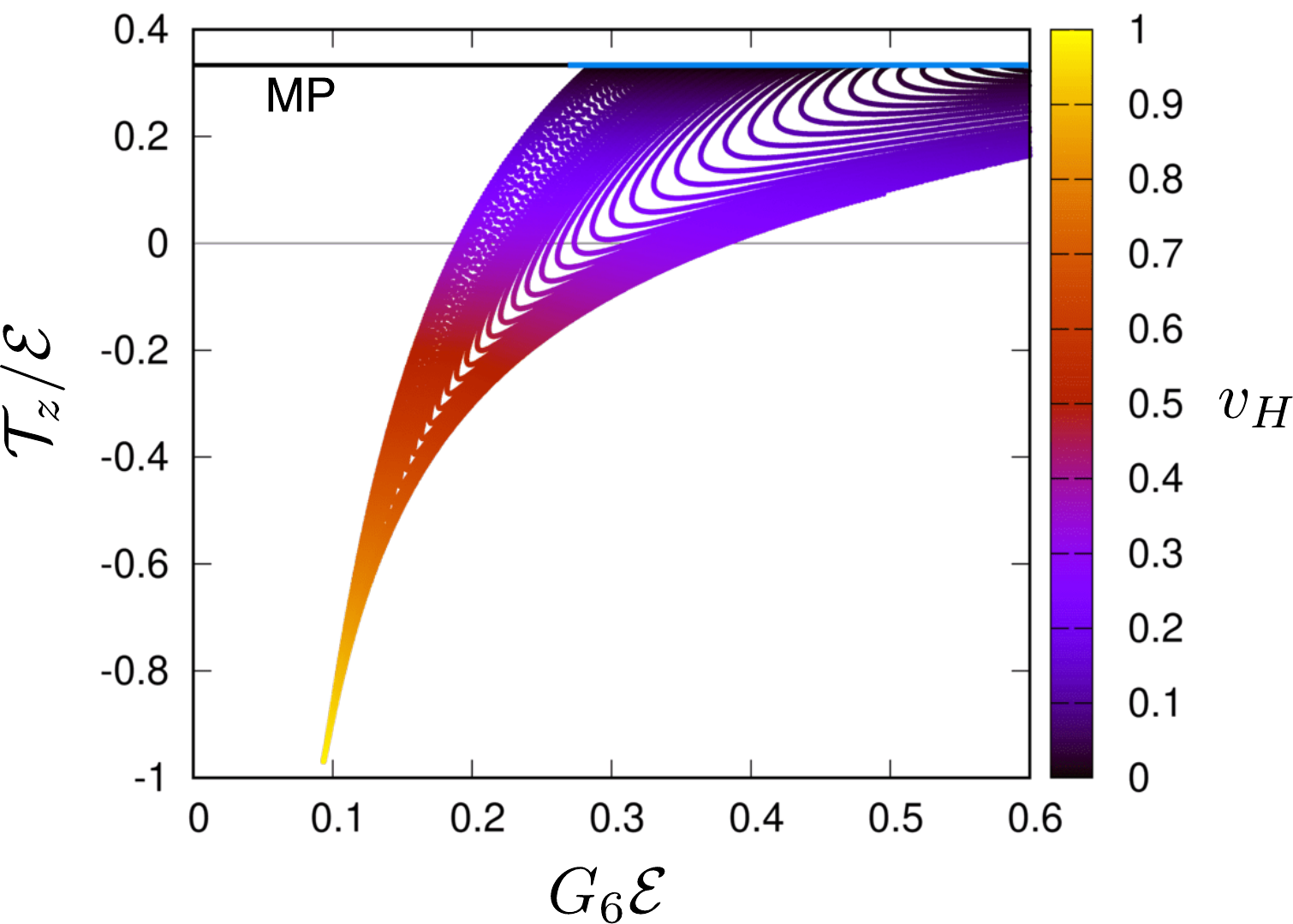}\label{fig:ETzvH}} \:\quad
\subfigure[$(\omega_H,v_H;\sigma_H)$.]{\includegraphics[scale=0.475]{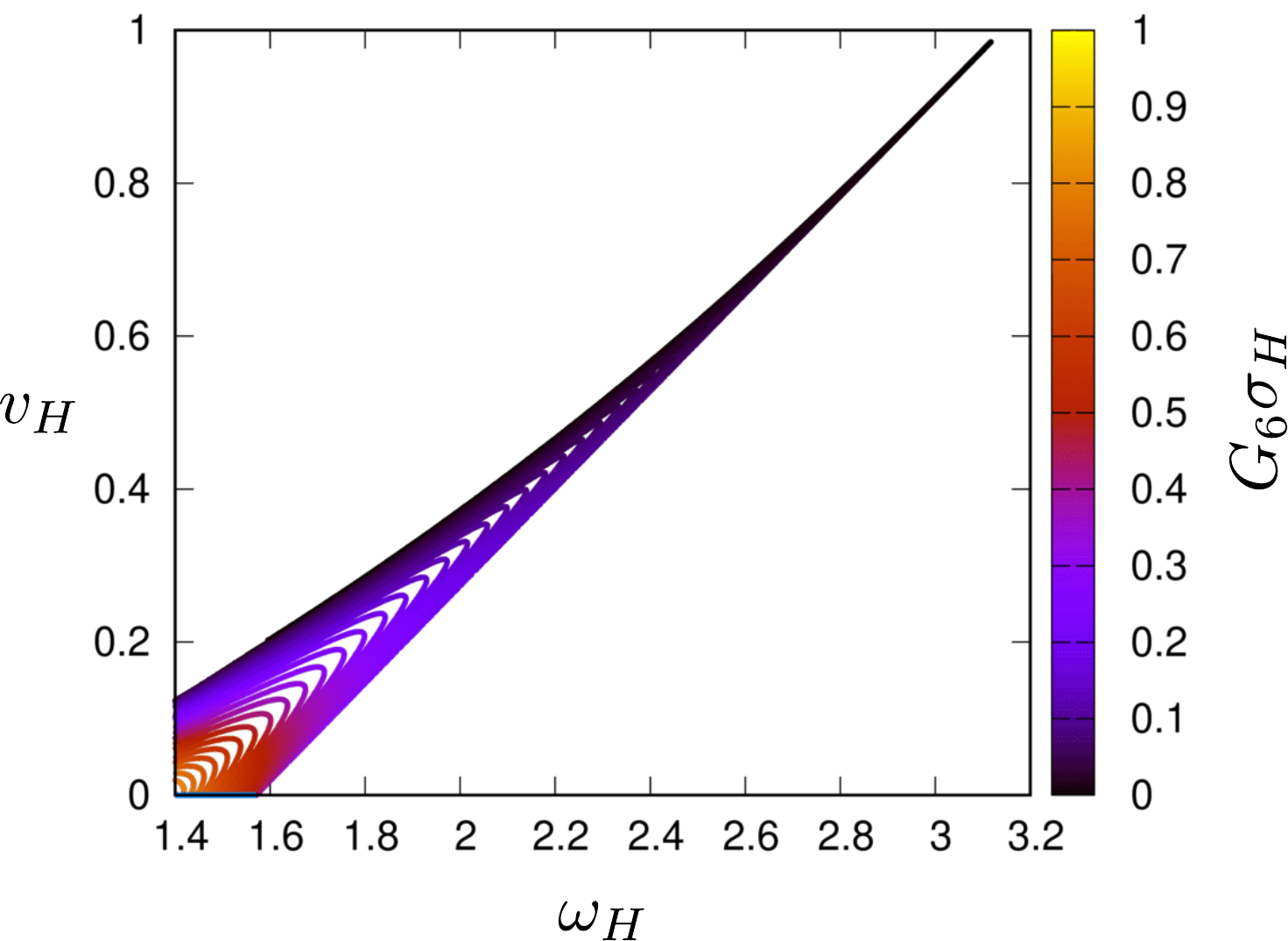}\label{fig:OmvHS}}
\caption{Thermodynamic quantities of helical black strings.}
\label{fig:thermoJ}
\end{figure}

Dimensionless thermodynamic quantities of helical black strings, as defined in \eqref{dimensionlessThermo}, are summarized in Fig.~\ref{fig:thermoJ} (for the solutions we obtained numerically\footnote{\label{foot:NumericConsiderations}From the analyses leading to Figs.~\ref{Fig:zeroModeGL} and~\ref{fig:stabilityDiag} one knows that the superradiant onset curve $cA$ runs over $\widetilde{\Omega}_H|_c \leq \widetilde{\Omega}_H \leq \widetilde{\Omega}_H|_A$ with $\widetilde{\Omega}_H|_c=3/5$ and $\widetilde{\Omega}_H|_A=\Omega_H^\mathrm{ext}=1/\sqrt{2}$ which corresponds to the energy range $\mathcal{E}|_c \leq \mathcal{E} \leq \mathcal{E}|_A$ with $G_6 \mathcal{E}|_c=27/(32\pi)\simeq 0.268574$ and $G_6 \mathcal{E}|_A= 45/(16 \pi)\simeq 0.895247$. This is the range of the blue superradiant onset curve in Fig.~\ref{fig:thermoJ}. Although, this onset starts at $\widetilde{\Omega}_H|_c/\Omega_H^\mathrm{ext}=3\sqrt{2}/5\simeq 0.848528$, in Fig.~\ref{fig:thermoJ} we only display helical strings near the onset for $ \Omega_H/\Omega_H^\mathrm{ext} \geq 0.865$, since for $3\sqrt{2}/5 \leq \Omega_H/\Omega_H^\mathrm{ext} < 0.865$ the asymptotic exponential decay \eqref{asymp_exp_power} is very weak, and the numerical analysis is difficult. In particular, we have evidence that the edge  $\widetilde{\Omega}_H \to \widetilde{\Omega}_H|_c^+$ is obtained in the limit $\eta_0 \to 1$, but in this limit the helical solution is distinct from  MPBS (which has $\eta=1$) and may not be described by a regular solution. Technical details for constructing these numerical solutions are further explained in appendix~\ref{app:tech} (in the case of the spherical gauge ansatz).}). In Figs.~\ref{fig:EJdiffOm} and \ref{fig:EST}, the blue, green, and red lines correspond to the onset of the instability of the MPBS (i.e. blue $cA$ curve in Fig.~\ref{fig:stabilityDiag}), confining cutoff for the asymptotic exponential decay of the MPBS perturbation (i.e. green line $cB$ in Fig.~\ref{fig:stabilityDiag}), and extremal MPBS, respectively.
Non-extremal MPBS exist below (above) the extremal red line in Fig.~\ref{fig:EJdiffOm} (Fig.~\ref{fig:EST}); otherwise MPBS describes
a naked singularity. In the `triangular-like' region bounded by these three curves (blue, green, red), the MPBS is superradiant unstable to the spin-2 tensor perturbation with the lowest  azimuthal quantum number $|m|=2$ (as detailed in the discussions of section~\ref{sec:fluctuation}, footnote~\ref{foot:sign-m} and of Figs.~\ref{Fig:zeroModeGL} and~\ref{fig:stabilityDiag}).

In Fig.~\ref{fig:EJdiffOm}, we plot the angular momentum  difference  as defined in~\eqref{DeltaJ}, $\Delta \mathcal{J}\equiv (\mathcal{J}-\mathcal{J}_{\hbox{\tiny ext\, MP}})|_{\hbox{\tiny same}\,\mathcal{E}}$, and the horizon angular velocity  $\omega_H$ (see the color code on the right column) as a function of the energy  $\mathcal{E}$.
As expected from the perturbative analysis, helical black strings exist at and above the blue superradiant onset line (where they merge with MPBS), but now, with the full nonlinear data, we see that they can extend well above for higher values of $\mathcal{J}$. In particular, helical black strings exist even in the parameter space where the MPBS is super-extremal and thus singular, i.e., where $\Delta \mathcal{J}>0$ (above the horizontal red line). Altogether, Fig.~\ref{fig:EJdiffOm} gives the phase space region $(\mathcal{E},\mathcal{J})$ where helical black strings exist, which is information relevant to discuss the properties of the black string system in the microcanonical phase diagram.

In Fig.~\ref{fig:EST}, we plot the entropy  $\sigma_H$ of helical strings, and their temperature  $\tau_H$ (see the color code on the right column), as a function of the energy  $\mathcal{E}$.
The entropy $\sigma_H$ can be very close to zero at nonzero $(\mathcal{E},\mathcal{J})$. In the zero entropy limit (bottom of the plot), the temperature $\tau_H$ is likely non-zero and finite.
Note that this zero entropy limit corresponds, in Fig.~\ref{fig:EJdiffOm}, to the left end of the upper boundary in the region $(\mathcal{E},\Delta\mathcal{J})$ of existence of the helical strings. Although this is not clear from Fig.~\ref{fig:EST}, we have explicitly checked (with analyses similar to the one in Fig.~\ref{fig:nearonsetPertNonLin} but for other values of $\mathcal{E}$ besides $G_6\mathcal{E}=0.35$) that, for a given $(\mathcal{E},\mathcal{J})$ where helical strings and MPBS co-exist, helical black strings {\it always} have higher entropy  than MPBS. This is in agreement with the perturbative findings summarized in Fig.~\ref{Fig:DeltaSigma} but extend them beyond the superradiant merger vicinity. Thus, helical black strings dominate the microcanonical phase diagram over MPBS (more discussion about this important property will be given in section~\ref{sec:numsol2}).

To further unravel properties of helical black strings,
in Fig.~\ref{fig:ETzvH} we plot the ratio of the tension along the string direction and the energy, $\mathcal{T}_z/\mathcal{E}$, and the horizon velocity $v_H$ (see the color code on the right column) as a function of the energy  $\mathcal{E}$. Here, the horizontal black line corresponds to the exact result for the MPBS,
$\mathcal{T}_z/\mathcal{E}=1/3$ as follows from \eqref{TenLE4MPBS}, and the superradiant merger line is the blue straight line with $0.268574 \lesssim G_6\mathcal{E} \lesssim 0.895247$ on this MP line (thus, also with $\mathcal{T}_z/\mathcal{E}=1/3$). We compare results for solutions with {\it no} momentum,  $P=0$ ({\it i.e.},~no Lorentz boost). In particular, this means that in Fig.~\ref{fig:OmvHS} the MPBS has $v_H=0$, but the helical strings have $v_H\neq 0$ (recall the discussions of~\eqref{asympcond} or~\eqref{NoPcond}). At the onset one has $v_H=0$ but the $v_H$ of helical strings grows substantially as we move away from the onset (see also the color in Fig.~\ref{fig:ETzvH}). The fact that our helical solutions have $P=0$ but $v_H\neq 0$ means that $v_H$ is an intrinsic velocity of the system that is generated spontaneously since it is required to support the symmetries of the helical black string. This corresponds to the configuration without a Lorentz boost (hence $P=0$); of course we can then generate boosted helical and MP black strings by applying a Lorentz boost to these fundamental solutions (which we are not doing in the presentation of any of our results).
Analysing Fig.~\ref{fig:ETzvH}, we conclude that the ratio $\mathcal{T}_z/\mathcal{E}$ of helical strings is always smaller than the MPBS ratio. In particular, as $v_H$ grows (say, in the region where $0.3 \lesssim v_H\leq 1$), one sees that $\mathcal{T}_z/\mathcal{E}$ can become negative with $ \mathcal{T}_z L/E\to -1$ as $v_H \to 1$.\footnote{It is known that the bare tension can be negative \cite{Hovdebo:2006jy} for a boosted black string, while the effective tension remains positive \cite{Kastor:2007wr} That situation is different from ours where the tension can be negative even for $P=0$.} This might suggest that helical strings can have some instability in this negative tension region of parameters, but we do not address this possibility here.
The tension to energy ratio reaches its minimum, $ \mathcal{T}_z L/E\to -1$, when the horizon velocity approaches the speed of light, $v_H \to 1$, where the entropy  also seems to approach zero (as far as we can numerically check), $\sigma_H\to 0$, and the angular velocity seems to reach its maximum value of $\omega_H \to \pi$ (consistent with the analytical result \eqref{omegaH_allowed}).
This limit also provides helical black strings with minimum energy and tension, $G_6 \mathcal{E}\simeq - G_6 \mathcal{T}_z \simeq 0.09$, and with maximum angular momenta and temperature, $G_6 \mathcal{J} \simeq 0.03$ and $\tau_H \simeq 0.29$ (where we used extrapolation of the data in Fig.~\ref{fig:thermoJ} to obtain these critical values).

Given the above properties of the phase diagram of helical strings, one naturally wonders whether there is room for the existence of regular Kaluza-Klein geons (i.e., horizonless strings whose centrifugal force balances self-gravitation) in the zero entropy limit of our helical strings (since this is a common feature in resonator and hairy black object solutions). We considered this possibility seriously, but to the best of our attempts our answer to this question is negative. In appendix~\ref{sec:geon}, we discuss the absence of Kaluza-Klein geons as well as the $v_H \to 1$ limit and pp-wave solutions in more detail. Although the horizon area vanishes in the $v_H \to 1$ limit, we have evidence that the geometry becomes singular in this limit. In particular, we have evaluated the Kretschmann curvature scalar at the horizon and find that it diverges in the $\sigma_H \to 0$ limit. This behaviour is thus similar to the zero horizon size limit of a Schwarzschild black hole, where no regular geon is obtained.

\subsection{Phase diagram of helical and resonator black strings}
\label{sec:numsol2}

Helical and MP black strings are not the only asymptotically Kaluza-Klein solutions ($\mathcal{M}^{1,4}\times S^1$) of 6-dimensional Einstein gravity. Indeed, as discussed in detail in \cite{Dias:2022mde,Dias:2022str}, the phase diagram of solutions also includes the so-called black resonator strings found in \cite{Dias:2022str}. Interestingly, the latter bifurcate from the MP black strings along the same superradiant onset curve as the helical black strings, for reasons detailed in section~4.4 of \cite{Dias:2022mde}. When they co-exist, we already know that resonator and helical black strings have higher entropy  than MP strings with the same energy and angular momenta (and same Kaluza-Klein length $L$). However, it is also important to ask which of the two solutions, helical or resonator strings, dominate the microcanonical ensemble. We find that, for fixed values of the energy $\mathcal{E}$ and angular momenta $\mathcal{J}$ where they co-exist, resonator black strings \cite{Dias:2022str} always have higher entropy  $\sigma_H$ than the helical black strings.

\begin{figure}[t]
\centering
\includegraphics[scale=0.35]{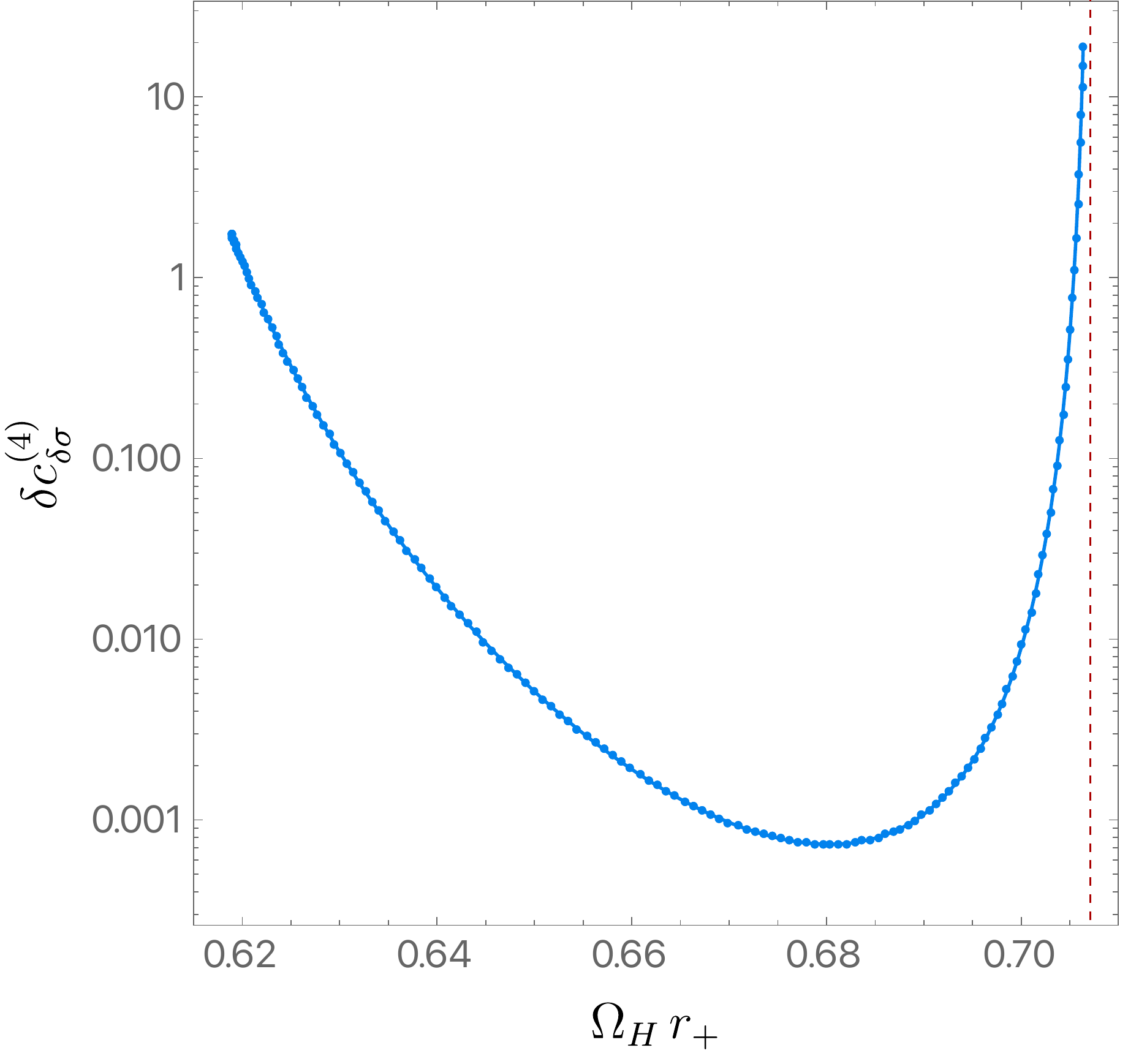}
\hspace{0.1cm}
\includegraphics[scale=0.29]{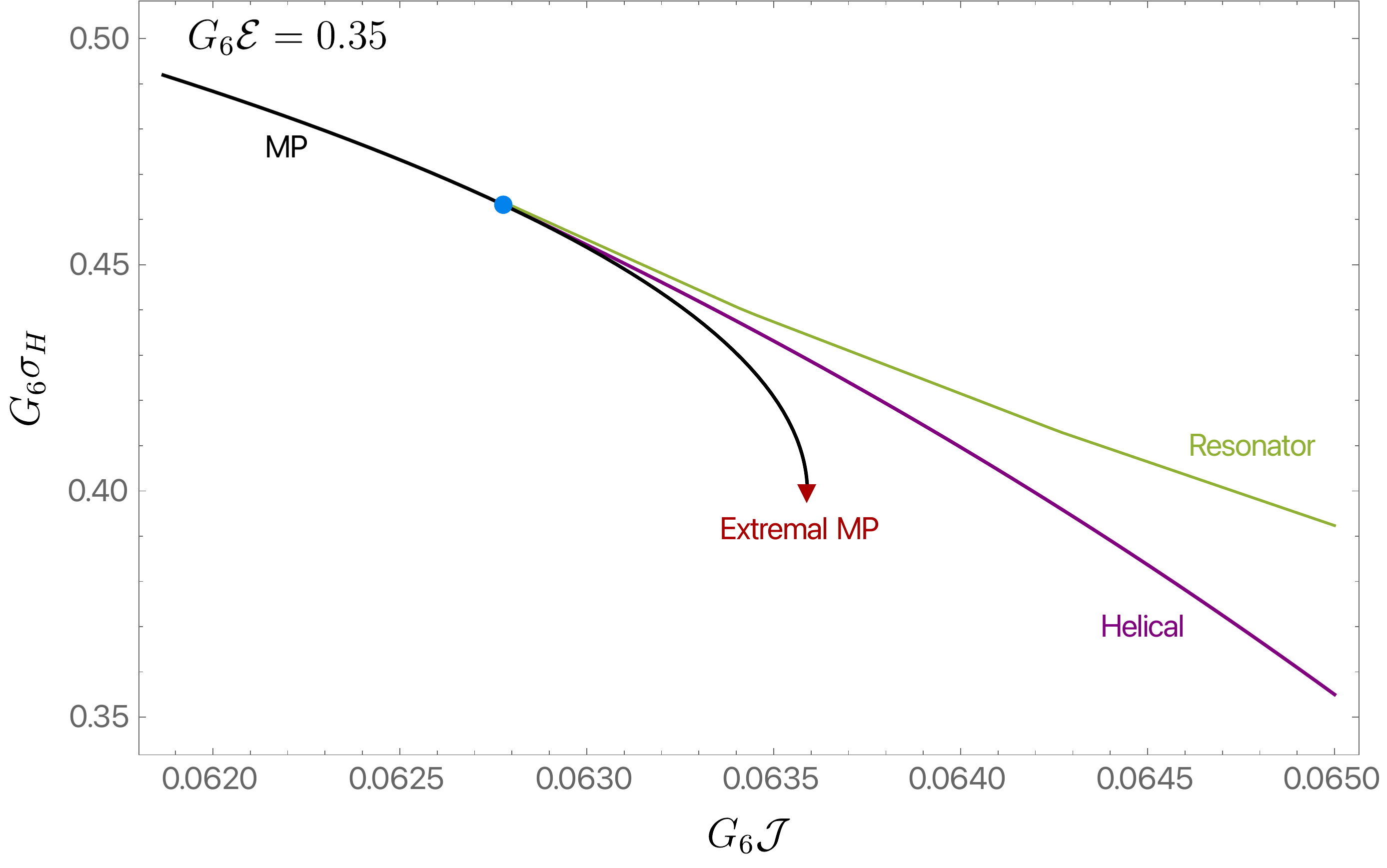}
\caption{{\bf Left panel:} Coefficient of the entropy  difference $\delta \sigma_H$ between the black resonator and helical strings near the superradiant merger ({\it i.e.}~as given by the perturbative analysis of section~\ref{sec:perturbative}) as a function of the horizon angular velocity along the merger curve (notice the logarithmic scale; the red dashed vertical corresponds to $\Omega_H=\Omega_H^\mathrm{ext}$). {\bf Right panel:} Comparison of the entropy $\Delta \sigma_H$ of the MP black strings (black curve), helical black strings (purple curve) and black resonator strings (green curve) with the same energy  $G_6 \mathcal{E}=0.35$ as a function of the angular momentum  not only near the merger (blue disk) but also away from it (nonlinear results). The blue disk where the three families meet is the superradiant onset/merger (it corresponds to the point in the left panel with $\tilde{\Omega}_H\simeq 0.64821697$) and the red dot at the endpoint of the MP curve marks the extremal MP black string.}
\label{fig:Resonator-Helical}
\end{figure}

This is explicitly demonstrated in Fig.~\ref{fig:Resonator-Helical}. In the left panel, we focus our attention only in the region of the phase diagram that is in a neighbourhood of the superradiant onset/merger line of the MP black string (recall this is the blue line $cA$ in Figs.~\ref{Fig:zeroModeGL} and~\ref{fig:stabilityDiag}). In this region we can simply use the perturbative helical string result  \eqref{EntropyDiff}-\eqref{EntropyDiff2} of section~\ref{sec:perturbative}, and borrow the resonator perturbative counterpart given in equations (4.14)-(4.15) of \cite{Dias:2022str}, as good approximations. In this comparison, we have to be cautious because the expansion and rotation parameters of the two solutions are not the same. To find the relation between the two,  we need to expand the expansion and rotation parameters $\{\epsilon,\tilde{a}\}|_{\rm res}$ of the resonator strings, as given in \cite{Dias:2022str}, as a function of the helical expansion and rotation parameters $\{\epsilon,\tilde{a}\}$ to find the relation that describes resonator strings with the same energy  $\mathcal{E}$ and angular momenta  $\mathcal{J}$ as the helical string (since we want to compare them in the microcanonical ensemble with fixed Kaluza-Klein length $L$). After completing this task, we can then compute the perturbative expansion of the entropy  difference $-$ let us call it  $\delta \sigma_H$ $-$ between resonator strings and the helical string with the same energy and angular momenta:
 \begin{eqnarray} \label{EntropyDiffResHel}
\delta \sigma_H&=&\left( \sigma_{H,\rm res}-\sigma_{H,\rm hel}\right)\big|_{{\rm same}\,(\mathcal{E},\mathcal{J})}\nonumber \\
&=&G_6\,\delta c_{\delta \sigma}^{(4)}\,\epsilon^4  +\mathcal{O}(\epsilon^6),
\end{eqnarray}
where $\epsilon$ is the helical expansion parameter introduced in section~\ref{sec:perturbative}.

The coefficient $\delta c_{\delta \sigma}^{(4)}$ of \eqref{EntropyDiffResHel} is plotted along the  superradiant onset/merger line $cA$ parametrized by $\widetilde{\Omega}_H|_c \leq \widetilde{\Omega}_H \leq 1/\sqrt{2}$  (see Figs.~\ref{Fig:zeroModeGL} and~\ref{fig:stabilityDiag}) in the left panel of Fig.~\ref{fig:Resonator-Helical}. Since one always has $\delta c_{\delta \sigma}^{(4)}>0$, we conclude that, around the merger line with the MP black strings, black resonator strings have higher entropy  than helical black strings with the same $\mathcal{E}$ and $\mathcal{J}$.

We may then ask if resonator strings still dominate the microcanonical phase diagram far away from the superradiant onset where the perturbative results no longer provide a good description/approximation. For that we need to use the full nonlinear helical solutions of Fig.~\ref{fig:thermoJ} and the nonlinear resonators in Fig.~6 of \cite{Dias:2022str}. We find that when they co-exist, resonator strings always have higher entropy  than the helical strings (with the same $\mathcal{E}$ and $\mathcal{J}$) no matter how far away they are from the superradiant onset. This is illustrated in the right panel of Fig.~\ref{fig:Resonator-Helical} for a particular family of black helical and resonator strings with $G_6 \mathcal{E}=0.35$. Resonator (green curve) and helical (purple curve) strings bifurcate from the MP black string (black curve) at $G_6 \mathcal{J}\simeq 0.06277479$ (blue disk) $-$ which corresponds to the point with $\tilde{\Omega}_H\simeq 0.64821697$ in the left panel $-$ and both extend to higher values of $\mathcal{J}$. As stated above, we clearly see that for a given $\mathcal{J}$ where the three (or two) solutions co-exist, the black resonator strings always have higher entropy  than the black helicoidal strings (and the MP strings).

\subsection{Stationarity of the helical black string revisited}
\label{sec:timelike}

As discussed previously, it can be shown that the helical black string spacetime is stationary. That is, the spacetime admits an asymptotically timelike Killing vector. In this section we argue that this property does not violate the rigidity theorem.

Let us consider a linear combination of $\partial_T$ and $\partial_Z$ as
\begin{equation}
\zeta \equiv c_T \partial_T + c_Z \partial_Z\, ,
\label{stat_zeta}
\end{equation}
where $c_T, c_Z$ are some constants. Using the asymptotic expansion \eqref{asymp_0th_power_sol}, we obtain as $r \to \infty$
\begin{equation}
\zeta^2 = g_{MN}\zeta^M \zeta^N = \frac{(4 c_T h_\infty + c_Z k_\infty)^2}{16} r^2 + (-c_T^2+c_Z^2) + \mathcal{O}(r^{-2})\, .
\label{normKVFinf_nonlinear}
\end{equation}
If $4 c_T h_\infty + c_Z k_\infty \neq 0$, this diverges, $\zeta^2 \to + \infty$. By setting $c_T=k_\infty$ and  $c_Z = - 4 h_\infty$, the behavior becomes
\begin{equation}
\zeta^2 \to - (k_\infty^2 - 16 h_\infty^2)  \quad (r \to \infty)\, .
\label{zetasq}
\end{equation}
For all numerical helical black strings, we find $\zeta^2<0$ (see Fig.~\ref{fig:bulk_z}). That is, $\zeta$ is an asymptotically timelike Killing vector, and hence this spacetime is stationary. Meanwhile, the $U(1)_\Psi$ rotational isometry is broken by the superradiant instability. Thus, the helical black string is a stationary non-axisymmetric spacetime.

The rigidity theorem for higher dimensional black holes states that a stationary black hole should have a Killing vector field other than an asymptotically timelike Killing vector \cite{Hollands:2006rj,Moncrief:2008mr,Hollands:2008wn}.
The point of the rigidity theorem is that the Killing horizon generator is spacelike in asymptotic infinity, and therefore a stationary black solution should be equipped with another Killing vector in addition to the asymptotically timelike Killing vector so that the latter can be given as a linear combination of the Killing horizon generator and the extra Killing vector.
For the helical black string geometry, while the axisymmetry $U(1)_\Psi$ is broken, we can use the translation symmetry generated by $\partial_Z$. The presence of this symmetry makes the helical black string solution consistent with the rigidity theorem.

In stationary spacetimes, we could use coordinates where the time translation Killing vector is asymptotically timelike. For the helical black string, the transformation from $(T,Z)$ to such coordinates $(\widetilde{T},\widetilde{Z})$ can be done by the Lorentz boost \eqref{lol_boost} with $\tanh b = - 4 h_\infty/k_\infty$,
\begin{equation}
\widetilde{T} = \frac{1}{\sqrt{k_\infty^2 - 16 h_\infty^2}} \left( k_\infty T + 4 h_\infty Z \right)\, , \quad
\widetilde{Z} = \frac{1}{\sqrt{k_\infty^2 - 16 h_\infty^2}} \left( k_\infty Z + 4 h_\infty T \right)\, .
\label{TZ_from_tauz}
\end{equation}
Their dual vectors are
\begin{equation}
\partial_{\widetilde{T}} = \frac{1}{\sqrt{k_\infty^2 - 16 h_\infty^2}} \left( k_\infty \partial_T - 4 h_\infty \partial_Z \right)\ , \quad
\partial_{\widetilde{Z}} = \frac{1}{\sqrt{k_\infty^2 - 16 h_\infty^2}} \left( k_\infty \partial_Z - 4 h_\infty \partial_T \right)\ .
\label{delTZ_from_tauz}
\end{equation}
The first one, $\partial_{\widetilde{T}}$, is nothing but the asymptotically timelike Killing vector \eqref{zetasq} normalised as $(\partial_{\widetilde{T}})^2|_{r \to \infty} \to -1$. The first equation in \eqref{delTZ_from_tauz} shows that the asymptotically timelike Killing vector $\partial_{\widetilde{T}}$ is a linear combination of the Killing horizon generator $\partial_{T}$ and the Killing vector $\partial_{Z}$ corresponding to the $U(1)_Z$ isometry. Conversely, this implies that the Killing horizon generator is formed as a linear combination of the asymptotically timelike Killing vector and the Killing vector of $U(1)_Z$.

The coordinates $(\widetilde{T},\widetilde{Z})$ are in the rotating frame at infinity and have their counterparts $(\tilde{t},\tilde{z})$ in the non-rotating frame at infinity. The transformation is analogous to \eqref{rot2nonrot}, but now the transformation of the angular coordinate $\Psi$ does not involve $\widetilde{T}$, giving
\begin{equation}
\tilde{t} = \widetilde{T}, \quad \tilde{z} = \widetilde{Z} \, , \quad
\tilde{\psi} =\Psi + \frac{\sqrt{k_\infty^2 - 16 h_\infty^2}}{2}\tilde{Z}\, .
\label{TZ_rot2nonrot}
\end{equation}
The dual vectors are related as
\begin{equation}
\partial_{\widetilde{T}} = \partial_{\tilde{t}}\, , \quad
\partial_{\widetilde{Z}} = \partial_{\tilde{z}} + \frac{\sqrt{k_\infty^2 - 16 h_\infty^2}}{2} \partial_\psi\, , \quad
\partial_\Psi = \partial_{\tilde{\psi}}\, .
\label{statdualvectors}
\end{equation}

In these coordinates, the bulk metric does not explicitly depend on $\tilde{t}$, while it explicitly does depend on $\tilde{z}$. Combining \eqref{rot2nonrot}, \eqref{TZ_from_tauz}, and \eqref{statdualvectors}, one can deduce that $\tilde{z}$ is precisely related to $\Theta$ in \eqref{Theta_ty} as
\begin{equation}
\tilde{z} = \frac{2}{\sqrt{k_\infty^2 - 16 h_\infty^2}} \Theta\, .
\end{equation}
The corresponding part of the bulk metric in the non-rotating frame at infinity \eqref{dSigma12} indicates that a $\tilde{z}$-shift is not an isometry. As shown in \eqref{statdualvectors}, the isometry $\partial_{\widetilde{Z}}$ is given by a linear combination of $\partial_{\tilde{z}}$ and $\partial_\psi$, both of which are broken translations but the particular combination realizes an isometry.

One is aware that using $(\tilde{t},\tilde{z})$ would simplify the bulk metric in the non-rotating frame at infinity. However, there is a drawback in that the helical black strings in these coordinates have nonzero boost $P \neq 0$ in general. This is clear from the Lorentz boost involved in the coordinate transformation \eqref{TZ_from_tauz}. When we consider black strings, we prefer to compare non-boosted solutions, where an interesting aspect in the helical black string geometry is the spontaneous generation of the (intrinsic) horizon velocity $v_H$ even for $P=0$ which is required to support the symetries of the helical black strings. Hence, in the non-rotating frame at infinity, we put priority to setting $P=0$, at the cost of allowing explicit $t$-dependence in the bulk metric as \eqref{dSigma12}. This is not a problem because thermodynamic quantities constructed from asymptotic infinity and horizon quantities are not time dependent.

\section{Conclusion}
\label{sec:conclusion}

In this paper, we constructed cohomogeneity-1 {\it helical black strings} with $\mathbb{R}_T \times U(1)_Z \times SU(2)$ isometries branching from the superradiant instability of the six-dimensional equal angular momenta Myers-Perry black string (MPBS). We showed that the helical black string  is a stationary non-axisymmetric spacetime since it has a Killing vector field that is timelike everywhere at the asymptotic boundary. The helical black string is supported by a nonzero intrinsic horizon velocity $v_H$ along the direction of the string even though the solution was no momentum along the string, $P=0$. So, this is an intrinsic horizon velocity required to support the helical symmetry of the helical strings and not a velocity that emerges after applying a Lorentz boost (which we can also do, but we do not); see discussions of~\eqref{asympcond} or~\eqref{NoPcond}. In a phase diagram of asimptotically Kaluza-Klein solutions, the helical black strings bifurcate from the MPBS at the onset of the superradiant instability and then they extend to higher values of angular momenta $\mathcal{J}$ beyond the region where MPBS can exist (the boundary of the latter is the extreme $T=0$ family of solutions with maximal $\mathcal{J}$); see e.g.~Fig.~\ref{fig:EJdiffOm}.
 The entropy  (i.e. dimensionless entropy measure in units of the string length $L$) of the helical black string is always higher than that of the MPBS with the same (dimensionless) energy $\mathcal{E}$ and angular momenta $\mathcal{J}$ when they coexist in the $\mathcal{E}-\mathcal{J}$ phase space (see e.g.~Figs.~\ref{fig:nearonsetPertNonLin} or~\ref{fig:Resonator-Helical} for illustrations of this property). This demonstrates that helical strings dominate the microcanonical phase diagram  over the MP strings, and indicates (using the second law of thermodynamics) that an unstable MPBS can dynamically evolve into a helical black string. It was observed that the small entropy limit $\sigma_H \to 0$ of the helical string corresponds to the maximal velocity limit $v_H \to 1$ where the horizon velocity reaches the speed of light.

In addition to the helical black string with $\mathbb{R}_T \times U(1)_Z \times SU(2)$ isometries, there is another black string solution with fewer isometries (namely $\mathbb{R}_T \times SU(2)$), coined {\it black resonator string} \cite{Dias:2022str}, that also branches-off from the same superradiant instability of the MPBS and has the same Kaluza-Klein asymptotics. In addition to $U(1)_\Psi$, the $U(1)_Z$ isometry is also broken in the black resonator strings and hence they are non-uniform string solutions. Remarkably, the black resonator string is a non-stationary spacetime in contrast to the helical black string. For the latter, the presence of the $U(1)_Z$ isometry makes it possible to form an asymptotically timelike Killing vector as a linear combination of the Killing horizon generator, which is asymptotically spacelike, and the $U(1)_Z$ translation Killing vector. For (dimensionless) energy $\mathcal{E}$ and angular momenta $\mathcal{J}$ where the three (helical, resonator and MP) black string solutions co-exist, the black resonator string always has the highest dimensionless entropy  (for a given $\{\mathcal{E},\mathcal{J}$\}), and the MPBS has the lowest entropy. This is illustrated in Fig.~\ref{fig:Resonator-Helical}. Therefore, the black resonator strings dominate the microcanonical ensemble (and helical black strings dominate over MPBS). This entropic property also suggests that it is possible to have a dynamical time evolution from helical black strings to black resonator strings or from a MPBS towards the resonator string (eventually with the helical strings being a metastable configuration).

This aspect is worth of a more detailed discussion. The fact that helical black strings remain cohomogeneity-1, yet break  rotational and time translation symmetries makes them uniquely simple objects for studying black strings that break these symmetries.  It would be considerably simpler to study their perturbations and basic time evolution properties than similar (5-dimensional Kerr or other) black strings which are typically cohomogeneity-2 or higher.  We leave this study for future work.

Nevertheless, we can briefly give a possible scenario for the nonlinear dynamics. 
For a fine-tuned initial superradiant perturbation~(\ref{MPBS_fluc_rot}) of the MPBS, in which there is exact translation symmetry generated by $\partial_Z$, the evolution would likely proceed towards a helical black string solution.
But from entropic arguments, the helical black string is further unstable to modes which break the translation symmetry $\partial_Z$ and would continue to evolve towards the black resonator string. Depending on the time scale of the instability of the helical black string, the helical black string may be realized as a transient state or simply as a short-lived metastable state. Interestingly, we should also note that comparing the $\mathcal{E}-\mathcal{J}$ region of existence of helical strings in Fig.~\ref{fig:EJdiffOm} with the region of existence of black resonator strings (Fig.~6 of \cite{Dias:2022str}) one sees that there is a small window of parameters (roughly, for $ 0.09 \lesssim \mathcal{E} \lesssim 0.2$) where helical black strings exist but there are no black resonators (at least with $|m|=2$). Such helical strings might then be stable (although $|m|> 2$ or other superradiant or Gregory-Laflamme perturbations might still make them unstable).
In this short discussion we have not considered the competition between the superradiant and Gregory-Laflamme instabilities in possible time evolutions of MP black strings, though this was discussed in \cite{Dias:2022str}.

The stability analysis of helical black strings would be an interesting direction to understand the dynamics of rotating black strings, especially given the high-symmetry of helical black strings, unlike other rotating black string solutions.  We would expect the Kerr string to exhibit similar physics, such as the existence of black resonator and helical strings that compete in the microcanonical ensemble, and a similar competition between superradiant and Gregory-Laflamme instabilities, but the Kerr case is significantly more difficult to study due to a lack of symmetry.  

We also tried to find the horizonless limit (zero entropy  limit) of the helical black string, which one might naively have expected to be a regular Kaluza-Klein geon. However, we did not find such solutions when considering the zero horizon radius limit of the cohomogenenity-1 helical black strings.
In more detail, the zero entropy limit of the helical black strings is accompanied with sending the horizon velocity  to the speed of light, $v_H \to 1$, as well as with approaching the maximal angular frequency, $\omega_H \to \pi$. Our numerical results suggest that in this limit the metric fields cease to be exponentially decaying at asymptotic infinity, meaning that the confinement mechanism for superradiant bound states is lost. (Obstacles for obtaining such Kaluza-Klein geons are further discussed in appendix~\ref{sec:geon}).
It should also be said that in \cite{Dias:2022str}, a non-uniform Kaluza-Klein geon with $\mathbb{R}_T \times SU(2)$ isometries was found that is {\it not} the zero horizon radius limit of the black resonator string. We cannot exclude the existence of a similar Kaluza-Klein geon with $\mathbb{R}_T \times U(1)_Z \times SU(2)$ isometries that is {\it not} the zero horizon radius limit of the helical string. However, we have not found evidence for the existence of this solution either (see appendix~\ref{sec:geon} for further discussions of our attempts).

\acknowledgments
The authors would like to thank Roberto Emparan, Takahisa Igata, Akihiro Ishibashi, and Masashi Kimura for useful discussions.
O.J.C.D. acknowledges financial support from the  STFC ``Particle Physics Grants Panel (PPGP) 2018" Grant No.~ST/T000775/1. OD acknowledges the Isaac Newton Institute, Cambridge, and the organizers of its long term programme ``Applicable resurgent asymptotics: towards a universal theory" during which this work was completed.
The work of T.I. was supported in part by JSPS KAKENHI Grant Number 19K03871.
The work of K.M. was supported in part by JSPS KAKENHI Grant Nos. 20K03976, 21H05186 and 22H01217.
B.W. acknowledges support from ERC Advanced Grant GravBHs-692951 and MEC grant FPA2016-76005-C2-2-P. J.~E.~S. has been partially supported by STFC consolidated grant ST/T000694/1. The authors also acknowledge the use of the IRIDIS High Performance Computing Facility, and associated support services at the University of Southampton, in the completion of this work.

\appendix

\section{Asymptotic solutions in the spherical gauge ansatz}

\subsection{Solutions near horizon}
\label{app:horexp}

In the context of the helical black string ansatz in the spherical gauge
of section~\ref{sec:ansatzSphericalGauge}, near the horizon, the metric components are expanded as in (\ref{Xexp}). Substituting these into the equations of motion~(\ref{EOM_f}-\ref{EOM_gamma}), we obtain the regular series expansion of the metric. For the expansion to be around the black hole horizon, $f_0=g_0=0$ are imposed. At the leading order, we obtain (\ref{horizon_4hkq_relation}), introducing one relation among 6 other coefficients. We regard $h_0$ to be fixed by (\ref{horizon_4hkq_relation}). Thus, we get 5 unfixed coefficients in the leading order.
In the next order, $(f_1,h_1,q_1)$ remain free, and the other coefficients satisfy
\begin{equation}
\begin{split}
 g_1=&\frac{2(\eta_0^2 - \eta_0 \beta_0 + 1)}{r_+ \eta_0}\, ,\\
 k_1=& \frac{2\{2r_+ h_1 q_1 \eta_0\beta_0\gamma_0(\eta_0^2 - \eta_0\beta_0 + 1)-f_1 k_0(\eta_0^2-1)^2\}}{r_+\beta_0\eta_0(\gamma_0 q_0 q_1 - f_1)(\eta_0^2 - \eta_0 \beta_0 + 1)}\, ,\\
\eta_1=&\frac{(\eta_0^2-1)^2\{r_+^2 \beta_0 k_0^2 (\eta_0^2 + 1)+16 \gamma_0 (\eta_0^2 - \eta_0\beta_0 + 1)\}}{ 8 r_+ \beta_0 \gamma_0(\eta_0^2 - \eta_0\beta_0 + 1)}\, ,\\
\gamma_1=&-\frac{r_+ f_1 k_0^2 (\eta_0^2 - 1)^2 + 8\eta_0 \gamma_0^2 q_1^2(\eta_0^2 - \eta_0 \beta_0 + 1)}{8 f_1 \eta_0 (\eta_0^2 - \eta_0 \beta_0 + 1)}\, ,\\
\beta_1=&\frac{1}{4r_+\eta_0^2(\gamma_0 q_0q_1-f_1)^2 (\eta_0^2 - \eta_0 \beta_0 + 1)^2}[
-4r_+^3  \eta_0^2 \beta_0^2 f_1 h_1^2 (\eta_0^2 - \eta_0 \beta_0 + 1)^2\\
&+4r_+^2 \eta_0 \beta_0 f_1 h_1 k_0 q_0 (\eta_0^2 - 1)^2 (\eta_0^2 - \eta_0 \beta_0 + 1)
-r_+ f_1 k_0^2 q_0^2 (\eta_0^2 - 1)^4 \\
&-8 (\eta_0^4 + \eta_0^3 \beta_0 - 2 \eta_0^2 \beta_0^2 - 2 \eta_0^2 + \eta_0 \beta_0 + 1) \eta_0 (\gamma_0 q_0 q_1 - f_1)^2 (\eta_0^2 - \eta_0 \beta_0 + 1)
]\, ,
\end{split}
\end{equation}
These are determined by 8-parameters $(f_1,\eta_0,\beta_0,h_1,k_0,\gamma_0,q_0,q_1)$ together with horizon radius $r_+$. Higher order coefficients in the series expansion are also fixed by these 8+1 parameters.

\subsection{Solutions near infinity}
\label{app:uvser}
We have already seen in the linear analysis that the perturbation $\delta\eta$ decays exponentially as \eqref{fluc_asymp} in the asymptotic infinity $r \to \infty$. To systematically incorporate this behaviour in the other fields beyond the linear order, we expand the fields in powers of exponentially decaying functions as
\begin{equation}
X(r) = \sum_{n=0}^{\infty} X_{(n)}(r) \left( \frac{e^{- \mu r}}{r^{3/2}} \right)^n\, ,
\label{asymp_exp_power}
\end{equation}
where the constant $\mu>0$ will be identified shortly. At each level $n$, we can evaluate $X_{(n)}(r)$ as a regular power series.

At $n=0$, we have $\eta_{(0)}=1$ identically. The asymptotic behaviour of the other fields takes the form
\begin{equation}
\begin{alignedat}{3}
f_{(0)} &= f_\infty + \frac{c_f}{r^2} + \cdots\, , \quad &
h_{(0)} &= h_\infty + \frac{c_h}{r^4} + \cdots\, , \quad &
k_{(0)} &= k_\infty + \frac{c_k}{r^4} + \cdots\, , \\
q_{(0)} &= q_\infty + \frac{c_q}{r^2} + \cdots\, , \quad &
\beta_{(0)} &= 1 + \frac{c_\beta}{r^4} + \cdots\, , \quad &
\gamma_{(0)} &= \gamma_\infty + \frac{c_\gamma}{r^2} + \cdots\, ,
\end{alignedat}
\label{asymp_0th_power_sol_app}
\end{equation}
where the 11 coefficients $(f_\infty,h_\infty,k_\infty,q_\infty,\gamma_\infty,c_f,c_h,c_k,c_q,c_\beta,c_\gamma)$ are not fixed by the asymptotic analysis and need to be determined by matching the bulk profile; for further context see also discussion of \eqref{asymp_0th_power_sol}. The series solution of $g$ is fixed by the equations of motion as
\begin{equation}
g_{(0)} = 1 + \left( \frac{c_f}{f_\infty} + \frac{c_\gamma}{\gamma_\infty} \right) \frac{1}{r^2} + \cdots\, .
\end{equation}

At $n=1$, the ``mass'' $\mu$ for the exponential decay is given by
\begin{equation}
\mu = \sqrt{\frac{k_\infty^2}{\gamma_\infty} - \frac{\left(  4 h_\infty - k_\infty q_\infty\right)^2}{f_\infty}}\, .
\label{asymp_m}
\end{equation}
With this exponent, the power series behaviour of $\eta_{(1)}$ is
\begin{equation}
\eta_{(1)} = c_\eta + \cdots\, ,
\end{equation}
while the rest of the fields vanish at $n=1$: $\hat{X}_{(1)}=0$ for $\hat{X} \equiv (f,g,h,k,q,\beta,\gamma)$. In fact, we find that $\hat{X}_{(n)}=0$ for all odd $n$ in higher orders. This order leaves $c_\eta$ as another coefficient that is not fixed by the asymptotic analysis. Thus we have 12 undetermined parameters before this order.

Coefficients in orders higher than $n$ are completely determined by these 12 coefficients.
At $n=2$, $\eta_{(2)}$ is analytically related to $\eta_{(1)}$ as
\begin{equation}
\eta_{(2)} = \frac{1}{2} \eta_{(1)}^2\, .
\end{equation}
The asymptotic solution hence becomes
\begin{equation}
\eta_{(2)} = \frac{c_\eta^2}{2} + \cdots\, .
\end{equation}
The coefficients for the other variables are also obtained as
\begin{equation}
\begin{split}
f_{(2)} &= \frac{c_\eta^2}{4 \mu^2} \left( 4 h_\infty - k_\infty q_\infty\right)^2 + \cdots\, , \quad
g_{(2)} = \frac{c_\eta^2}{4} + \cdots\, , \quad
h_{(2)} = \frac{4 h_\infty c_\eta^2}{\mu^2} \frac{1}{r^2} + \cdots\, , \\
k_{(2)} &= \frac{4 k_\infty c_\eta^2}{\mu^2} \frac{1}{r^2} + \cdots\, , \quad
q_{(2)} = - \frac{k_\infty c_\eta^2}{4 \gamma_\infty \mu^2} \left( 4 h_\infty - k_\infty q_\infty\right) + \cdots\, , \\
\beta_{(2)} &= -\frac{5 c_\eta^2}{\mu^2} \frac{1}{r^2} + \cdots\, , \quad
\gamma_{(2)} = -\frac{k_\infty^2 c_\eta^2}{4 \mu^2} + \cdots\, .
\end{split}
\end{equation}

In summary, the regular asymptotic expansion in $r\to\infty$ has the following structure:
\begin{align}
\hat{X}(r) &= \hat{X}_{(0)}(r) + \hat{X}_{(2)}(r) \frac{e^{- 2 \mu r}}{r^3} + \cdots\, , \label{asymp_expansion_Xhat}\\
\eta(r) &= 1 + \eta_{(1)}(r) \frac{e^{- \mu r}}{r^{3/2}} + \eta _{(2)}(r) \frac{e^{- 2 \mu r}}{r^3} + \cdots\, , \label{asymp_expansion_eta}
\end{align}
where $\hat{X}_{(n)}(r)$ and $\eta_{(n)}(r)$ can be obtained by solving power series expansion. Solutions for $n \le 2$ are explicitly given above.

Naively, there are quite a few parameters in the series solutions at the horizon and in asymptotic infinity. However, many of these are fixed by either scaling symmetries as we will argue next.

\subsection{Marginal solution}
\label{app:marginal}

The exponential decay \eqref{asymp_expansion_Xhat} or \eqref{asymp_expansion_eta} becomes power law when $\mu \to 0$. This limit occurs when the confinement by the Kaluza-Klein mass becomes absent, and it would lead to a marginally bounded helical solutions (that may be called {\it warm holes} following \cite{Dias:2021vve}). Here, we argue that no such marginally bounded solutions exist, by asymptotic analysis.

If $\mu=0$, the form of the expansion at asymptotic behavior is different from  \eqref{asymp_expansion_Xhat} and \eqref{asymp_expansion_eta}. Solving $\mu=0$ for \eqref{asymp_m} with $f_\infty=\gamma_\infty=1$ and $q_\infty=0$, we obtain $k_\infty = 4 h_\infty$. If this condition is satisfied, the asymptotic series becomes power law. Let us focus on $\eta$ first. If $k_\infty = 4 h_\infty$, instead of \eqref{asymp_expansion_eta}, we have
\begin{equation}
\eta(r) = \eta_+ (r^{-1+\sqrt{\Delta_0}} + \cdots) + \eta_- (r^{-1-\sqrt{\Delta_0}} + \cdots)\, .
\label{eta_power}
\end{equation}
where $\eta_\pm$ are constants and
\begin{equation}
\Delta_0 = 9- 16 \pi G_6 (\mathcal{E}+2\mathcal{P}-\mathcal{T}_z) \, .
\end{equation}
If $\Delta_0>0$, the solution with $r^{-1-\sqrt{\Delta_0}}$ gives a normalizable behavior, whereas the other one is unphysical and removed. However, for all numerical results, we find $\Delta_0<0$, ruling out the existence of normalizable solutions with a power law tail.

\section{Scaling transformations}
\label{app:scaling0}

\subsection{Scaling symmetry}
\label{app:scaling}

The cohomogeneity-1 metric \eqref{screw_metric} has four scaling symmetries in the $(T,Z)$-plane: $Z \to Z+ c_1, \, T \to T+ c_2, \, t \to c_3 T, \, Z \to c_4 Z$. Among these, the second one turns out to be a bit complicated. Therefore, we will instead use the Lorentz boost, which in fact can be formed by combining the four. The transformation rules of the fields for the other three are given by
\begin{align}
Z &\to Z+ c_1\, , \quad h \to h - \frac{c_1}{4} k\, , \quad q \to q-c_1\, ,
\label{scaling1} \\
T &\to c_3 T\, , \quad f \to \frac{f}{c_3^2}\ , \quad h \to \frac{h}{c_3}\, , \quad q \to \frac{q}{c_3}\, ,
\label{scaling3} \\
Z &\to c_4 Z\, , \quad \gamma \to \frac{\gamma}{c_4^2}\, , \quad k \to \frac{k}{c_4}\, , \quad q \to c_4 q\, .
\label{scaling4}
\end{align}

The Lorentz boost in the $(T,Z)$-plane is given by
\begin{equation}
T \to \cosh b \, T - \sinh b \, Z\, , \quad
Z \to \cosh b \, Z - \sinh b \, T\, ,
\label{lol_boost}
\end{equation}
where $b$ is the boost parameter. The boost \eqref{lol_boost} transforms the metric components as
\begin{equation}
\begin{split}
f &\to f \frac{\gamma}{\gamma_b}\ , \quad
\gamma \to \left( \cosh b + q \sinh b \right)^2 \gamma - f \sinh^2 b \equiv \gamma_b\ , \\
h &\to h \cosh b + \frac{k}{4} \sinh b\ , \quad
k \to k \cosh b + 4 h \sinh b\ , \\
q &\to \frac{\gamma}{\gamma_b} \left( q \cosh(2b) + \frac{1}{2} \sinh(2b) \left( 1+q^2-\frac{f}{\gamma} \right) \right)\ .
\end{split}
\label{boost_components}
\end{equation}

\subsection{Scaling of asymptotic coefficients}
\label{app:boost}

We use the scaling symmetries to remove redundancies in the asymptotic spherical gauge solutions, as described next. In our numerical treatment, raw numerical solutions we obtain have $(f_\infty,\gamma_\infty,q_\infty) \neq (1,1,0)$ in general (see appendix~\ref{app:tech}). Using the scaling transformations~(\ref{scaling1}-\ref{scaling4}), we can set the asymptotic values of the metric components as $(f_\infty,\gamma_\infty,q_\infty)=(1,1,0)$. Accordingly, coefficients in the asymptotic expansion of the fields in (\ref{scaling1}-\ref{scaling4}) are also scaled.

We then apply the Lorentz boost \eqref{boost_components}. The boost does not change the leading asymptotic behaviour of $f, \gamma, q$ once $(f_\infty,\gamma_\infty,q_\infty)=(1,1,0)$ are fixed. Meanwhile, their subleading coefficients as well as $h$ and $k$ are affected. Substituting the asymptotic solutions \eqref{asymp_0th_power_sol_app} with $(f_\infty,\gamma_\infty,q_\infty)=(1,1,0)$ to \eqref{boost_components}, we find that the Lorentz boost changes the asymptotic coefficients as
\begin{equation}
\begin{split}
h_\infty &\to h_\infty \cosh b + \frac{k_\infty}{4} \sinh b\, , \quad
k_\infty \to k_\infty \cosh b + 4 h_\infty \sinh b\, , \\
c_f &\to c_f \cosh^2 b - c_\gamma \sinh^2 b - 2 c_q \cosh b \sinh b\, , \\
c_h &\to c_h \cosh b + \frac{c_k}{4} \sinh b\, , \quad
c_k \to c_k \cosh b + 4 c_h \sinh b\, ,\\
c_q &\to c_q \left( \cosh^2 b + \sinh^2 b \right) + \left( c_\gamma - c_f \right) \cosh b \sinh b \equiv c'_q\, , \\
c_\gamma &\to c_\gamma \cosh^2 b - c_f \sinh^2 b + 2 c_q \cosh b \sinh b\, .
\end{split}
\label{boost_series_coeff}
\end{equation}
Simultaneously, coefficients in the horizon expansion are also boosted as
\begin{equation}
\begin{split}
f_1 &\to \frac{f_1}{\xi_b^2}\, , \quad
h_0 \to \frac{k_0}{4} \left( q_0 \cosh b + \sinh b \right)\, , \\
h_1 &\to h_1 \left(\cosh b - \frac{\gamma_0 q_1}{f_1 - \gamma_0 q_0 q_1} \sinh b
+ \frac{f_1 k_0 (\eta_0 - \eta_0^{-1})^2}{r_+^2 \beta_0 g_1 (f_1 - \gamma_0 q_0 q_1)} \sinh b \right)\, , \\
k_0 &\to k_0 \xi_b\, , \quad
q_0 \to \frac{1}{\xi_b} \left( q_0 \cosh b + \sinh b \right)\, , \quad
q_1 \to \frac{1}{\xi_b^2} q_1 - \frac{f_1}{\gamma_0 \xi_b^3} \sinh b\, , \quad
\gamma_0 \to \gamma_0 \xi_b\, ,
\end{split}
\label{boost_series_horizon}
\end{equation}
where we defined $\xi_b \equiv \cosh b + q_0 \sinh b$. These four scalings fix the boundary condition in the asymptotic infinity to be locally flat without boost \eqref{asympcond}.

\section{Technical details for constructing helical black strings in the spherical gauge ansatz}
\label{app:tech}

As discussed in section~\ref{sec:rhser} and appendix~\ref{app:horexp}, 8 parameters $(f_1$, $\eta_0$, $\beta_0$, $h_1$, $k_0$, $\gamma_0$, $q_0$, $q_1)$ are undetermined in the Taylor expansion near the horizon, as well as $r_+$. For efficient numerical calculations, we fix 4 of these coefficients by the scaling symmetries and Lorentz boost.
By the three scalings (\ref{scaling1}-\ref{scaling4}), we can set without loss of generality
\begin{equation}
 f_1=\gamma_0=1\, ,\quad q_0=0\, ,
\label{condbyscaling}
\end{equation}
where $h_0=0$ also follows from \eqref{horizon_4hkq_relation}.
The Lorentz boost~(\ref{boost_components}) is used to adjust the input value of $q_1$.
While $q_1$ can be any value, it is efficient to tune $q_1$ so that $P$ is small because it reduces numerical errors.\footnote{In our numerical calculations in the spherical gauge ansatz, a bare numerical solution has $P \neq 0$ in general. We then apply the Lorentz boost as appendix~\ref{app:boost} to transform the solution to $P=0$. In this procedure, numerical errors are badly enhanced if a solution with large $P$ is unboosted to $P=0$. This can happen if we always fix $q_1$ to some value. Therefore, it turns out effective to adjust $q_1$ beforehand so that numerical calculations are done at small $P$.} The specific procedure to adjust $q_1$ will be explained shortly.
Meanwhile, we can set $r_+=1$ without loss of generality.
In the end of the day, 4 parameters $(\eta_0, \beta_0, h_1, k_1)$ are unfixed in the horizon series expansion.

Among these, we need to fix two by matching the boundary conditions in asymptotic infinity. Practically, we use $(\eta_0,h_1)$ as controllable parameters and determine the other two $(k_0,\beta_0)$ by shooting methods to satisfy the boundary condition $\eta, \beta \to 1$. Thus, helical black strings are a two-parameter family of solutions.

In practice, to find the helical strings in the spherical gauge ansatz~\eqref{screw_metric}, we carry out numerical calculations as follows:
\begin{enumerate}
\item We pick up a MPBS at the onset of instability. From this, a sequence of helical black strings branches off. For numerics, we rescale the MPBS solution to satisfy (\ref{condbyscaling}) by the scaling transformations and read off the rescaled value of $h_1=h_1^\textrm{onset}$. Meanwhile, we have $\eta_0=1$ trivially for the MPBS.
\item We slightly vary the controllable parameters as $(\eta_0,h_1)=(1-\epsilon_\eta,h_1^\textrm{onset}-\epsilon_h)$, where we choose $\epsilon_h=10\epsilon_\eta$, though this can be arbitrary.\footnote{We vary both parameters $(\eta_0,h_1)$ to try to efficiently cover the parameter space.} With these parameters, we tune $(\beta_0, k_1)$ by shooting methods so that $\eta, \beta\to 1$ ($r\to\infty$).
\item The resulting numerical solution has $f_\infty \neq 1$, $\gamma_\infty \neq 1$, and $q_\infty \neq 0$, as well as $P\neq 0$. This bare solution is rescaled by the three scalings~(\ref{scaling1}-\ref{scaling4}) so that $f_\infty=\gamma_\infty=1$ and $q_\infty=0$. After that, it is unboosted with the boost parameter
\begin{equation}
\tanh b = \frac{c_f - c_\gamma + \sqrt{\left( c_f - c_\gamma \right)^2 - 4 c_q}}{2 c_q}\, ,
\label{lol_boost_num_xsol}
\end{equation}
where the quantities on the right hand side have been read out before the boost.
\item We slightly increase the value of $\epsilon_\eta$ and repeat the shooting. For an initial guess of the next step, the previous solution, which satisfies $P=0$, is rescaled (\ref{scaling1}-\ref{scaling4}) such that (\ref{condbyscaling}) is satisfied. Because the solution with $P=0$ is used as the initial guess, the amount of boost $P$ in the next step's bare numerical result can be kept small.
\item We repeat the steps 3 and 4 as long as numerical accuracy is satisfactory.
\end{enumerate}

In this way, we were able to obtain numerical data for $\Omega_H/\Omega_H^\mathrm{ext}|_\textrm{onset} \ge 0.865$. For $\Omega_H/\Omega_H^\mathrm{ext}|_\textrm{onset} < 0.865$, the asymptotic exponent $m$ \eqref{asymp_m} is very small at and around the onset, and numerical calculations are considerably harder. Our numerical results are hence limited to $\Omega_H/\Omega_H^\mathrm{ext}|_\textrm{onset} \ge 0.865$ (as stated in footnote~\ref{foot:NumericConsiderations}), but we believe our data cover a reasonably wide range of the onset frequencies. It appears, however, that the results obtained by adjusting the controllable parameters as stated above did not cover the parameter space nicely, especially in the region where the entropy  $\sigma_H$ is small. Hence, in addition, we did supplemental calculations by changing $(\eta_0,h_1)$ differently to cover a wider parameter range, especially where the entropy is small.

\section{On-shell action and quasi-local stress tensor}
\label{app:counterterm}

\subsection{Quasi-local stress tensor using the counterterm method}
\label{app:stresstensor}

We derive the Brown-York quasi-local stress tensor \cite{Brown:1992br} by regularizing and renormalizing the action in asymptotically flat spacetime. This machinery has been developed in the AdS/CFT duality and known as the ``holographic renormalization'' \cite{Balasubramanian:1999re,deHaro:2000vlm,Bianchi:2001kw}. This idea can also be applied to asymptotically flat spacetime \cite{Kraus:1999di}.

The on-shell Einstein-Hilbert action supplemented with the Gibbons-Hawking term diverges. Therefore, we first regularize it. Let us introduce a cutoff at finite $r=r_\Lambda$. Let $M$ and $\partial M$ denote the regularized spacetime manifold and the cutoff surface. The regularized Einstein-Hilbert action with the Gibbons-Hawking term is given by
\begin{equation}
S_\mathrm{reg} = \frac{1}{16 \pi G_6} \int_M \mathrm{d}^6x \sqrt{-g} R + \frac{1}{8 \pi G_6} \int_{\partial M} \mathrm{d}^5x \sqrt{-\gamma} K\, ,
\label{S_reg}
\end{equation}
where $K \equiv K_{ab} \gamma^{ab}$ is the trace of the extrinsic curvature $K_{ab}$ with respect to the induced metric $\gamma_{ab}$ on $\partial M$, and $a,b$ denote the five-dimensional coordinates other than the radial direction $r$. To be precise, the extrinsic curvature is given by
\begin{equation}
K_{ab} = \frac{1}{2} \delta_a^A \delta_b^B \left( \nabla_A n_B + \nabla_B n_A \right)\, ,
\end{equation}
where $n^A$ is an outward unit normal satisfying $g_{AB} n^A n^B = 1$, and $\gamma^{ab}$ is the pull back of $\gamma^{AB} = g^{AB}-n^A n^B$. Because our metric satisfies $g_{A r}=0$, we can take $n_A \mathrm{d}x^A = 1/\sqrt{g^{rr}} \, \mathrm{d}r$. Because $R=0$, the variation of \eqref{S_reg} is given by the contribution from $\partial M$ as
\begin{equation}
\delta S_\mathrm{reg} =  \frac{1}{8 \pi G_6} \int_{\partial M} \mathrm{d}^5x \sqrt{-\gamma} \, \frac{1}{2} \left( K_{ab} - K \gamma_{ab} \right) \delta \gamma^{ab}\, ,
\label{varS_reg}
\end{equation}
where we dropped total derivative terms for $\partial M$. However, $S_\mathrm{reg}$ diverges in $r_\Lambda \to \infty$, and therefore it needs to be ``renormalised''.

We introduce counterterms to compensate the diverging behaviour of the action. Because the black string extends in one direction, we need to choose the coefficient of the counterterm as if we have a four dimensional cutoff surface. It turns out that the following counterterm cancels the divergence \cite{Kraus:1999di}, and this is sufficient for our purpose:\footnote{Other counterterms were also proposed in \cite{Mann:2005yr}.}
\begin{equation}
S_\mathrm{ct} = - \frac{1}{8 \pi G_6} \int_{\partial M} \mathrm{d}^5x \sqrt{-\gamma} \sqrt{\frac{3}{2} \mathcal{R}}\, ,
\label{S_ct}
\end{equation}
where $\mathcal{R}$ is the Ricci scalar made of $\gamma_{ab}$. The variation is
\begin{equation}
\delta S_\mathrm{ct}
= - \frac{1}{8 \pi G_6} \int_{\partial M} \mathrm{d}^5x \sqrt{-\gamma} \, \frac{1}{2} \sqrt{\frac{3}{2\mathcal{R}}}\left(\mathcal{R}_{ab} -\mathcal{R} \gamma_{ab} \right) \delta \gamma^{ab}\, .
\end{equation}
The renormalised action is given by sending $r_\Lambda \to \infty$ for the sum of \eqref{S_reg} and \eqref{S_ct} as
\begin{equation}
S_\mathrm{ren} = \lim_{r_\Lambda \to \infty} \left(S_\mathrm{reg} + S_\mathrm{ct} \right)\, .
\end{equation}

The renormalised quasi-local stress energy tensor $T_{ab}$ is then given by
\begin{align}
8 \pi G_6 T_{ab}
= - \frac{2}{\sqrt{-\gamma}} \frac{\delta S_\mathrm{ren}}{\delta \gamma^{ab}}
= - K_{ab} + K \gamma_{ab} + \sqrt{\frac{3}{2\mathcal{R}}}\left(\mathcal{R}_{ab} -\mathcal{R} \gamma_{ab} \right)\, .
\end{align}
Using the boundary series expansion~(\ref{asymp_0th_power_sol}), we obtain the quasi-local stress tensor~(\ref{Tij_result_can}).

\section{Horizonless limit of helical black strings}
\label{sec:geon}

In our numerical construction of the helical black strings, we found that these solutions can approach $\sigma_H \to 0$. This raises a question whether a regular horizonless soliton exists, which may be denoted as a Kaluza-Klein geon. In this appendix, we discuss the horizonless limit and collect evidences that regular Kaluza-Klein geons do not seem to be obtained in the zero horizon radius limit of helical black strings. We also argue/speculate that this limit looks more like a pp-wave.

\subsection{Helical black string versus pp-wave}
\label{sec:ppwave}

As we have seen in Fig.~\ref{fig:thermoJ}, the zero entropy $\sigma_H \to 0$ limit of helical strings is accompanied by the limit that the horizon velocity along the string approaches the speed of light $v_H \to 1$. This suggests that this limit may have something to do with a pp-wave.

Let us begin with the boosted Myers-Perry black string in the rotating frame at infinity, which can be obtained by applying the Lorentz boost \eqref{boost_components} to the MPBS \eqref{MPBS_sol}. In the spherical gauge, the metric has the same form as \eqref{screw_metric}, and now the metric components are given by
\begin{equation}
\begin{split}
f &= \frac{\bar{f}}{\gamma}\, , \quad
h = \bar{h} \cosh b\, , \quad
k = 4 \bar{h} \sinh b\, , \\
q &= \cosh b \, \sinh b \, \frac{\bar{f}}{\gamma}\, , \quad
\gamma = 1 + \sinh^2 b \bar{f}\, , \\
\end{split}
\label{boost_MPBS_components}
\end{equation}
where $\bar{f},\bar{h}$ are from the original non-boosted MPBS,
\begin{equation}
\bar{f} = 1 - \frac{2 \mu r^2 }{r^4 + 2 \mu a^2}\, , \quad
\bar{h} = \Omega_H - \frac{2 \mu a}{r^4 \beta}\, .
\end{equation}
Meanwhile, $g,\eta,\beta$ are unaffected by the boost ({\it i.e.}~$g,\beta$ are still given by \eqref{MPBS_sol}, and $\eta=1$).

Taking the scaling horizonless limit of the boosted MPBS gives a pp-wave with rigid rotation. For that, we send $b \to \infty$ together with $r_+ \to 0$, while keeping the rotation finite. In \eqref{boost_MPBS_components}, we rewrite $(b,\Omega_H)$ with new parameters $(c_\ast,\Omega_\ast)$ as
\begin{equation}
b = \log \left( \frac{c_\ast r_+}{2}\right)\, , \quad \Omega_H = c_\ast r_+ \Omega_\ast\, ,
\end{equation}
Then, sending $r_+ \to 0$ while keeping $(c_\ast,\Omega_\ast)$ fixed gives a metric in the form \eqref{screw_metric} with
\begin{equation}
\begin{split}
f(r)&=\frac{c_\ast^2 r^2}{1+c_\ast^2 r^2}\, , \quad g(r)=\eta(r)=\beta(r)=1\, , \quad h(r)=\Omega_\ast\, , \quad k(r)=-4 \Omega_\ast\ ,\\
q(r)&=-\frac{1}{1+c_\ast^2 r^2}=f(r)-1\, , \quad \gamma(r)=1+\frac{1}{c_\ast^2 r^2}=f(r)^{-1}\, .
\end{split}
\label{boost_scalinglimit}
\end{equation}
In this limit, the Kretschmann scalar is identically zero. Hence, the geometry is regular everywhere.

The metric \eqref{boost_scalinglimit} is nothing but a pp-wave with a rigid rotation. In the light-cone coordinates defined as $(u,v)=(T-Z,T+Z)$, the metric is written as
\begin{equation}
\mathrm{d}s^2 = -\mathrm{d}u \, \mathrm{d}v + \frac{\mathrm{d}u^2}{c_\ast^2 r^2} + \mathrm{d}r^2
+ \frac{r^2}{4} \left[ \Sigma_1^2 + \Sigma_2^2 + \left( \Sigma_3 +2 \Omega_\ast \mathrm{d}u \right)^2 \right]\, .
\label{boost_pp_metric}
\end{equation}
The wavefront propagates along $\partial_v$. In this metric, the rigid rotation is redundant and can be absorbed by changing coordinate frames to the non-rotating frame at infinity. This simply can be done by setting $\Omega_\ast=0$, and the plane wave metric of \cite{Horne:1991cn} is reproduced.

\begin{figure}[t]
\centering
\subfigure[$f,q$]{\includegraphics[scale=0.35]{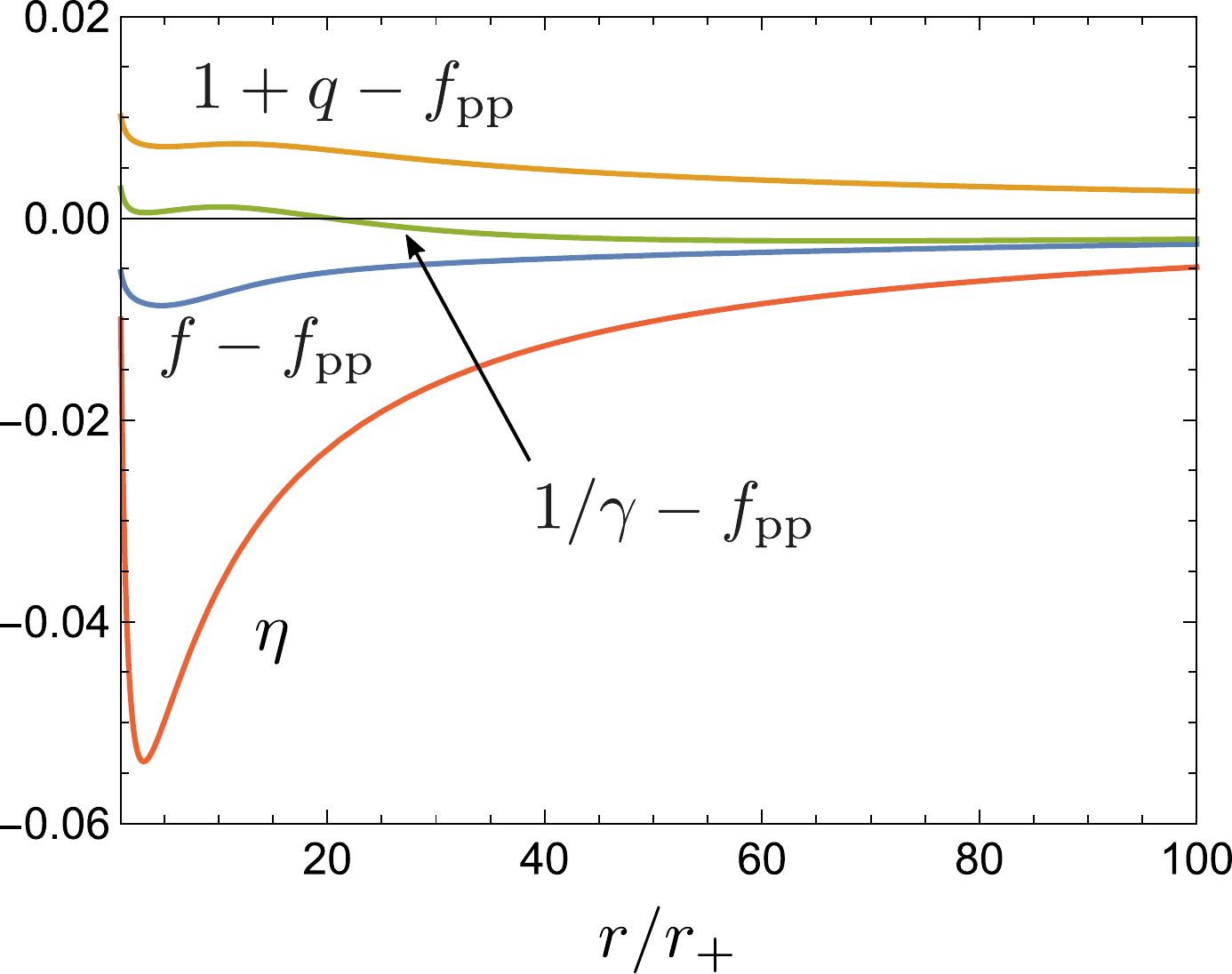}\label{fig:pp_f}}
\subfigure[$\alpha$]{\includegraphics[scale=0.35]{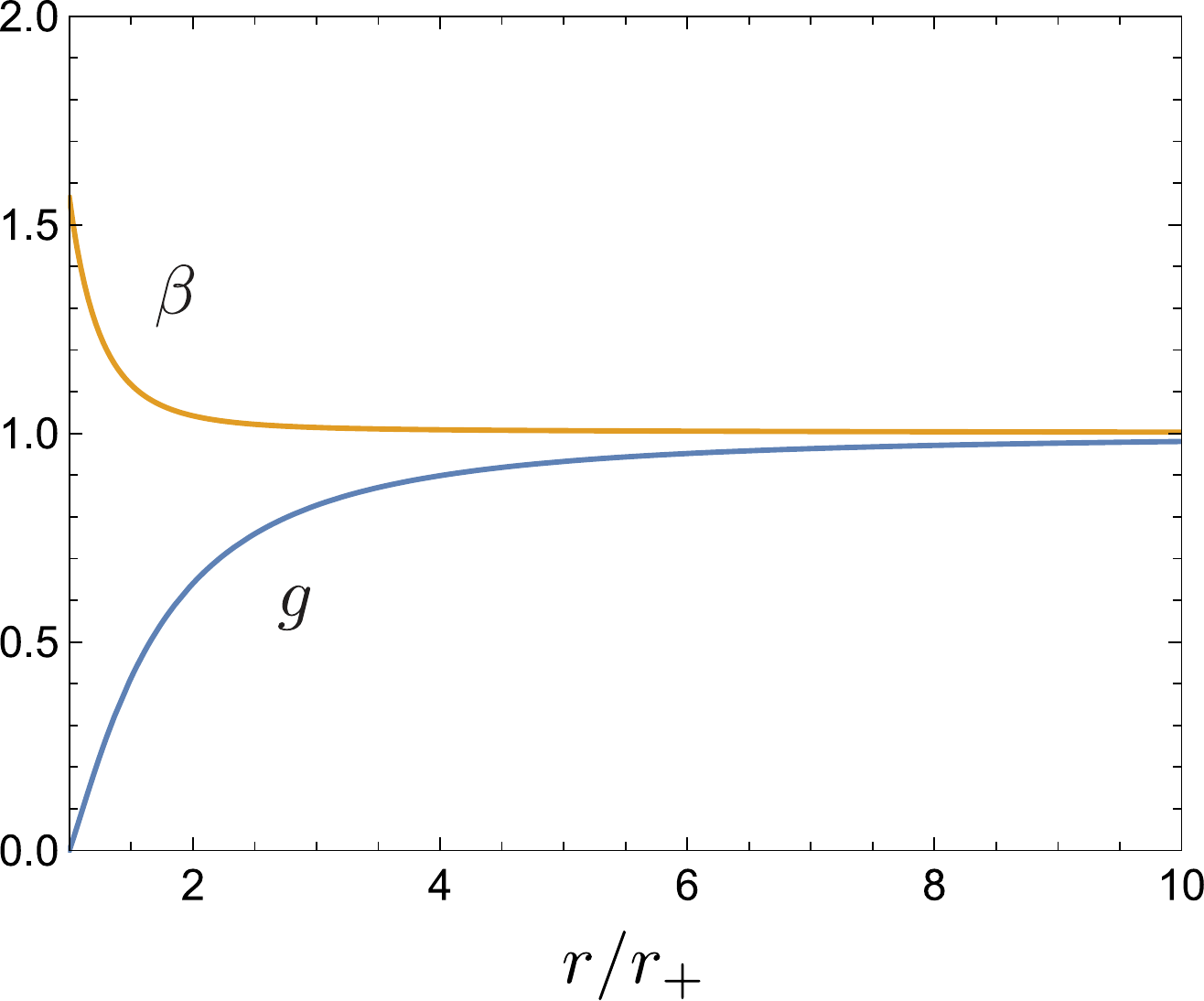}\label{fig:pp_g}}
\subfigure[$g,\beta$]{\includegraphics[scale=0.35]{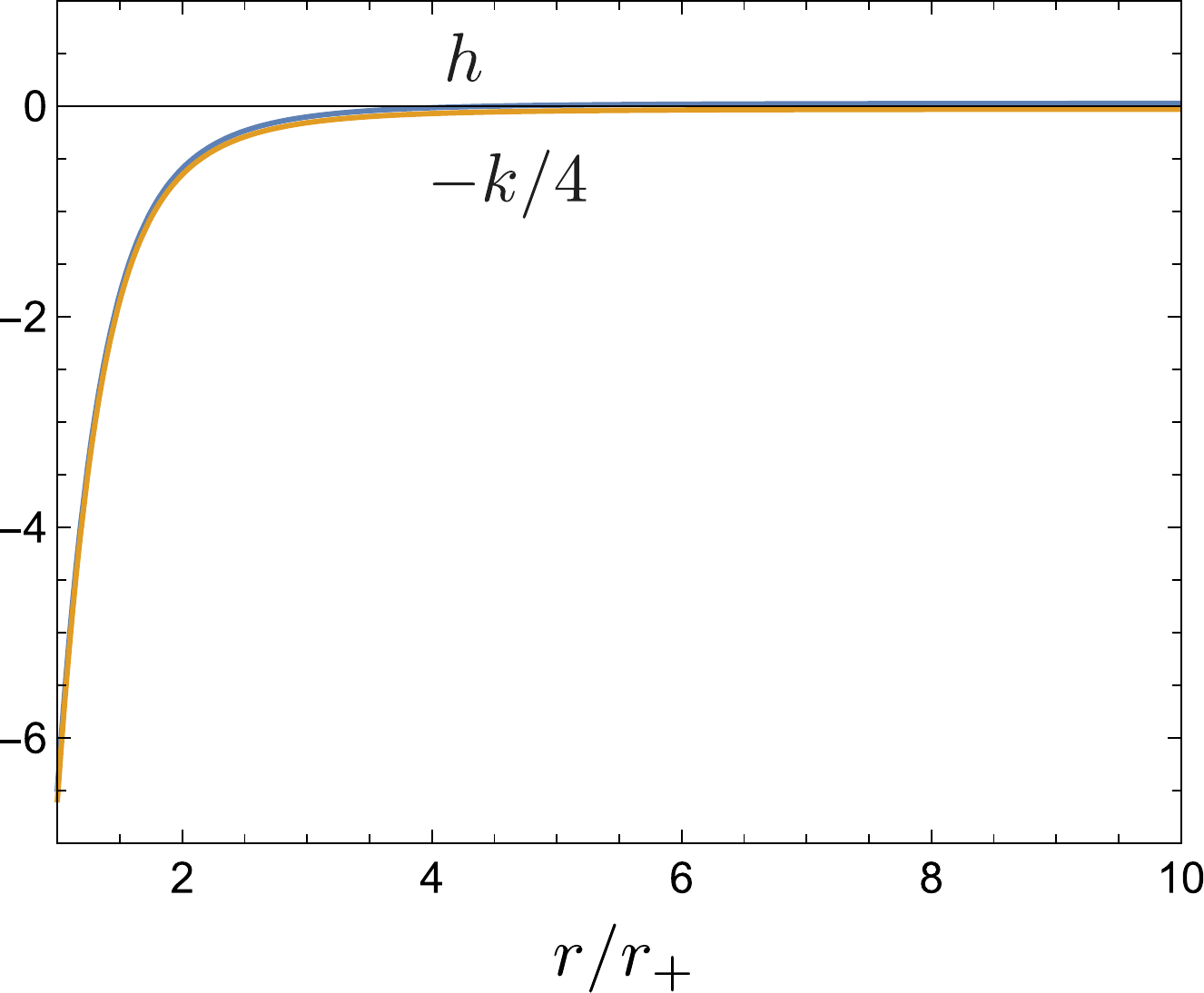}\label{fig:pp_h}}
\caption{Bulk profile for a helical black string with a high velocity along the string, $(G_6 \mathcal{E},G_6\mathcal{J},G_6\sigma_H,\omega_H,v_H) \simeq (0.093,0.030,0.00038,3.12,0.98)$, which has $\eta_0=0.99$. In~\ref{fig:pp_f}, the difference from $f_{\mathrm{pp}} \equiv f$ of the pp-wave with $c_\ast=0.073$ \eqref{boost_scalinglimit} is plotted.}
\label{fig:bulk_pp}
\end{figure}

So, for $v_H \simeq 1$ and $\sigma_H \simeq 0$, the helical black string is comparable to a pp-wave. In Fig.~\ref{fig:bulk_pp}, we show the bulk profiles of the helical black string for an illustrative case with $(G_6 \mathcal{E},G_6\mathcal{J},G_6\sigma_H,\omega_H,v_H) \simeq (0.093,0.030,0.00038,3.12,0.98)$, which also has $\eta_0=0.99$. As seen in Fig.~\ref{fig:pp_f}, $(f,q,\gamma)$ are indeed quite close to the pp-wave profile $f_{\mathrm{pp}} \equiv f$ with $c_*=0.073$.

From our numerical results, we find that the limit $v_H \to 1$, $\sigma_H \to 0$, $\omega_H \to \pi$ corresponds to the limit of vanishing confinement $m \to 0$, where there are no bounded solutions as discussed in appendix~\ref{app:marginal}. It is then likely that a solution with $\eta(r) =1$ is the only possibility in the limit we are considering. While the helical black string profile shown in Fig.~\ref{fig:bulk_pp} has finite deformation $\eta(r) \neq 1$, such a deformation should probably disappear and one should approach $\eta(r) \to 1$ as $v_H \to 1$. If so, the  $v_H \simeq 1$ and $\sigma_H \simeq 0$ limit of a helical black string might simply be a pp-wave.

\subsection{Asymptotic analysis around the origin for Kaluza-Klein geons}

In conjunction with the discussion in appendix~\ref{sec:ppwave}, here we describe our attempts to directly obtain nontrivial regular Kaluza-Klein geons in the horizonless limit of the helical black strings.

First of all, we did not find any nontrivial normal modes for the perturbation \eqref{MPBS_fluc2} on the $\mathcal{M}^{1,4}\times S^1$ space or pp-wave \eqref{boost_pp_metric} backgrounds. This already suggests that Kaluza-Klein geons with the same symmetries of the helical string would not be a nonlinear back-reaction of normal modes of such backgrounds.

However, there is still the possibility that horizonless Kaluza-Klein solutions exist but are not perturbatively connected to the above backgrounds. To explore this possibility, we assume $r_+=0$ and consider the asymptotic series near $r=0$.

Requiring regularity at $r=0$, we can obtain the following asymptotic expansion:
\begin{equation}
\begin{alignedat}{2}
f(r)&=f_0 + O(r^6)\ , \quad &
g(r)&=1 - \beta_2 r^2 + O(r^4)\ , \\
h(r)&=h_0 + \frac{h_0 \eta_2^2}{2} r^4 + O(r^6)\ , \quad &
k(r)&=k_0 + \frac{k_0 \eta_2^2}{2} r^4 + O(r^6)\ , \\
q(r)&=q_0 + O(r^6)\ , \quad &
\eta(r)&=1 + \eta_2 r^2 + O(r^4)\ , \\
\beta(r)&=1 + \beta_2 r^2 + O(r^4)\ , \quad &
\gamma(r)&=\gamma_0 + O(r^6)\ .
\end{alignedat}
\label{geoncenterbc}
\end{equation}
This has 7 independent parameters $(f_0,h_0,k_0,q_0,\eta_2,\beta_2,\gamma_0)$. We can use one of them to control the overall scale (the Kaluza-Klein compactification scale $L$). Meanwhile, the conditions that need be satisfied in $r\to\infty$ are 6. Therefore,  if a geon solution exists, it should be a 0-parameter family and not continuously connected to the MPBS. Numerically, we found only the MPBS as the solution when the boundary condition \eqref{geoncenterbc} is imposed. That is, $\eta_2=0$. We did not find discrete solutions separated from MPBS.

Instead, if we assume the boundary condition that is the same as the pp-wave ({\it i.e.}~$f(r) \sim r^2$ and $\gamma(r) \sim r^{-2}$ as $r \to 0$), we obtain
\begin{equation}
\begin{alignedat}{2}
f(r)&=f_2 r^2 -\gamma_0 q_2^2 + O(r^6)\ , \quad &
g(r)&=1 - \beta_2 r^2 + O(r^4)\ , \\
h(r)&=\frac{k_0 q_0}{4} + \frac{k_0 q_0 \eta_2^2}{8} r^4 + O(r^6)\ , \quad &
k(r)&=k_0 + \frac{k_0 \eta_2^2}{2} r^4 + O(r^6)\ , \\
q(r)&=q_0 + q_2 r^2 + O(r^4)\ , \quad &
\eta(r)&=1 + \eta_2 r^2 + O(r^4)\ , \\
\beta(r)&=1 + \beta_2 r^2 + O(r^4)\ , \quad &
\gamma(r)&=\frac{f_2}{q_2^2} \frac{1}{r^2} + \gamma_0 + O(r^2)\ .
\end{alignedat}
\end{equation}
This is also specified by 7 parameters $(f_2,k_0,q_0,q_2,\eta_2,\beta_2,\gamma_0)$. One of them is used to set the overall scale, and we end up with a 0-parameter family again. This no-go result is consistent with the fact that, in our numerical search, we only find the pp-wave \eqref{boost_scalinglimit}.

\bibliography{bib_c1bs}

\end{document}